\documentclass[fleqn,usenatbib]{mnras}

\usepackage[margin=5pt]{subfig}
\usepackage{comment}
\usepackage{longtable}
\usepackage{caption}
\usepackage{threeparttable}
\usepackage{booktabs}
\usepackage[figuresright]{rotating}
\usepackage{textcase}
\newcommand{\be}{\begin{equation}}
\newcommand{\ee}{\end{equation}}
\newcommand{\bea}{\begin{eqnarray}}
\newcommand{\eea}{\end{eqnarray}}
\newcommand{\hunit}{$\rm{km \ s^{-1} \ Mpc^{-1}}$}
\newcommand{\lcdm}{$\Lambda$CDM}
\newcommand{\pcdm}{$\phi$CDM}
\usepackage{array}
\makeatletter
\newcommand{\thickhline}{%
    \noalign {\ifnum 0=`}\fi \hrule height 1pt
    \futurelet \reserved@a \@xhline
}
\newcolumntype{"}{@{\hskip\tabcolsep\vrule width 1pt\hskip\tabcolsep}}
\makeatother

\newcommand{\hiig}{H\,\textsc{ii}G}
\newcommand{\hii}{H\,\textsc{ii}}
\newcommand{\Om}{\Omega_{m0}}
\newcommand{\Ok}{\Omega_{k0}}
\newcommand{\wX}{w_{\rm X}}
\newcommand{\om}{$\Omega_{m0}$}
\newcommand{\ok}{$\Omega_{k0}$}
\newcommand{\wx}{$w_{\rm X}$}
\newcommand{\mq}{Mg\,\textsc{ii} QSO}
\newcommand{\obh}{\Omega_{b}h^2}
\newcommand{\och}{\Omega_{c}h^2}
\newcommand{\onh}{\Omega_{\nu}h^2}
\newcommand{\obhs}{$\Omega_{b}h^2$}
\newcommand{\ochs}{$\Omega_{c}h^2$}

\usepackage[T1]{fontenc}
\usepackage{scalerel}
\usepackage{tikz}
\usetikzlibrary{svg.path}

\definecolor{orcidlogocol}{HTML}{A6CE39}
\tikzset{
  orcidlogo/.pic={
    \fill[orcidlogocol] svg{M256,128c0,70.7-57.3,128-128,128C57.3,256,0,198.7,0,128C0,57.3,57.3,0,128,0C198.7,0,256,57.3,256,128z};
    \fill[white] svg{M86.3,186.2H70.9V79.1h15.4v48.4V186.2z}
                 svg{M108.9,79.1h41.6c39.6,0,57,28.3,57,53.6c0,27.5-21.5,53.6-56.8,53.6h-41.8V79.1z M124.3,172.4h24.5c34.9,0,42.9-26.5,42.9-39.7c0-21.5-13.7-39.7-43.7-39.7h-23.7V172.4z}
                 svg{M88.7,56.8c0,5.5-4.5,10.1-10.1,10.1c-5.6,0-10.1-4.6-10.1-10.1c0-5.6,4.5-10.1,10.1-10.1C84.2,46.7,88.7,51.3,88.7,56.8z};
  }
}

\newcommand\orcidicon[1]{\href{https://orcid.org/#1}{\mbox{\scalerel*{
\begin{tikzpicture}[yscale=-1,transform shape]
\pic{orcidlogo};
\end{tikzpicture}
}{|}}}}

\usepackage{hyperref} %<--- Load after everything else

\DeclareRobustCommand{\VAN}[3]{#2}
\let\VANthebibliography\thebibliography
\def\thebibliography{\DeclareRobustCommand{\VAN}[3]{##3}\VANthebibliography}

\usepackage{graphicx}	% Including figure files
\usepackage{amsmath}	% Advanced maths commands
\usepackage{amssymb}	% Extra maths symbols
\usepackage{fnpct}
\title[Non-CMB cosmological parameter constraints]{Using lower-redshift, non-CMB, data to constrain the Hubble constant and other cosmological parameters}

 \author[S. Cao \& B. Ratra]{
 Shulei Cao$^{\orcidicon{0000-0003-2421-7071}}$,$^{1}$\thanks{E-mail: shulei@phys.ksu.edu}
 Bharat Ratra$^{\orcidicon{0000-0002-7307-0726}1}$\thanks{E-mail: ratra@phys.ksu.edu}
 \\
 $^{1}$Department of Physics, Kansas State University, 116 Cardwell Hall, Manhattan, KS 66506, USA\\
 }

\date{Accepted XXX. Received YYY; in original form ZZZ} 

\pubyear{2022}

\begin{document}
\label{firstpage}
\pagerange{\pageref{firstpage}--\pageref{lastpage}}
\maketitle

\begin{abstract}
We use updated Hubble parameter and baryon acoustic oscillation data, as well as other lower-redshift Type Ia supernova, Mg\,\textsc{ii} reverberation-measured quasar, quasar angular size, H\,\textsc{ii} starburst galaxy, and Amati-correlated gamma-ray burst data, to jointly constrain cosmological parameters in six cosmological models. The joint analysis provides model-independent determinations of the Hubble constant, $H_0=69.7\pm1.2$ $\rm{km \ s^{-1} \ Mpc^{-1}}$, and the current non-relativistic matter density parameter, $\Omega_{m0}=0.295\pm0.017$. These error bars are factors of 2.2 and 2.3 larger than the corresponding error bars in the flat \lcdm\ model from \textit{Planck} TT,TE,EE+lowE+lensing cosmic microwave background anisotropy data. Based on the deviance information criterion (DIC), the flat \lcdm\ model is most favored but mild dark energy dynamics and a little spatial curvature are not ruled out.
\end{abstract}


\begin{keywords}
cosmological parameters -- dark energy -- cosmology: observations -- gamma-ray bursts
\end{keywords}
%%%%%%%%%%%%%%%%%%%%%%%%%%%%%%%%%%%%%%%%%%%%%%%%%%

%%%%%%%%%%%%%%%%% BODY OF PAPER %%%%%%%%%%%%%%%%%%

\section{Introduction} \label{sec:intro}

The expansion of the Universe is currently accelerating. This is well-supported by many observations but the underlying theory remains obscure. If general relativity is valid on cosmological scales, a dark energy that has negative pressure is thought to be responsible for the accelerated cosmological expansion. In the well-known spatially-flat \lcdm\ model \citep{peeb84}, dark energy is a cosmological constant $\Lambda$ and contributes $\sim70\%$ of the current cosmological energy budget \citep[see, e.g.][]{Farooq_Ranjeet_Crandall_Ratra_2017, scolnic_et_al_2018, planck2018b, eBOSS_2020}. However, potential observational discrepancies \citep[see, e.g.][]{DiValentinoetal2021a,PerivolaropoulosSkara2021,Abdallaetal2022} motivate consideration of other cosmological models besides flat \lcdm. In our analyses here we also allow for non-zero spatial curvature\footnote{The \textit{Planck} TT,TE,EE+lowE+lensing cosmic microwave background (CMB) anisotropy data favor positive spatial curvature over flatness \citep{planck2018b}.} as well as dark energy dynamics.

Many observations have been used to compare the goodness of fit of cosmological models and determine cosmological parameter constraints. These include CMB anisotropy data \citep[see, e.g.][]{planck2018b} that largely probe the high-redshift, $z\sim1100$, Universe, as well as lower-$z$ cosmological measurements that we make use of here, such as reverberation-measured $\mathrm{H}\beta$ quasar (QSO) and Mg\,\textsc{ii} QSO observations that reach to $z \sim 1.9$ \citep[see, e.g.][]{Czernyetal2021, Zajaceketal2021, Yuetal2021, Khadkaetal_2021a, Khadkaetal2021c}\footnote{Current $\mathrm{H}\beta$ QSO data probe to $z \sim 0.9$ and the resulting cosmological parameter constraints from these data are in $\sim 2\sigma$ tension with those from better-established cosmological probes \citep{Khadkaetal2021c} so we do not use these data in our analyses here.}, Hubble parameter [$H(z)$] data that reach to $z\sim2$ \citep[see, e.g.][]{moresco_et_al_2016, Farooq_Ranjeet_Crandall_Ratra_2017, Ryanetal2019,  CaoRyanRatra2022}, type Ia supernova (SN Ia) observations that reach to $z\sim 2.3$ \citep[see, e.g.][]{scolnic_et_al_2018, DES_2019d}, baryon acoustic oscillation (BAO) measurements that reach to $z\sim 2.3$ \citep[see, e.g.][]{eBOSS_2020, CaoRyanRatra2022}, \hii\ starburst galaxy apparent magnitude data that reach to $z \sim 2.4$ \citep[see, e.g.][]{Mania_2012, Chavez_2014, GM2021, CaoRyanRatra2022, Johnsonetal2022, Mehrabietal2022}, QSO angular size (QSO-AS) measurements that reach to $z \sim 2.7$ \citep[see, e.g.][]{Cao_et_al2017b, Ryanetal2019, CaoRyanRatra2020, CaoRyanRatra2022, Zhengetal2021, Lian_etal_2021}, QSO flux observations that reach to $z \sim 7.5$ \citep{RisalitiLusso2015, RisalitiLusso2019, KhadkaRatra2020a, KhadkaRatra2020b, KhadkaRatra2021, KhadkaRatra2022, Lussoetal2020, Yangetal2020, ZhaoXia2021, Lietal2021, Lian_etal_2021, Luongoetal2021, Rezaeietal2022, DainottiBardiacchi2022}\footnote{We do not use these data in this paper since the latest \cite{Lussoetal2020} QSO flux compilation assumes a UV--X-ray correlation model that is invalid above $z \sim 1.5-1.7$ \citep{KhadkaRatra2021, KhadkaRatra2022}.}, and gamma-ray burst (GRB) data that reach to $z \sim 8.2$ \citep[see, e.g.][]{Wang_2016, Wangetal_2021, Dainottietal2016, Dainottietal2017, Dainottietal2020, Dirirsa2019, Amati2019, KhadkaRatra2020c, Huetal_2021, Daietal_2021, Demianskietal_2021, Khadkaetal_2021b, LuongoMuccino2021, CaoKhadkaRatra2021, CaoDainottiRatra2022, CaoDainottiRatra2022b, Liuetal2022, DainottiNielson2022}\footnote{Only a subset containing 118 Amati-correlated GRBs are suitable for cosmological purposes \citep{KhadkaRatra2020c, Caoetal_2021, Khadkaetal_2021b}, and these are the Amati-correlated GRBs we use in our analyses here.}.

In this paper, we use most of the aforementioned non-CMB data sets to jointly constrain cosmological parameters. In \cite{CaoRyanRatra2021}, by using $H(z)$ + BAO + SN data (SN refers to Pantheon and DES-3yr SN Ia data, discussed in Sec.\ \ref{sec:data} below), we estimated summary values of the current non-relativistic matter density parameter $\Om=0.294\pm0.020$ and the Hubble constant $H_0=68.8\pm1.8$ \hunit. In \cite{CaoRyanRatra2022}, by using $H(z)$ + BAO + SN + QSO-AS + \hiig\ data, summary values of $\Om=0.293\pm0.021$ and $H_0=69.7\pm1.2$ \hunit\ were obtained. Compared to our earlier analysis, the addition of QSO-AS and \hiig\ data results in similar constraints on \om\ with a slightly larger $1\sigma$ uncertainty and more restrictive ($1\sigma$ uncertainty reduced by 50\%) $H_0$ constraints, with a higher central value of $H_0$ ($0.42\sigma$ higher). 

In our analysis here we improve on our earlier work by more correctly accounting for the neutrinos. We also use updated BAO and $H(z)$ data and now also include \mq\ and A118 GRB data.\footnote{We also examined constraints from mutually consistent Platinum + A101 GRB data used in \cite{CaoDainottiRatra2022} and jointly analyzed them with QSO-AS, \hiig, and \mq\ data. Cosmological constraints from the joint QSO-AS + \hiig\ + \mq\ + Platinum + A101 data are similar to those from the QSO-AS + \hiig\ + \mq\ + A118 data, so we decided to perform further analyses with the latter that constrain fewer non-cosmological parameters.} Here the joint analyses of $H(z)$ + BAO + SN + QSO-AS + \hiig\ + \mq\ + A118 data provide model-independent values of $\Om=0.295\pm0.017$ and $H_0=69.7\pm1.2$ \hunit. Our $H_0$ measurement is in better agreement with the median statistics $H_0$ estimate of \cite{chenratmed} than with the local expansion rate $H_0$ estimate of \cite{Riess_2021}. Flat \lcdm\ is favored the most, but mild dark energy dynamics or a little spatial curvature energy density is not ruled out. Although here we use updated $H(z)$ and BAO data, and add \mq\ and A118 data, the constraint on $H_0$ is identical to that of \cite{CaoRyanRatra2022}\footnote{\mq\ and A118 data do not have the power to constrain $H_0$ and updated $H(z)$ and BAO data we use here provide similar constraints to those from older BAO and $H(z)$ data.}, whereas the new constraint on \om\ is more restrictive ($1\sigma$ uncertainty reduced by $\sim24$\%) and $\sim0.15\sigma$ higher.

This paper is organized as follows. In Sec.\ \ref{sec:model} we introduce the cosmological models/parametrizations used in our analyses. In Sec.\ \ref{sec:data} we describe the data sets used in our analyses, with the methods we use summarized in Sec.\ \ref{sec:analysis}. We discuss our cosmological parameter constraints results in Sec.\ \ref{sec:results} and summarize our conclusions in Sec.\ \ref{sec:conc}.

\section{Cosmological models}
\label{sec:model}

In this paper, we use various combinations of data to constrain cosmological model parameters in six spatially-flat and non-flat dark energy cosmological models.\footnote{For recent determinations of constraints on spatial curvature, see \citet{Chen_et_al_2016}, \citet{Ranaetal2017}, \citet{Oobaetal2018a, Oobaetal2018b}, \citet{Yuetal2018}, \citet{ParkRatra2019a, ParkRatra2019b}, \citet{Wei2018}, \citet{DESCollaboration2019}, \citet{Lietal2020}, \citet{Handley2019}, \citet{EfstathiouGratton2020}, \citet{DiValentinoetal2021b}, \citet{Vagnozzietal2020, Vagnozzietal2021}, \citet{KiDSCollaboration2021}, \citet{ArjonaNesseris2021}, \citet{Dhawanetal2021}, \citet{Renzietal2021}, \citet{Gengetal2022}, \citet{WeiMelia2022}, \citet{MukherjeeBanerjee2022}, and references therein.} Using a number of  different models allows us to determine which results are less dependent on the model used to derive them. The expansion rate, $E(z, \textbf{\emph{p}})$, as a function of redshift $z$ and the cosmological parameters $\textbf{\emph{p}}$ in a given cosmological model, is defined as $E(z, \textbf{\emph{p}})\equiv H(z, \textbf{\emph{p}})/H_0$, with $H(z, \textbf{\emph{p}})$ being the Hubble parameter. The expansion rate is used to compute cosmological-parameter-dependent predictions in the cosmological models we study. In these cosmological models, as in \cite{CaoDainottiRatra2022}, we assume one massive and two massless neutrino species, with the effective number of relativistic neutrino species $N_{\rm eff} = 3.046$ and the total neutrino mass $\sum m_{\nu}=0.06$ eV. Therefore, here the current value of the non-relativistic neutrino physical energy density parameter, $\onh=\sum m_{\nu}/(93.14\ \rm eV)$, is not a free parameter, and along with the current values of the observationally-constrained baryonic (\obhs) and cold dark matter (\ochs) physical energy density parameters, \om\ is derived as $\Om = (\onh + \obh + \och)/{h^2}$, where $h$ is the Hubble constant in units of 100 \hunit.

In the \lcdm\ models the expansion rate function is
\be
\label{eq:EzL}
    E(z, \textbf{\emph{p}}) = \sqrt{\Om\left(1 + z\right)^3 + \Ok\left(1 + z\right)^2 + \Omega_{\Lambda}},
\ee
where $\Omega_{\Lambda} = 1 - \Om - \Ok$ is the cosmological constant dark energy density parameter, $\Ok$ is the current value of the spatial curvature energy density parameter, and $\Ok = 0$ implies flat spatial hypersurfaces. The cosmological parameters $\textbf{\emph{p}}=\{H_0, \obh\!, \och\}$ and $\textbf{\emph{p}}=\{H_0, \obh\!, \och\!, \Ok\}$ are constrained in the flat and non-flat \lcdm\ models, respectively. Note that when $H_0$ and \obhs\ are fixed in analyses of some of the data sets we use, $\textbf{\emph{p}}$ change accordingly.

In the XCDM parametrizations, 
\be
\label{eq:EzX}
\resizebox{0.475\textwidth}{!}{%
    $E(z, \textbf{\emph{p}}) = \sqrt{\Om\left(1 + z\right)^3 + \Ok\left(1 + z\right)^2 + \Omega_{{\rm X}0}\left(1 + z\right)^{3\left(1 + \wX\right)}},$%
    }
\ee
where \wx\ is the X-fluid equation of state parameter, and $\Omega_{{\rm X}0} = 1 - \Om - \Ok$ is the current value of the X-fluid dynamical dark energy density parameter. The cosmological parameters $\textbf{\emph{p}}=\{H_0, \obh\!, \och\!, \wX\}$ and $\textbf{\emph{p}}=\{H_0, \obh\!, \och\!, \wX, \Ok\}$ are constrained in the flat and non-flat XCDM parametrizations, respectively. When $\wX = -1$ the XCDM parametrization reduces to the \lcdm\ model.

In the \pcdm\ models \citep{peebrat88,ratpeeb88,pavlov13}\footnote{For recent determinations of constraints on \pcdm\, see \cite{Zhaietal2017}, \cite{ooba_etal_2018b, ooba_etal_2019}, \cite{park_ratra_2018, park_ratra_2019b, park_ratra_2020}, \cite{Sangwanetal2018}, \cite{SolaPercaulaetal2019}, \citet{Singhetal2019}, \cite{UrenaLopezRoy2020}, \cite{SinhaBanerjee2021}, \cite{Xuetal2021}, \cite{deCruzetal2021}, \cite{Jesusetal2021}, and references therein.},
\be
\label{eq:Ezp}
    E(z, \textbf{\emph{p}}) = \sqrt{\Om\left(1 + z\right)^3 + \Ok\left(1 + z\right)^2 + \Omega_{\phi}(z,\alpha)},
\ee
where
\be
\label{Op}
\Omega_{\phi}(z,\alpha)=\frac{1}{6H_0^2}\bigg[\frac{1}{2}\dot{\phi}^2+V(\phi)\bigg],
\ee
is the scalar field ($\phi$) dynamical dark energy density parameter and is determined by numerically solving the Friedmann equation \eqref{eq:Ezp} and the equation of motion of the scalar field
\be
\label{em}
\ddot{\phi}+3H\dot{\phi}+V'(\phi)=0.
\ee 
An inverse power-law scalar field potential energy density is assumed as
\be
\label{PE}
V(\phi)=\frac{1}{2}\kappa m_p^2\phi^{-\alpha}.
\ee
In the preceding equations an overdot and a prime denote a derivative with respect to time and $\phi$, respectively, $m_p$ is the Planck mass, $\alpha$ is a positive constant (when $\alpha=0$ \pcdm\ reduces to \lcdm), and $\kappa$ is a constant that is determined by the shooting method in the Cosmic Linear Anisotropy Solving System (\textsc{class}) code \citep{class}. The cosmological parameters $\textbf{\emph{p}}=\{H_0, \obh\!, \och\!, \alpha\}$ and $\textbf{\emph{p}}=\{H_0, \obh\!, \och\!, \alpha, \Ok\}$ are constrained in the flat and non-flat \pcdm\ models, respectively.

\begin{table}
\centering
\scriptsize
\begin{threeparttable}
\caption{Updated $H(z)$ data.}\label{tab:hz}
\setlength{\tabcolsep}{5mm}{
\begin{tabular}{lcc}
\toprule
$z$ & $H(z)$\tnote{a} & Reference\\
\midrule
0.07 & $69.0\pm19.6$ & \cite{73}\\
0.09 & $69.0\pm12.0$ & \cite{69}\\
0.12 & $68.6\pm26.2$ & \cite{73}\\
0.17 & $83.0\pm8.0$ & \cite{69}\\
0.179 & $75.0\pm4.0$ & \cite{70}\\
0.199 & $75.0\pm5.0$ & \cite{70}\\
0.2 & $72.9\pm29.6$ & \cite{73}\\
0.27 & $77.0\pm14.0$ & \cite{69}\\
0.28 & $88.8\pm36.6$ & \cite{73}\\
0.352 & $83.0\pm14.0$ & \cite{70}\\
0.3802 & $83.0\pm13.5$ &  \cite{moresco_et_al_2016}\\
0.4 & $95.0\pm17.0$ & \cite{69}\\
0.4004 & $77.0\pm10.2$ &  \cite{moresco_et_al_2016}\\
0.4247 & $87.1\pm11.2$ &  \cite{moresco_et_al_2016}\\
0.4497 & $92.8\pm12.9$ &  \cite{moresco_et_al_2016}\\
0.47 & $89.0\pm50.0$ & \cite{15}\\
0.4783 & $80.9\pm9.0$ &  \cite{moresco_et_al_2016}\\
0.48 & $97.0\pm62.0$ & \cite{71}\\
0.593 & $104.0\pm13.0$ & \cite{70}\\
0.68 & $92.0\pm8.0$ & \cite{70}\\
0.75 & $98.8\pm33.6$ & \cite{Borghi_etal_2022}\\
0.781 & $105.0\pm12.0$ & \cite{70}\\
0.875 & $125.0\pm17.0$ & \cite{70}\\
0.88 & $90.0\pm40.0$ & \cite{71}\\
0.9 & $117.0\pm23.0$ & \cite{69}\\
1.037 & $154.0\pm20.0$ & \cite{70}\\
1.3 & $168.0\pm17.0$ & \cite{69}\\
1.363 & $160.0\pm33.6$ & \cite{72}\\
1.43 & $177.0\pm18.0$ & \cite{69}\\
1.53 & $140.0\pm14.0$ & \cite{69}\\
1.75 & $202.0\pm40.0$ & \cite{69}\\
1.965 & $186.5\pm50.4$ & \cite{72}\\
\bottomrule
\end{tabular}}
\begin{tablenotes}[flushleft]
\item[a] \hunit.
\end{tablenotes}
\end{threeparttable}
\end{table}

\section{Data}
\label{sec:data}

In this paper we use updated $H(z)$ and BAO data, as well as other data sets, to constrain cosmological parameters. These are summarized next.

\begin{table}
\centering
\scriptsize
\begin{threeparttable}
\caption{Updated BAO data.}\label{tab:bao}
\setlength{\tabcolsep}{1.3mm}{
\begin{tabular}{lccc}
\toprule
$z$ & Measurement\tnote{a} & Value & Reference\\
\midrule
$0.122$ & $D_V\left(r_{s,{\rm fid}}/r_s\right)$ & $539\pm17$ & \cite{Carter_2018}\\
$0.38$ & $D_M/r_s$ & 10.23406 & \cite{eBOSSG_2020}\tnote{b}\\
$0.38$ & $D_H/r_s$ & 24.98058 & \cite{eBOSSG_2020}\tnote{b}\\
$0.51$ & $D_M/r_s$ & 13.36595 & \cite{eBOSSG_2020}\tnote{b}\\
$0.51$ & $D_H/r_s$ & 22.31656 & \cite{eBOSSG_2020}\tnote{b}\\
$0.698$ & $D_M/r_s$ & 17.85823691865007 & \tnote{c}\\
$0.698$ & $D_H/r_s$ & 19.32575373059217 & \tnote{c}\\
$0.81$ & $D_A/r_s$ & $10.75\pm0.43$ & \cite{DES_2019b}\\
$1.48$ & $D_M/r_s$ & 30.6876 & \tnote{d}\\
$1.48$ & $D_H/r_s$ & 13.2609 & \tnote{d}\\
$2.334$ & $D_M/r_s$ & 37.5 & \tnote{e}\\
$2.334$ & $D_H/r_s$ & 8.99 & \tnote{e}\\
\bottomrule
\end{tabular}}
\begin{tablenotes}[flushleft]
\item[a] $D_V$, $r_s$, $r_{s, {\rm fid}}$, $D_M$, $D_H$, and $D_A$ have units of Mpc.
\item[b] The four measurements from \cite{eBOSSG_2020} are correlated; see equation \eqref{CovM2} for their correlation matrix.
\item[c] The two measurements from \cite{eBOSSG_2020} and \cite{eBOSSL_2021} are correlated; see equation \eqref{CovM3} for their correlation matrix.
\item[d] The two measurements from \cite{eBOSSQ_2020} and \cite{eBOSSQ_2021} are correlated; see equation \eqref{CovM4} for their correlation matrix.
\item[e] The two measurements from \cite{duMas2020} are correlated; see equation \eqref{CovM1} for their correlation matrix.
\end{tablenotes}
\end{threeparttable}
\end{table}

\begin{itemize}

\item[]{$\textbf{ \emph{H(z)}}$ \bf data}. There are 32 $H(z)$ measurements listed in Table \ref{tab:hz}, spanning the redshift range $0.07 \leq z \leq 1.965$. Compared with what is given in table 1 of \cite{Ryan_1}, the updated $H(z)$ data here have one additional data point from \cite{Borghi_etal_2022}.

\item[]{\bf BAO data}. There are 12 BAO measurements listed in Table \ref{tab:bao}, spanning the redshift range $0.122 \leq z \leq 2.334$. The covariance matrices for given BAO data are summarized below.

The covariance matrix $\textbf{C}$ for BAO data from \cite{duMas2020} is
\be
\label{CovM1}
    % \textbf{C}=
    \begin{bmatrix}
    1.3225 & -0.1009 \\
    -0.1009 & 0.0380
    \end{bmatrix},
\ee
for BAO data from \cite{eBOSSG_2020} $\textbf{C}$ is
\be
\label{CovM2}
    % \textbf{C}=
    \resizebox{\columnwidth}{!}{%
    $\begin{bmatrix}
    0.02860520 & -0.04939281 & 0.01489688 & -0.01387079\\
    -0.04939281 & 0.5307187 & -0.02423513 & 0.1767087\\
    0.01489688 & -0.02423513 & 0.04147534 & -0.04873962\\
    -0.01387079 & 0.1767087 & -0.04873962 & 0.3268589
    \end{bmatrix},$%
    }
\ee
for BAO data from \cite{eBOSSG_2020} and \cite{eBOSSL_2021} $\textbf{C}$ is
\be
\label{CovM3}
    % \textbf{C}=
    \begin{bmatrix}
    0.1076634008565565 & -0.05831820341302727\\
    -0.05831820341302727 & 0.2838176386340292 
    \end{bmatrix},
\ee
and for BAO data from \cite{eBOSSQ_2020} and \cite{eBOSSQ_2021} $\textbf{C}$ is
\be
\label{CovM4}
    % \textbf{C}=
    \begin{bmatrix}
    0.63731604 & 0.1706891\\
    0.1706891 & 0.30468415
    \end{bmatrix}.
\ee

\item[]{\bf SN Ia data}. As in \cite{CaoRyanRatra2022}, we use SN Ia data that consist of 1048 Pantheon \citep{scolnic_et_al_2018} and 20 binned DES 3yr \citep{DES_2019d} SNe Ia, spanning the redshift ranges $0.01 < z < 2.3$ and $0.015 \leq z \leq 0.7026$, respectively.

\item[]{\bf QSO angular size (QSO-AS) data}. There are 120 QSO-AS measurements listed in table 1 of \cite{Cao_et_al2017b}, spanning the redshift range $0.462 \leq z \leq 2.73$. The measured quantities are $z$ and the angular size $\theta(z)$ with the characteristic linear size of QSOs in the sample, $l_{\rm m}$, as a free parameter to be constrained. The angular size $\theta(z)=l_{\rm m}/D_A(z)$, where $D_A(z)$ is the angular diameter distance. A detailed description of the use of these data can be found in \cite{CaoRyanRatra2022}.

\item[]{\bf \hiig\ data}. There are 181 \hiig\ measurements listed in table A3 of \cite{GM2021}, with 107 low-$z$ data from \cite{Chavez_2014} recalibrated by \cite{G-M_2019}, spanning the redshift range $0.0088 \leq z \leq 0.16417$, and 74 high-$z$ data spanning the redshift range $0.63427 \leq z \leq 2.545$. The measured quantities are $z$, \hiig\ flux $F(\mathrm{H}\beta)$, and velocity dispersion $\sigma$.

\item[]{\bf \mq\ sample}. The \mq\ sample consists of 78 QSOs listed in table A1 of \cite{Khadkaetal_2021a}, spanning the redshift range $0.0033 \leq z \leq 1.89$. \mq\ data obey the radius-luminosity ($R-L$) relation and the measured quantities are the time delay $\tau$ and QSO flux $F_{3000}$ measured at 3000 \(\text{\r{A}}\).

\item[]{\bf A118 sample}. The A118 sample includes 118 long GRBs listed in table 7 of \cite{Khadkaetal_2021b}, spanning the redshift range $0.3399 \leq z \leq 8.2$. A118 data obey the Amati (or $E_{\rm p}-E_{\rm iso}$) correlation and the measured quantities are $z$, rest-frame spectral peak energy $E_{\rm p}$, and measured bolometric fluence $S_{\rm bolo}$, computed in the standard rest-frame energy band $1-10^4$ keV.\footnote{As noted in \citet{Liuetal2022}, the $E_{\rm p}$ value for GRB081121 reported in table 5 of \cite{Dirirsa2019}, and used in our analysis here, is incorrect. One should instead use the correct value provided in table 4 of \cite{Wang_2016}, $E_{\rm p}=871\pm123$ keV. However, since this data point has negligible effect on the cosmological-model and GRB-correlation parameter constraints and the conclusions remain unchanged after correcting it, we do not revise our Amati-correlated GRB results here and in \cite{Caoetal_2021,CaoKhadkaRatra2021,CaoDainottiRatra2022}. In future analyses we will use the correct \cite{Wang_2016} value.}

\item[]{\bf Platinum + A101 sample}. The Platinum sample includes 50 long GRBs listed in table A1 of \cite{CaoDainottiRatra2022}, spanning the redshift range $0.553 \leq z \leq 5.0$. The A101 sample includes 101 long GRBs with common GRBs between the Platinum and the A118 samples excluded, spanning the redshift range $0.3399 \leq z \leq 8.2$. The Platinum GRBs obey the three-dimensional Dainotti correlation and the measured quantities are $z$, characteristic time scale $T^{*}_{X}$, the measured $\gamma$-ray energy flux $F_{X}$ at $T^{*}_{X}$, the prompt peak flux $F_{\rm peak}$ over a 1 s interval, and the X-ray spectral index of the plateau phase $\beta^{\prime}$. 

\end{itemize}

\section{Data Analysis Methodology}
\label{sec:analysis}

In this paper we determine constraints on the cosmological model parameters, and non-cosmological parameters related to different data sets, by maximizing the likelihood function, $\mathcal{L}$. These analyses are performed by using the Markov chain Monte Carlo (MCMC) code \textsc{MontePython} \citep{Audren:2012wb}, with the physics coded in the \textsc{class} code. In Table \ref{tab:priors}, we list the flat prior ranges of the constrained free parameters.

The detailed descriptions for the likelihood functions of $H(z)$, BAO, \hiig, QSO-AS, and SN Ia data can be found in \cite{CaoRyanRatra2020, CaoRyanRatra2021, Caoetal_2021}, whereas those of Platinum, A118/A101, and \mq\ data can be found in \cite{CaoDainottiRatra2022} and \cite{Khadkaetal_2021a}. One can also find the definitions of the Akaike Information Criterion (AIC) and the Bayesian Information Criterion (BIC) as well as the deviance information criterion (DIC) in \cite{CaoDainottiRatra2022}.\footnote{Unlike AIC and BIC, DIC estimates the effective number of free parameters.} We compute $\Delta \mathrm{AIC}$, $\Delta \mathrm{BIC}$, and $\Delta \mathrm{DIC}$ differences for the other five cosmological models relative to the flat \lcdm\ reference model values. Negative (positive) values of $\Delta \mathrm{AIC}$, $\Delta \mathrm{BIC}$, or $\Delta \mathrm{DIC}$ indicate that the model under investigation fits the data compilation better (worse) than does the reference model. Relative to the model with minimum AIC(BIC/DIC), $\Delta \mathrm{AIC(BIC/DIC)} \in (0, 2]$ is defined to be weak evidence against the model under investigation, $\Delta \mathrm{AIC(BIC/DIC)} \in (2, 6]$ is positive evidence against the model under investigation, $\Delta \mathrm{AIC(BIC/DIC)} \in (6, 10] $ is strong evidence against the model under investigation, and $\Delta \mathrm{AIC(BIC/DIC)}>10$ is very strong evidence against the model under investigation.

\begin{table}
\centering
\resizebox{\columnwidth}{!}{%
\begin{threeparttable}
\caption{Flat priors of the constrained parameters.}
\label{tab:priors}
\begin{tabular}{lcc}
\toprule
Parameter & & Prior\\
\midrule
 & Cosmological Parameters & \\
\midrule
$H_0$\,\tnote{a} &  & [None, None]\\
\obhs\,\tnote{b} &  & [0, 1]\\
\ochs\,\tnote{c} &  & [0, 1]\\
\ok &  & [-2, 2]\\
$\alpha$ &  & [0, 10]\\
\wx &  & [-5, 0.33]\\
\midrule
 & Non-Cosmological Parameters & \\
\midrule
$k$ &  & [0, 5]\\
$b_{\mathrm{\textsc{m}}}$ &  & [0, 10]\\
$\sigma_{\rm int}$ &  & [0, 5]\\
$a$ &  & [-5, 5]\\
$b_{\mathrm{\textsc{p}}}$ &  & [-5, 5]\\
$C_{o}$ &  & [-50, 50]\\
$\beta$ &  & [0, 5]\\
$\gamma$ &  & [0, 300]\\
$l_{\rm m}$ &  & [None, None]\\
\bottomrule
\end{tabular}
\begin{tablenotes}[flushleft]
\item [a] \hunit. In the \mq\ + A118 case, $H_0$ is set to be 70 \hunit, while in other cases, the prior range is irrelevant (unbounded).
\item [b] In the \mq\ + A118 case, \obhs\ is set to be 0.0245, i.e. $\Omega_{b}=0.05$.
\item [c] In the \mq\ + A118 case, $\Om\in[0,1]$ is ensured.
\end{tablenotes}
\end{threeparttable}%
}
\end{table}

\section{Results}
\label{sec:results}

\begin{figure*}
\centering
 \subfloat[]{%
    \includegraphics[width=0.5\textwidth,height=0.5\textwidth]{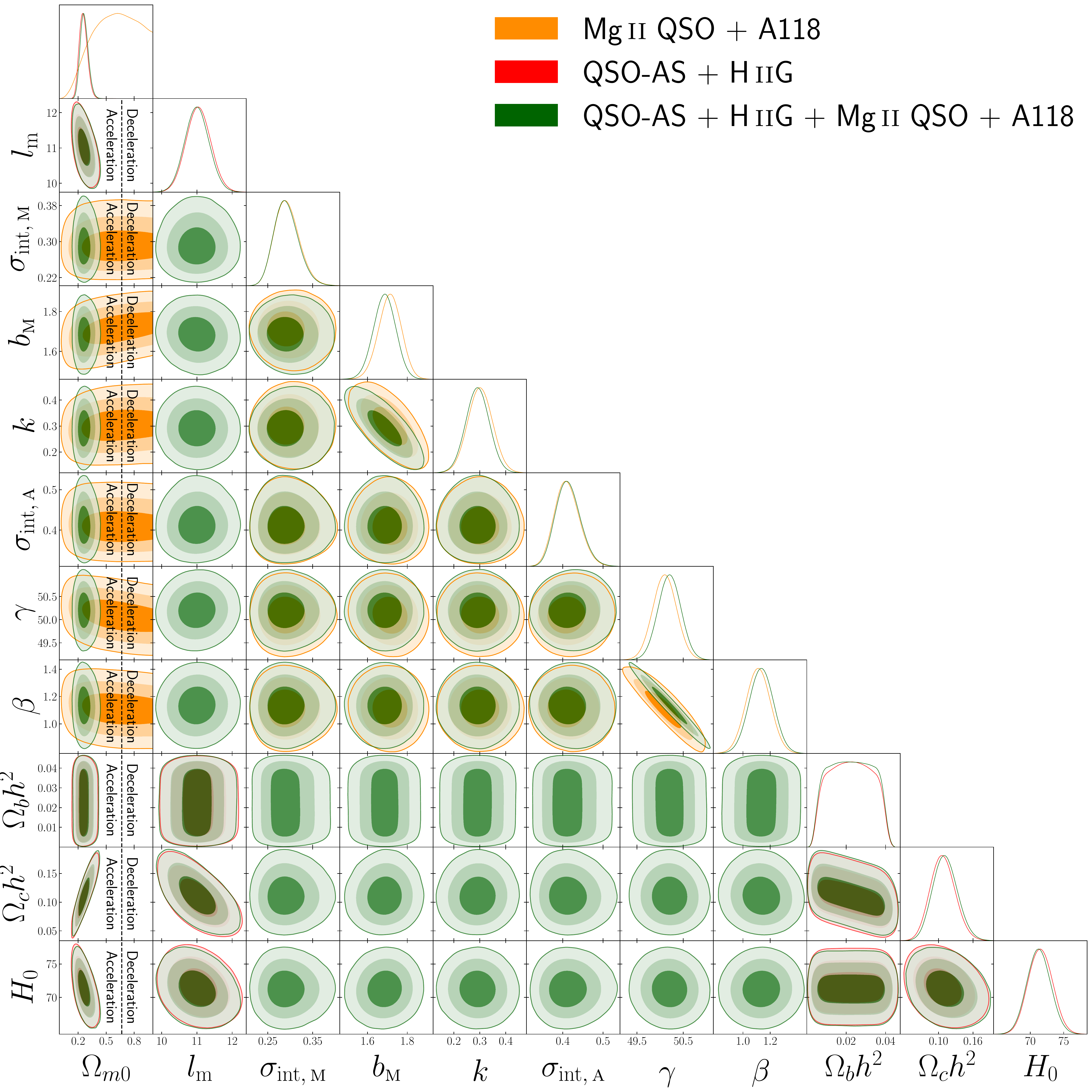}}
 \subfloat[]{%
    \includegraphics[width=0.5\textwidth,height=0.5\textwidth]{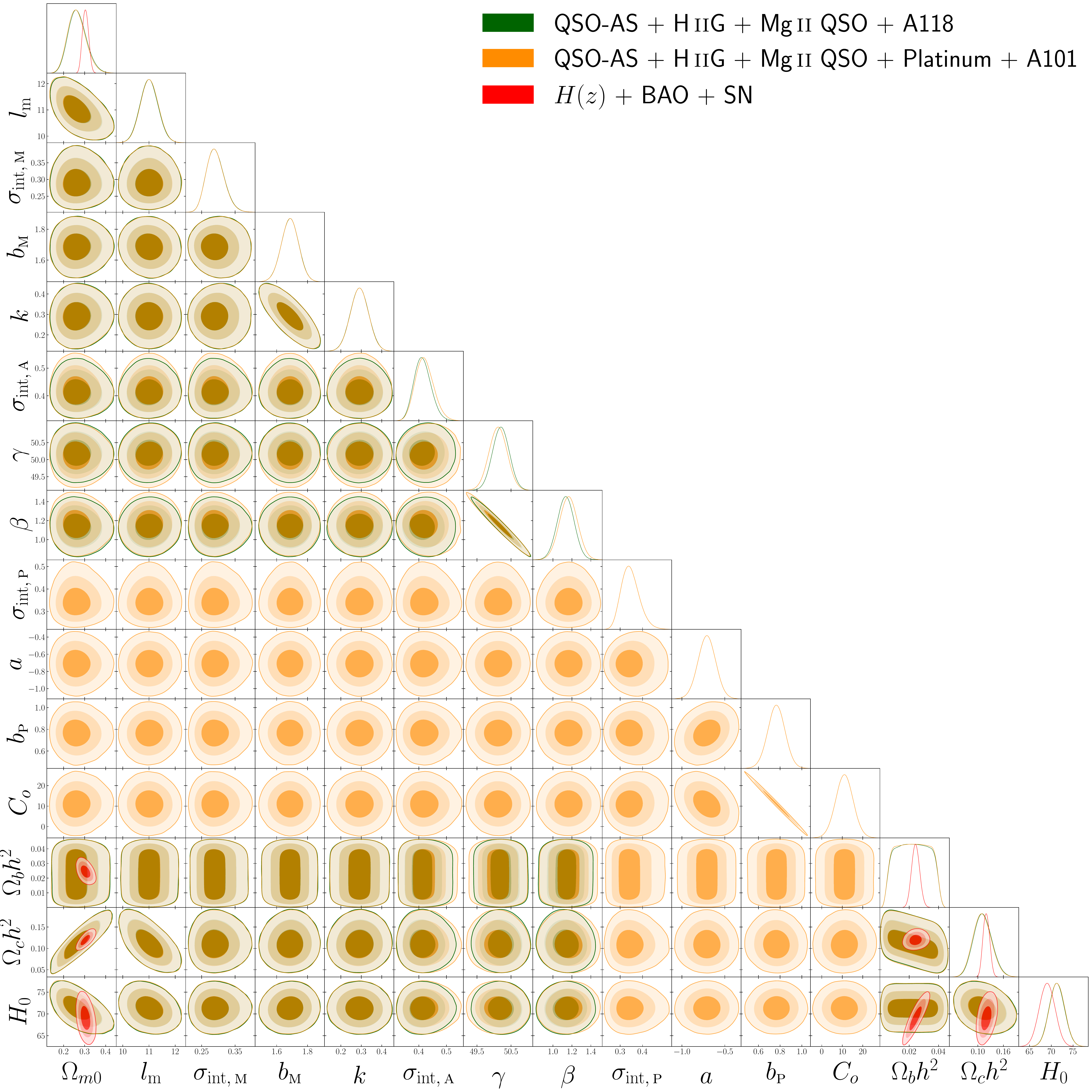}}\\
 \subfloat[]{%
    \includegraphics[width=0.5\textwidth,height=0.5\textwidth]{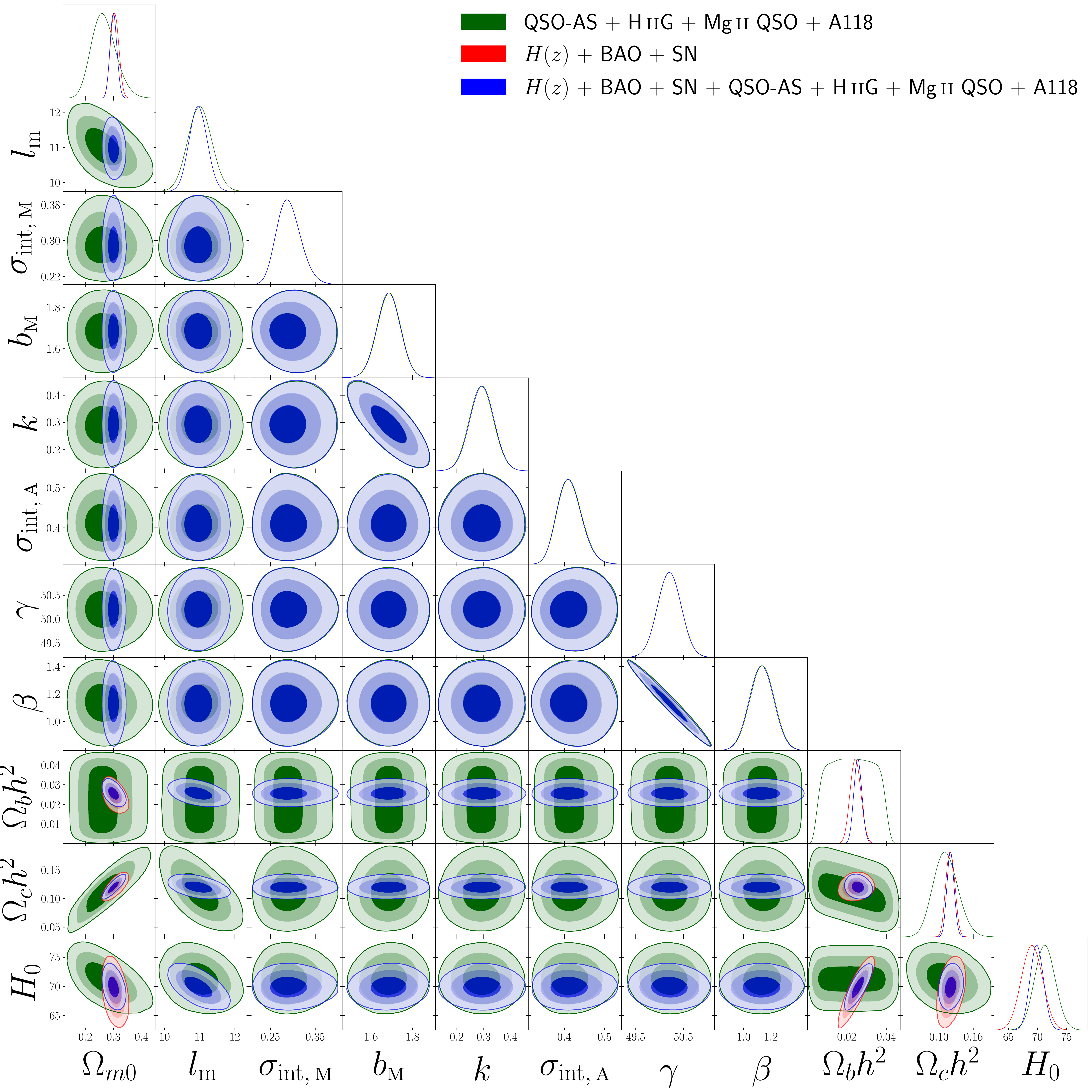}}
 \subfloat[]{%
    \includegraphics[width=0.5\textwidth,height=0.5\textwidth]{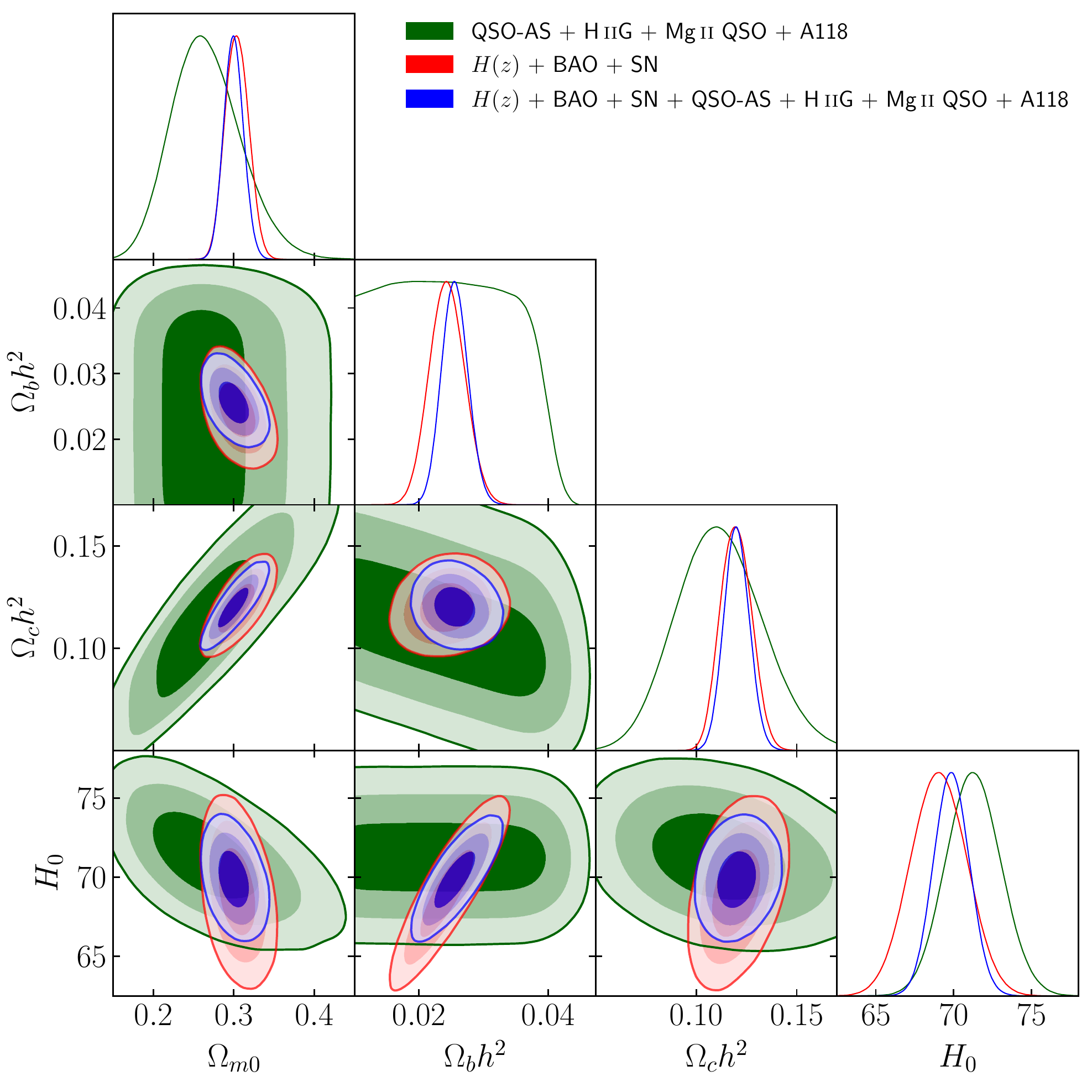}}\\
\caption{One-dimensional likelihood distributions and 1$\sigma$, 2$\sigma$, and 3$\sigma$ two-dimensional likelihood confidence contours for flat \lcdm\ from various combinations of data. The zero-acceleration black dashed lines in some (a) panels divide the parameter space into regions associated with currently-accelerating (left) and currently-decelerating (right) cosmological expansion.}
\label{fig1}
\end{figure*}

\begin{figure*}
\centering
 \subfloat[]{%
    \includegraphics[width=0.5\textwidth,height=0.5\textwidth]{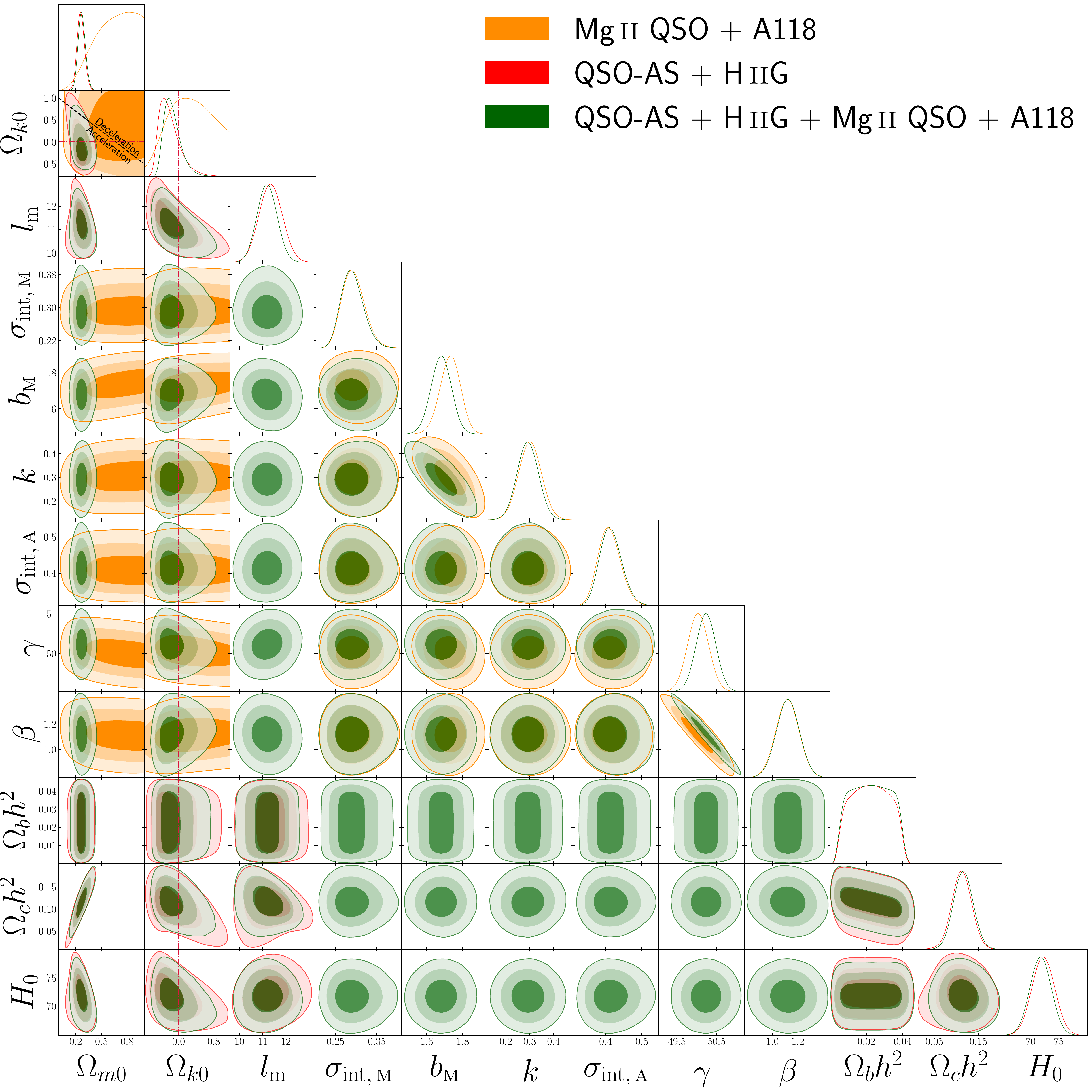}}
 \subfloat[]{%
    \includegraphics[width=0.5\textwidth,height=0.5\textwidth]{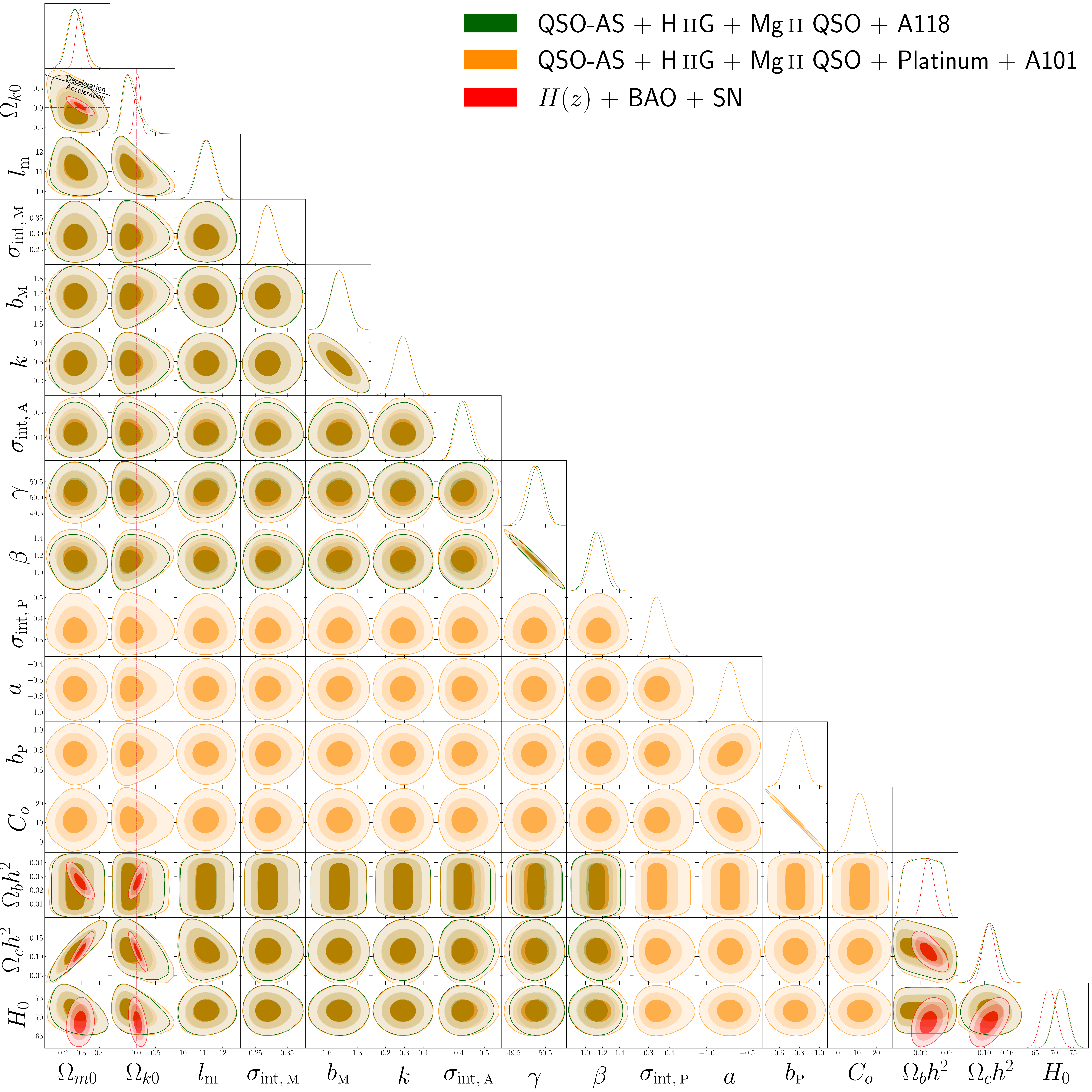}}\\
 \subfloat[]{%
    \includegraphics[width=0.5\textwidth,height=0.5\textwidth]{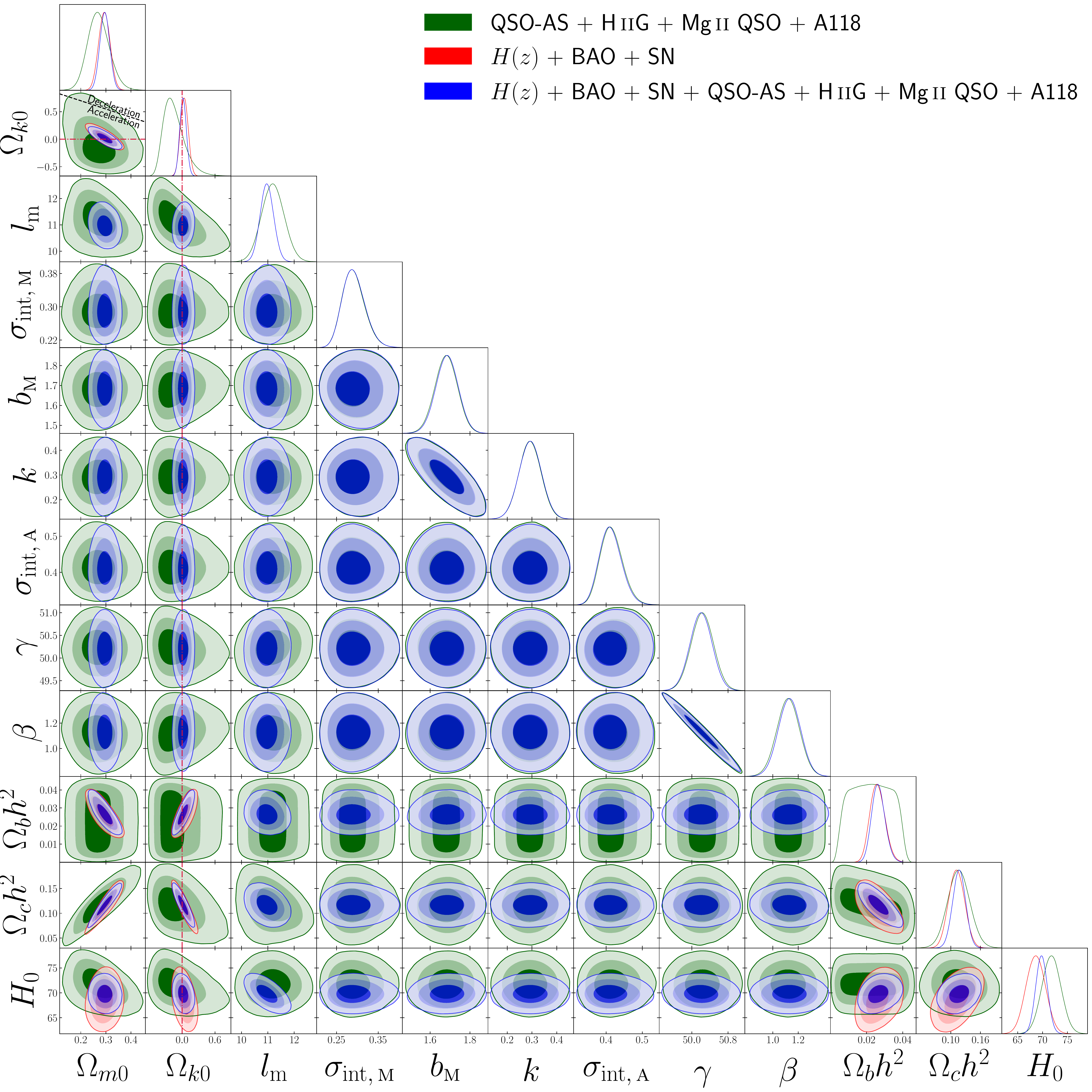}}
 \subfloat[]{%
    \includegraphics[width=0.5\textwidth,height=0.5\textwidth]{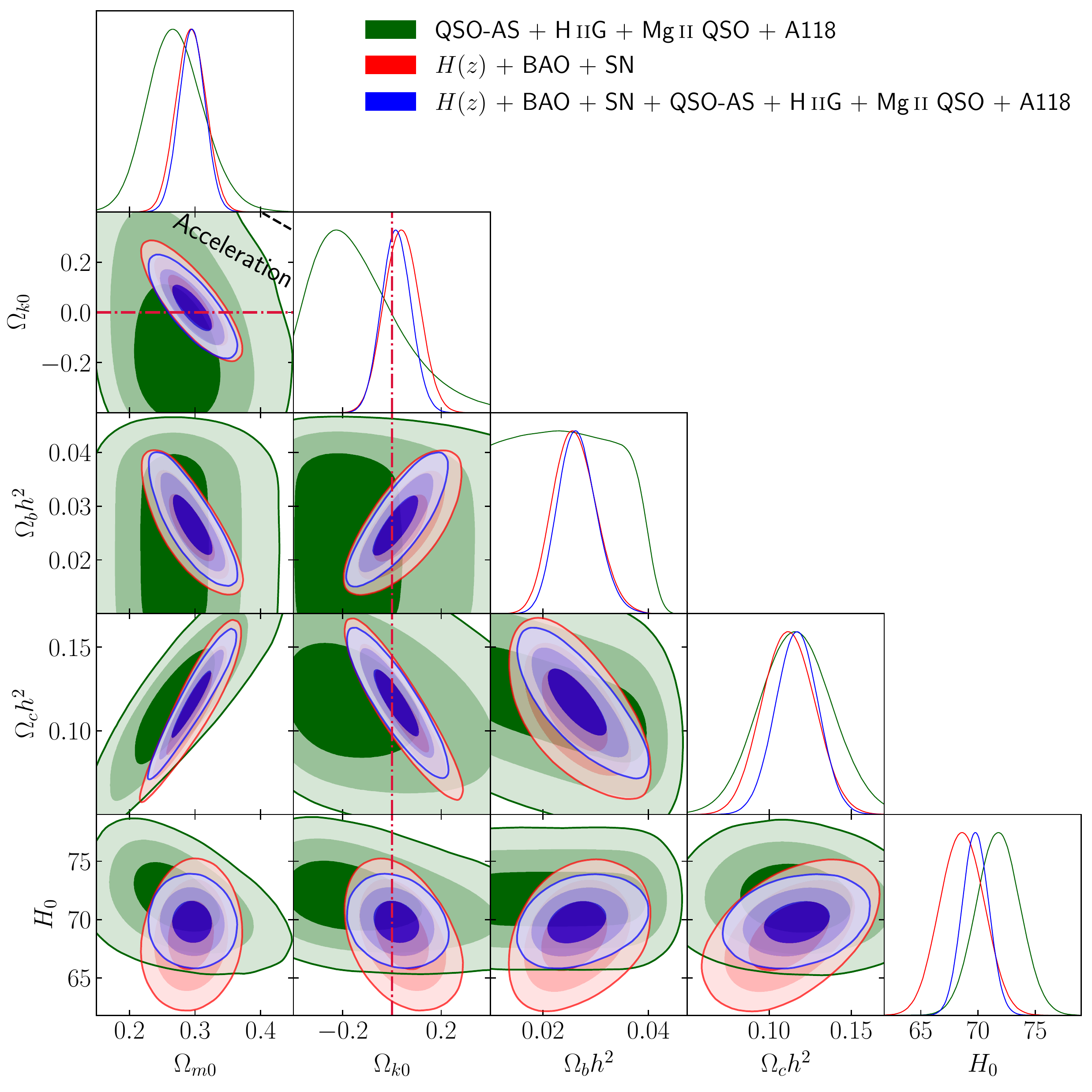}}\\
\caption{Same as Fig.\ \ref{fig1} but for non-flat \lcdm. The zero-acceleration black dashed lines divide the parameter space into regions associated with currently-accelerating (below left) and currently-decelerating (above right) cosmological expansion.}
\label{fig2}
\end{figure*}

\begin{figure*}
\centering
 \subfloat[]{%
    \includegraphics[width=0.5\textwidth,height=0.5\textwidth]{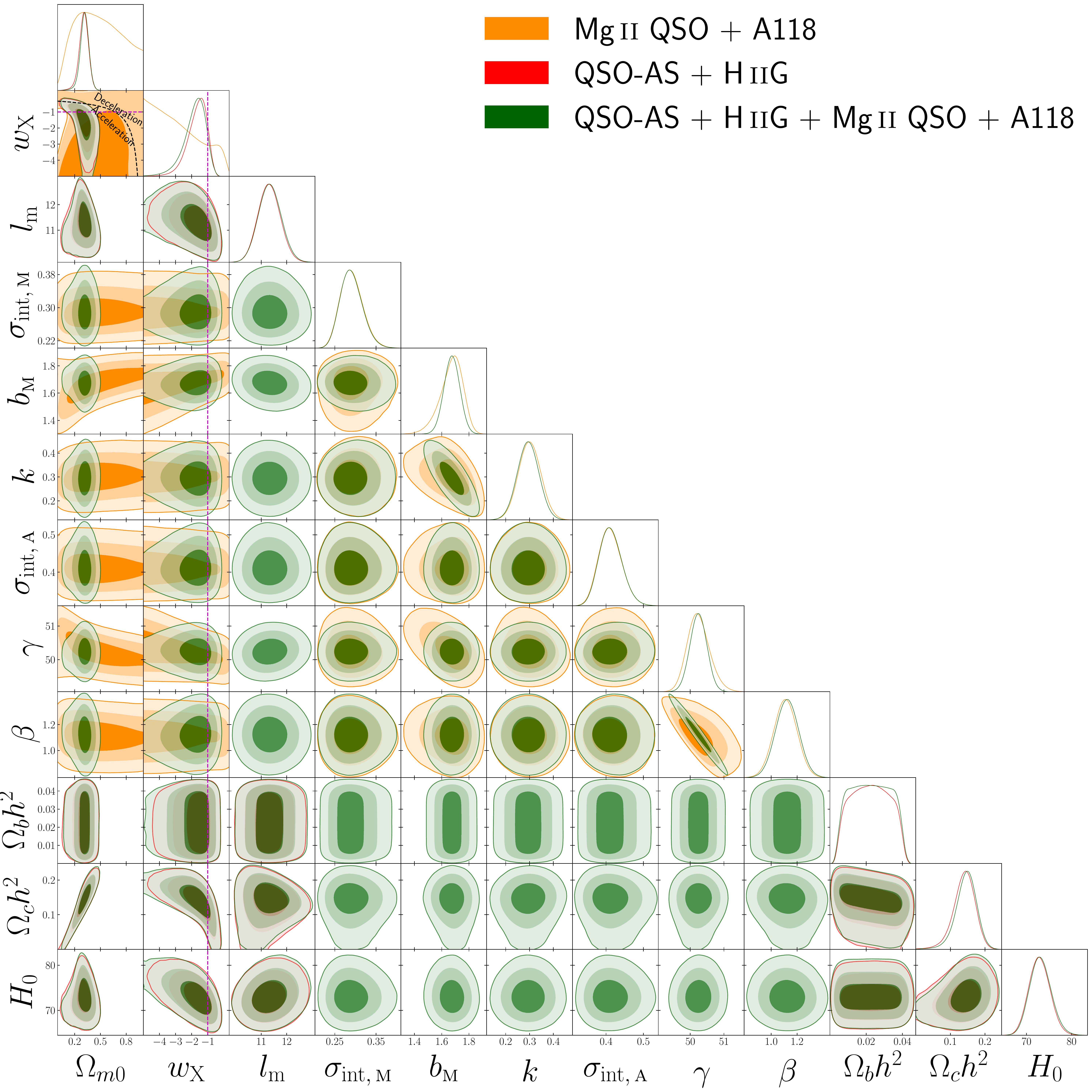}}
 \subfloat[]{%
    \includegraphics[width=0.5\textwidth,height=0.5\textwidth]{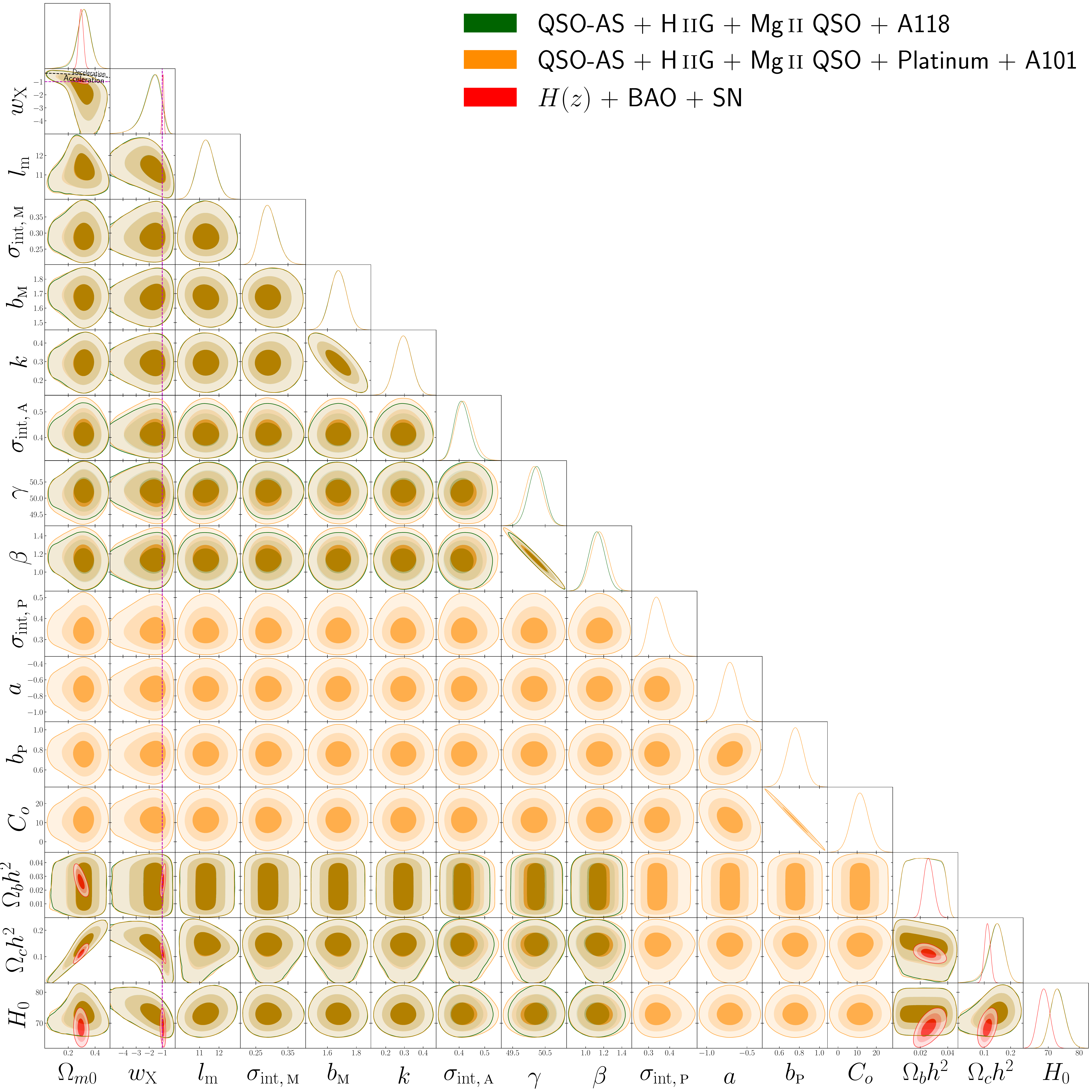}}\\
 \subfloat[]{%
    \includegraphics[width=0.5\textwidth,height=0.5\textwidth]{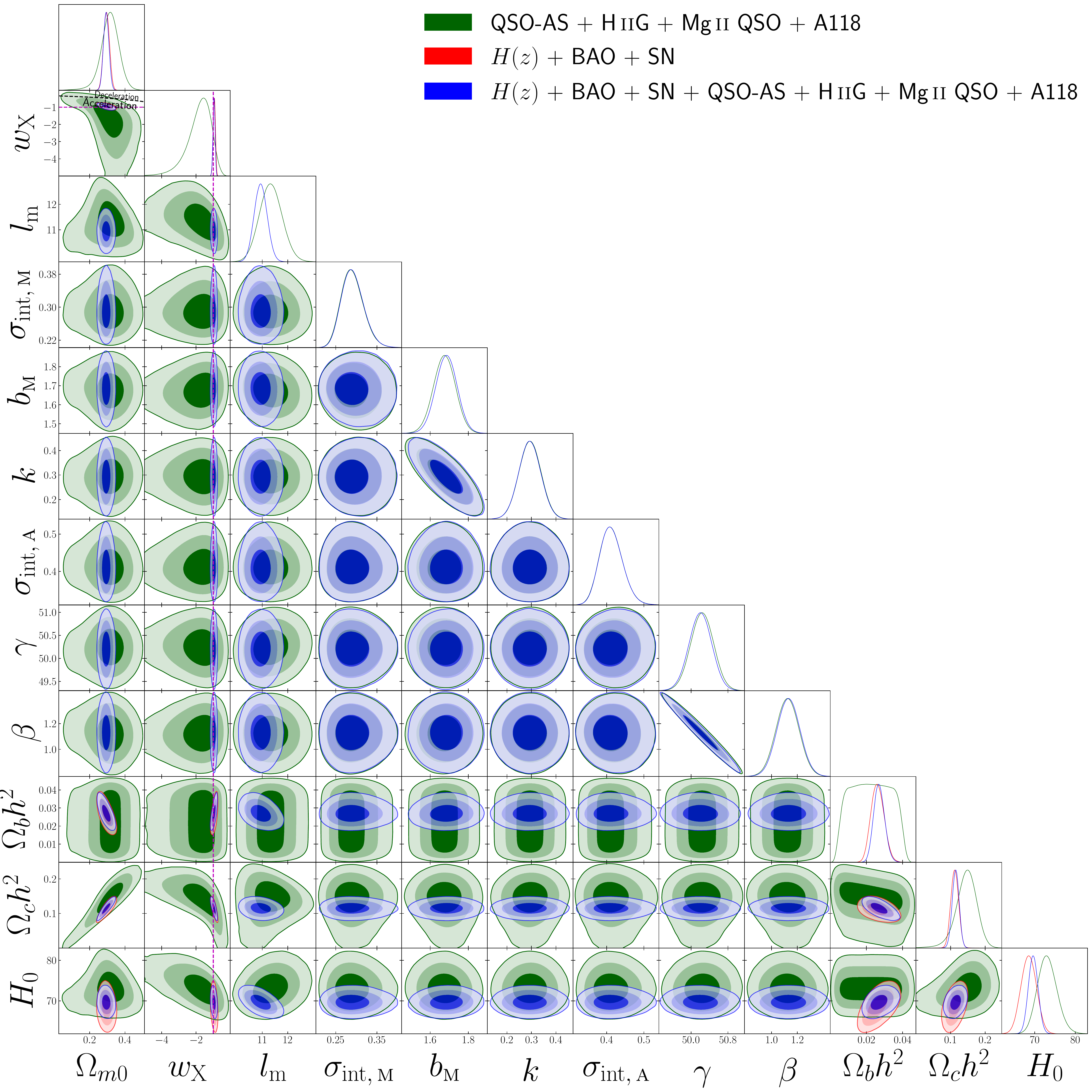}}
 \subfloat[]{%
    \includegraphics[width=0.5\textwidth,height=0.5\textwidth]{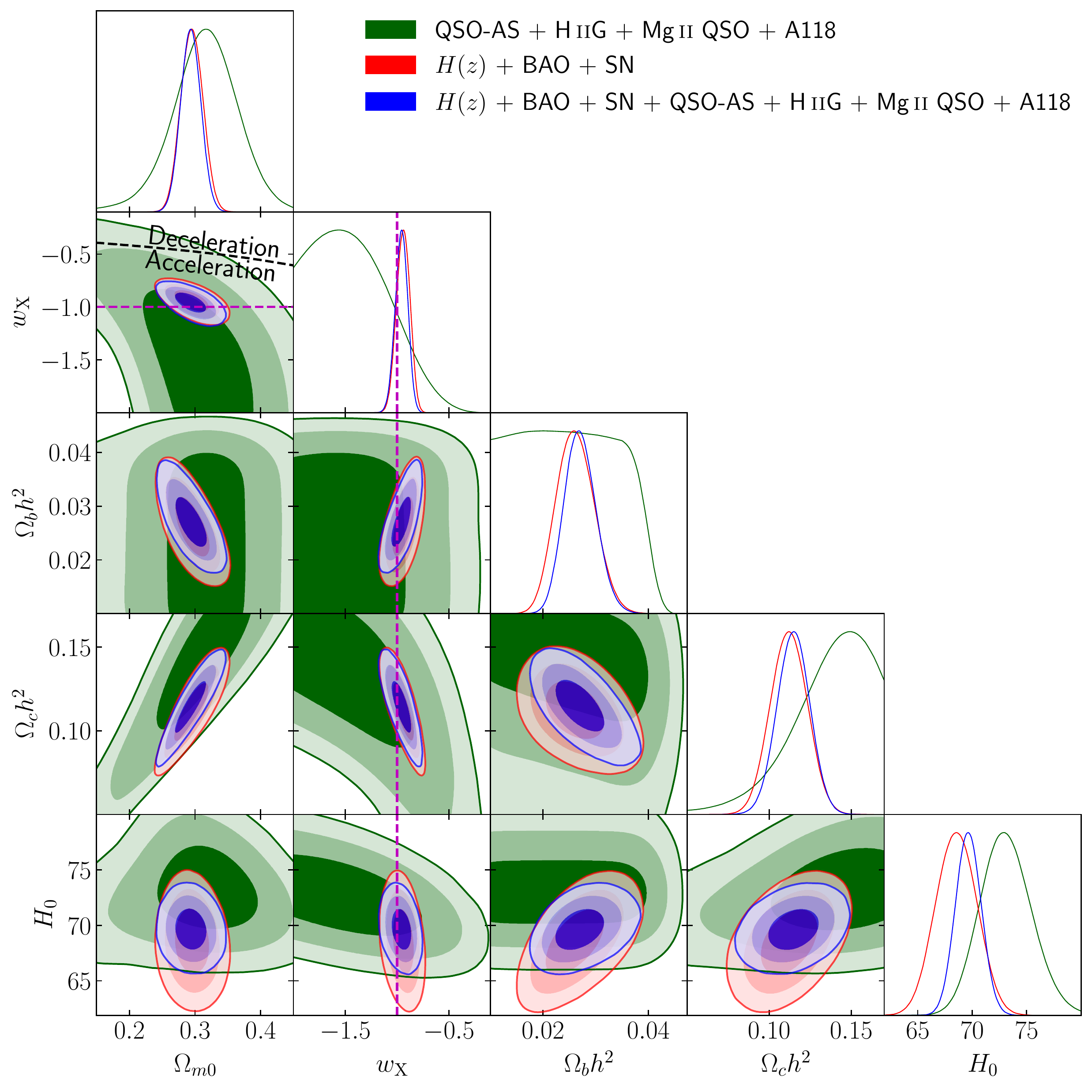}}\\
\caption{One-dimensional likelihood distributions and 1$\sigma$, 2$\sigma$, and 3$\sigma$ two-dimensional likelihood confidence contours for flat XCDM from various combinations of data. The zero-acceleration black dashed lines divide the parameter space into regions associated with currently-accelerating (either below left or below) and currently-decelerating (either above right or above) cosmological expansion. The magenta dashed lines represent $w_{\rm X}=-1$, i.e.\ flat \lcdm.}
\label{fig3}
\end{figure*}

\begin{figure*}
\centering
 \subfloat[]{%
    \includegraphics[width=0.5\textwidth,height=0.5\textwidth]{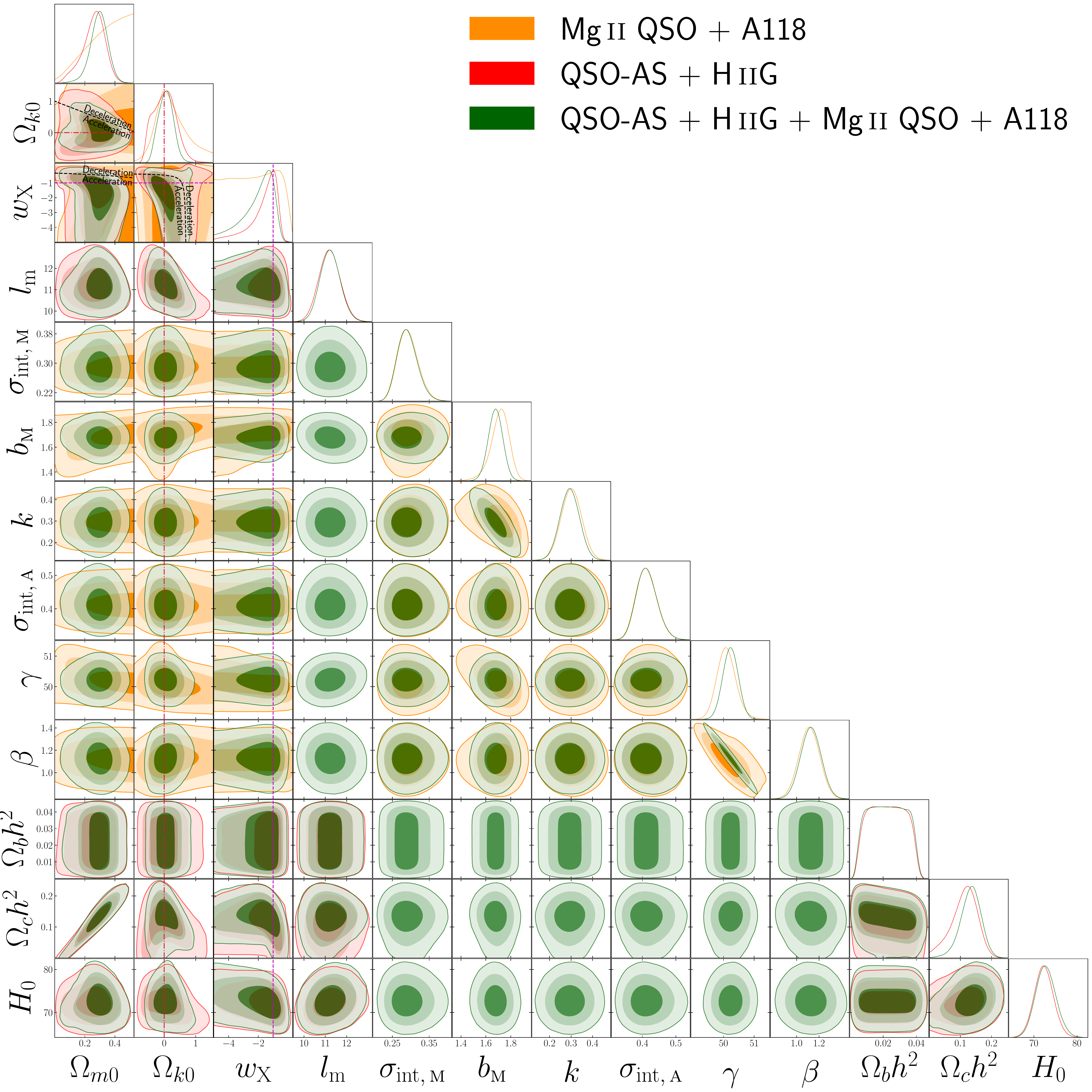}}
 \subfloat[]{%
    \includegraphics[width=0.5\textwidth,height=0.5\textwidth]{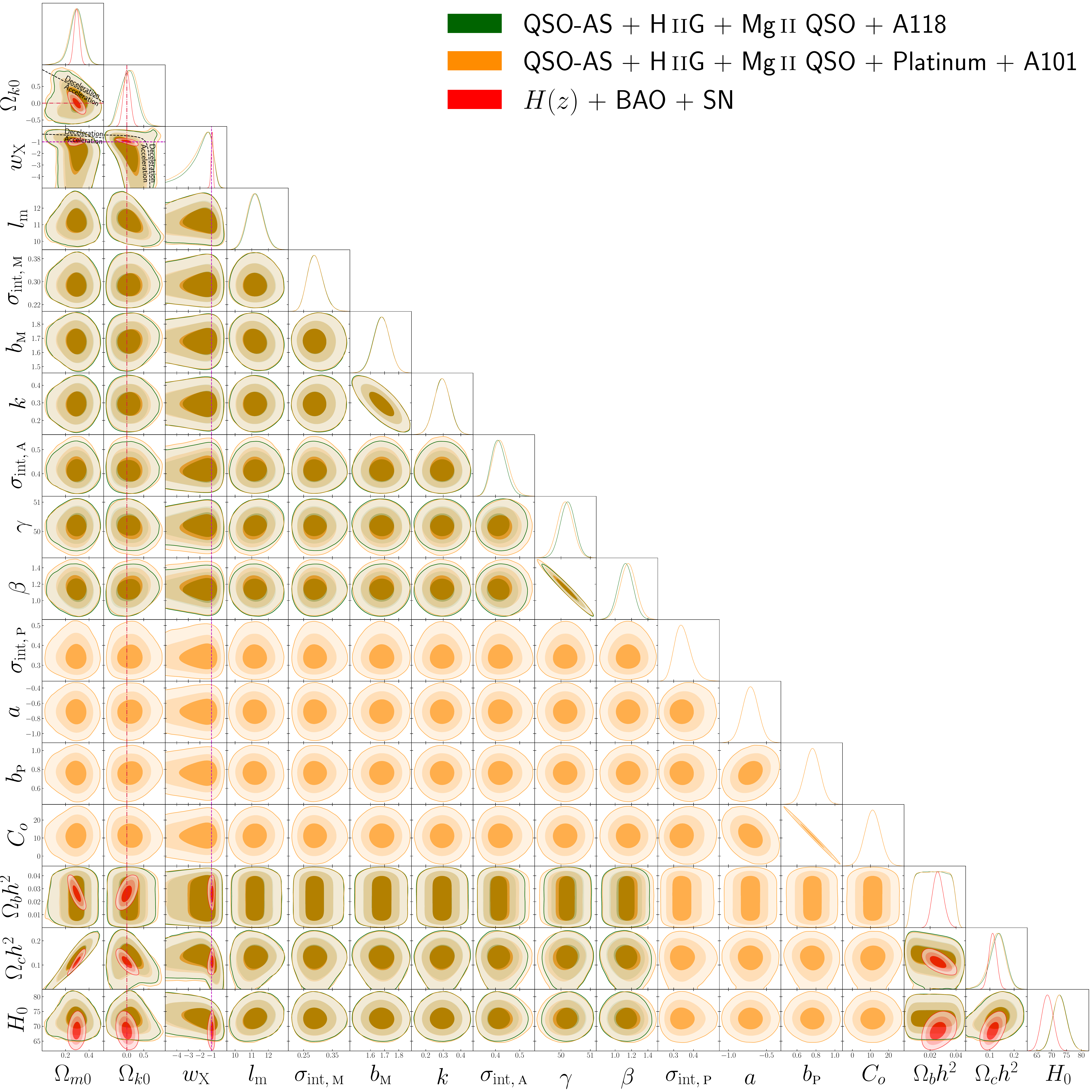}}\\
 \subfloat[]{%
    \includegraphics[width=0.5\textwidth,height=0.5\textwidth]{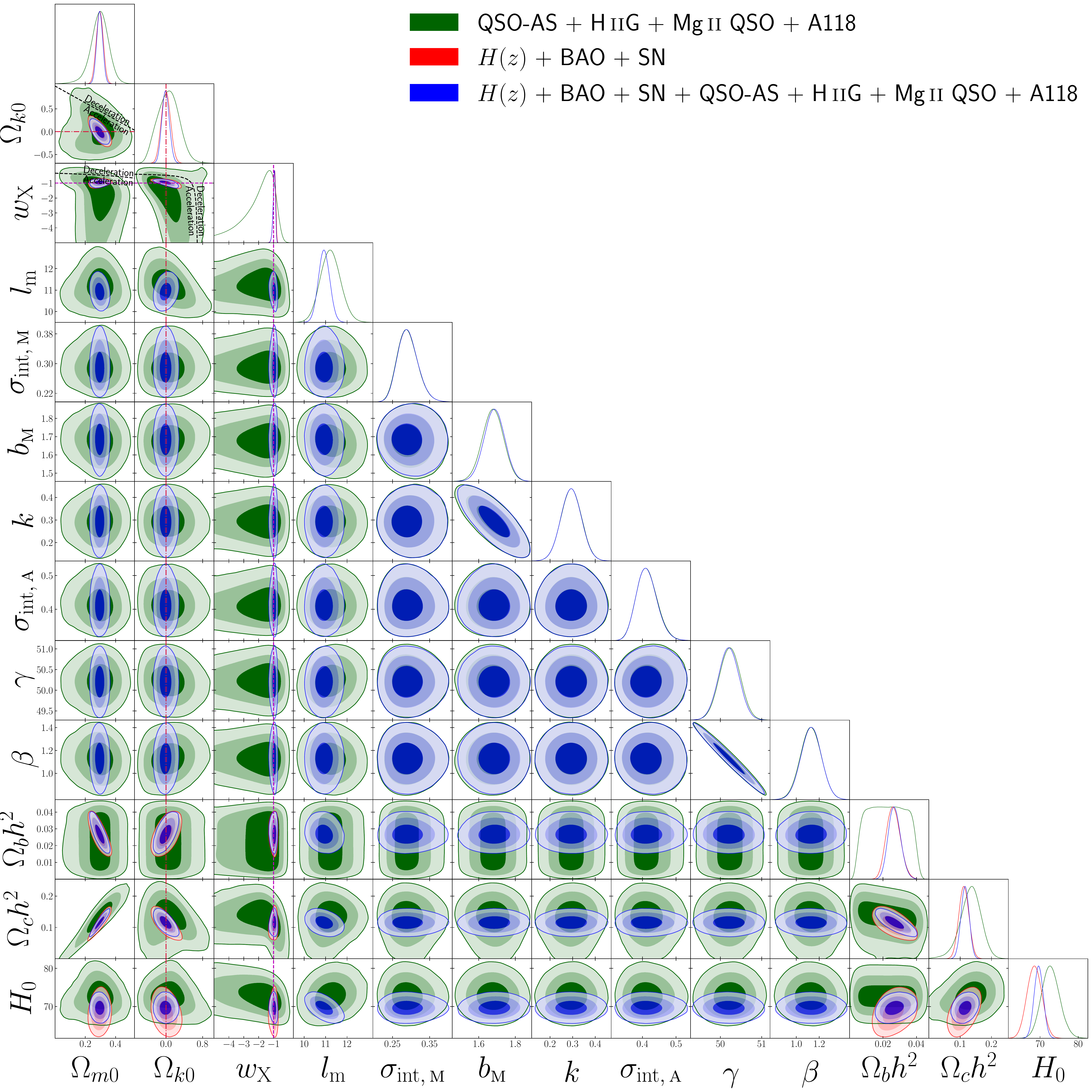}}
 \subfloat[]{%
    \includegraphics[width=0.5\textwidth,height=0.5\textwidth]{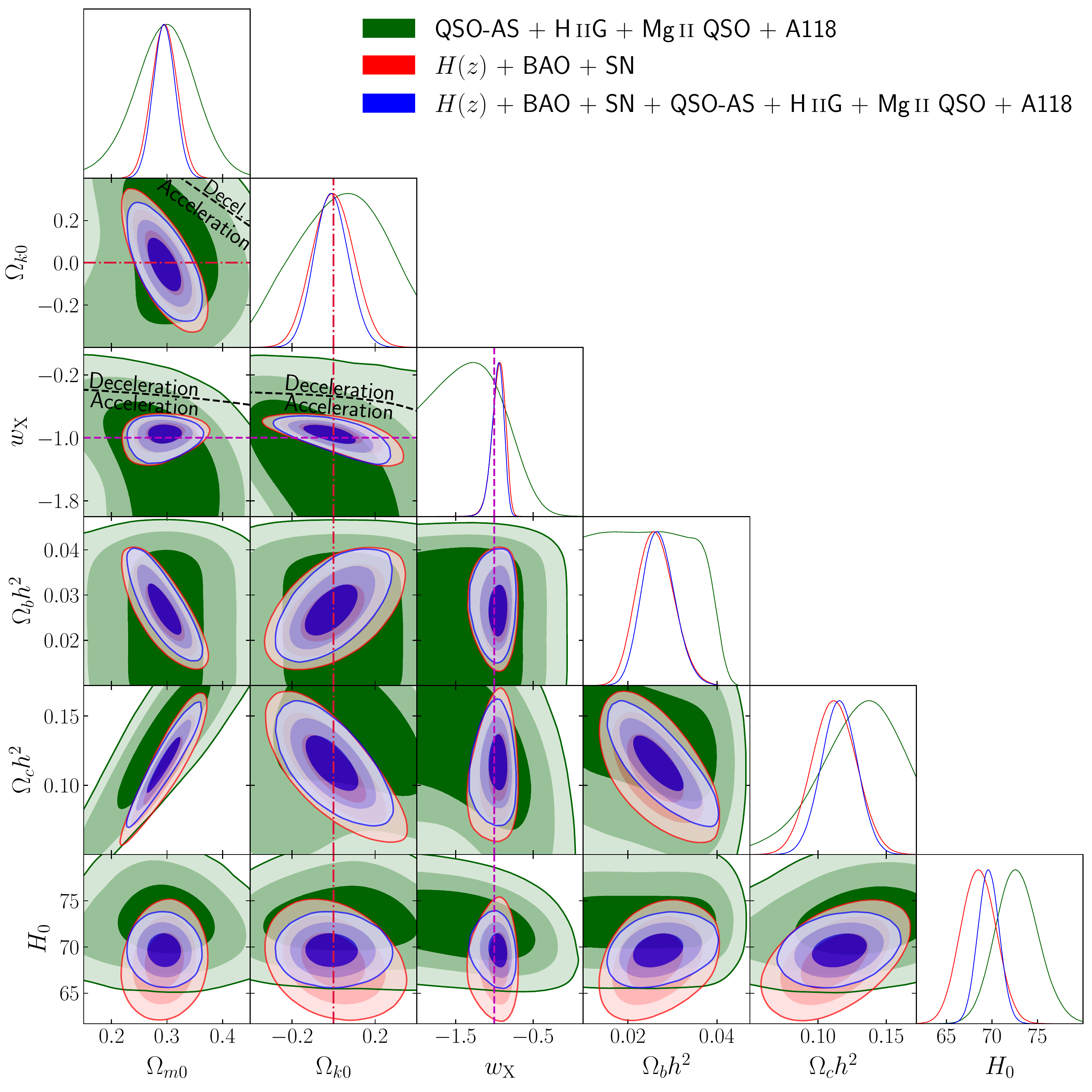}}\\
\caption{Same as Fig.\ \ref{fig3} but for non-flat XCDM. The zero-acceleration black dashed lines are computed for the third cosmological parameter set to the $H(z)$ + BAO data best-fitting values listed in Table \ref{tab:BFP}, and divide the parameter space into regions associated with currently-accelerating (either below left or below) and currently-decelerating (either above right or above) cosmological expansion. The crimson dash-dot lines represent flat hypersurfaces, with closed spatial hypersurfaces either below or to the left. The magenta dashed lines represent $w_{\rm X}=-1$, i.e.\ non-flat \lcdm.}
\label{fig4}
\end{figure*}

\begin{figure*}
\centering
\centering
 \subfloat[]{%
    \includegraphics[width=0.5\textwidth,height=0.5\textwidth]{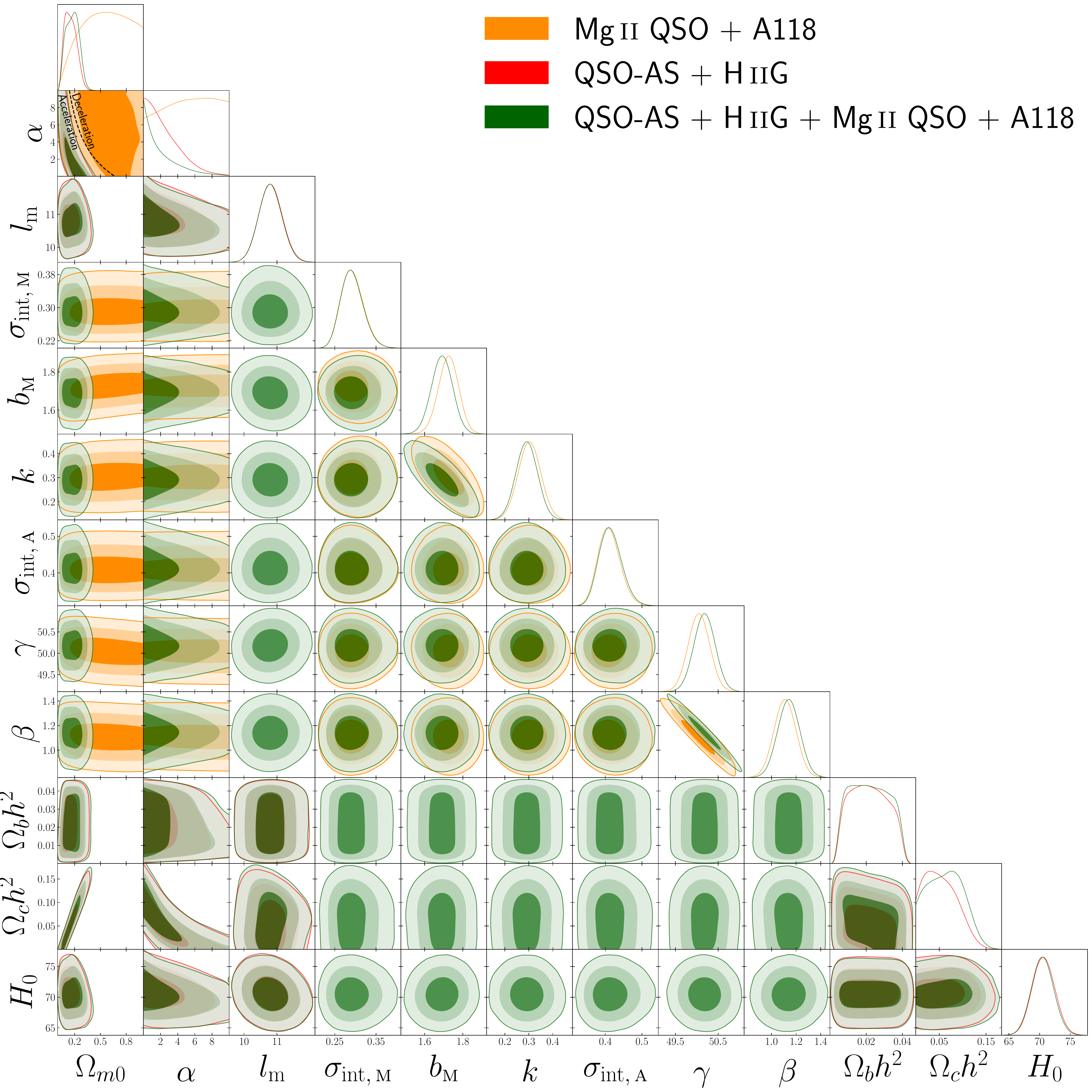}}
 \subfloat[]{%
    \includegraphics[width=0.5\textwidth,height=0.5\textwidth]{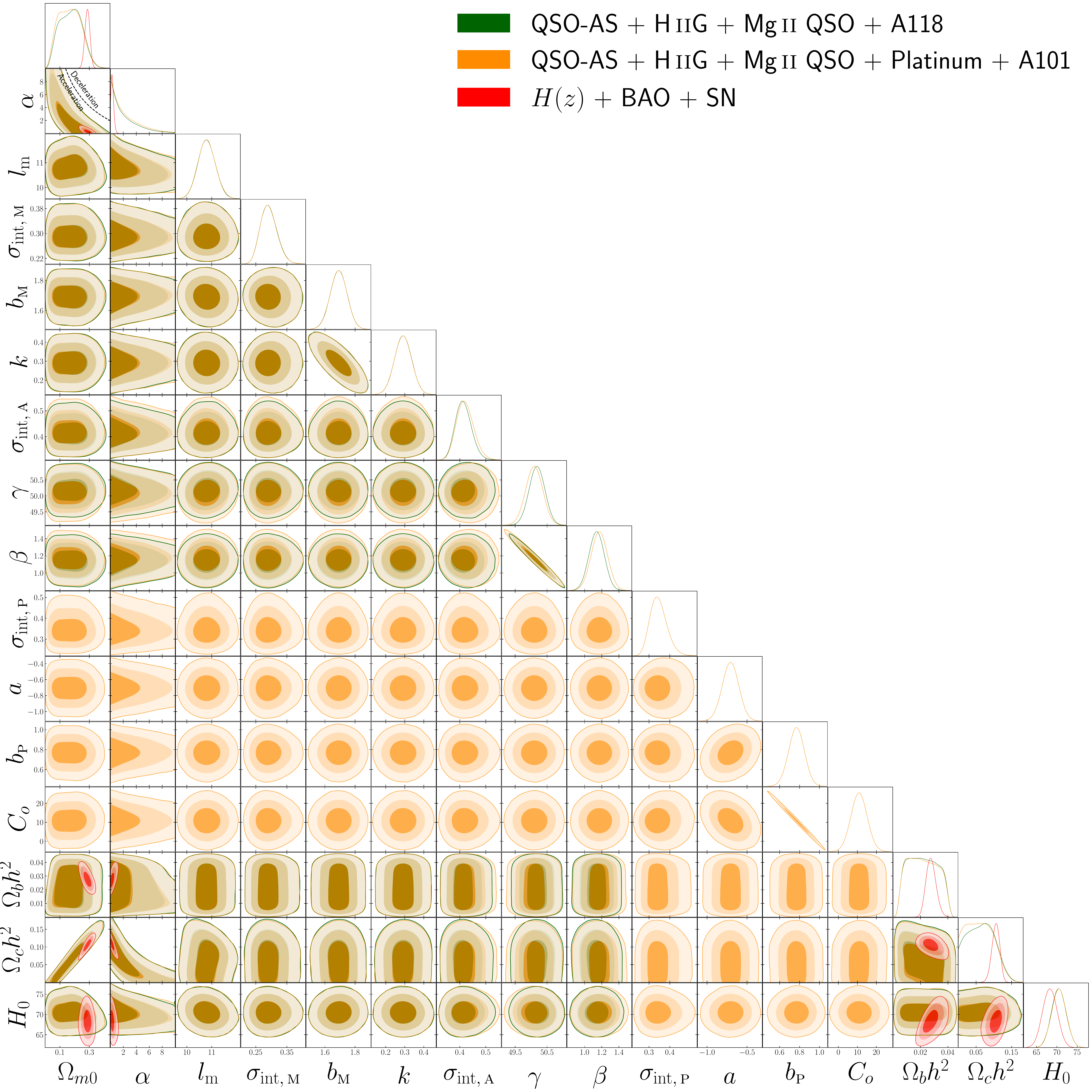}}\\
 \subfloat[]{%
    \includegraphics[width=0.5\textwidth,height=0.5\textwidth]{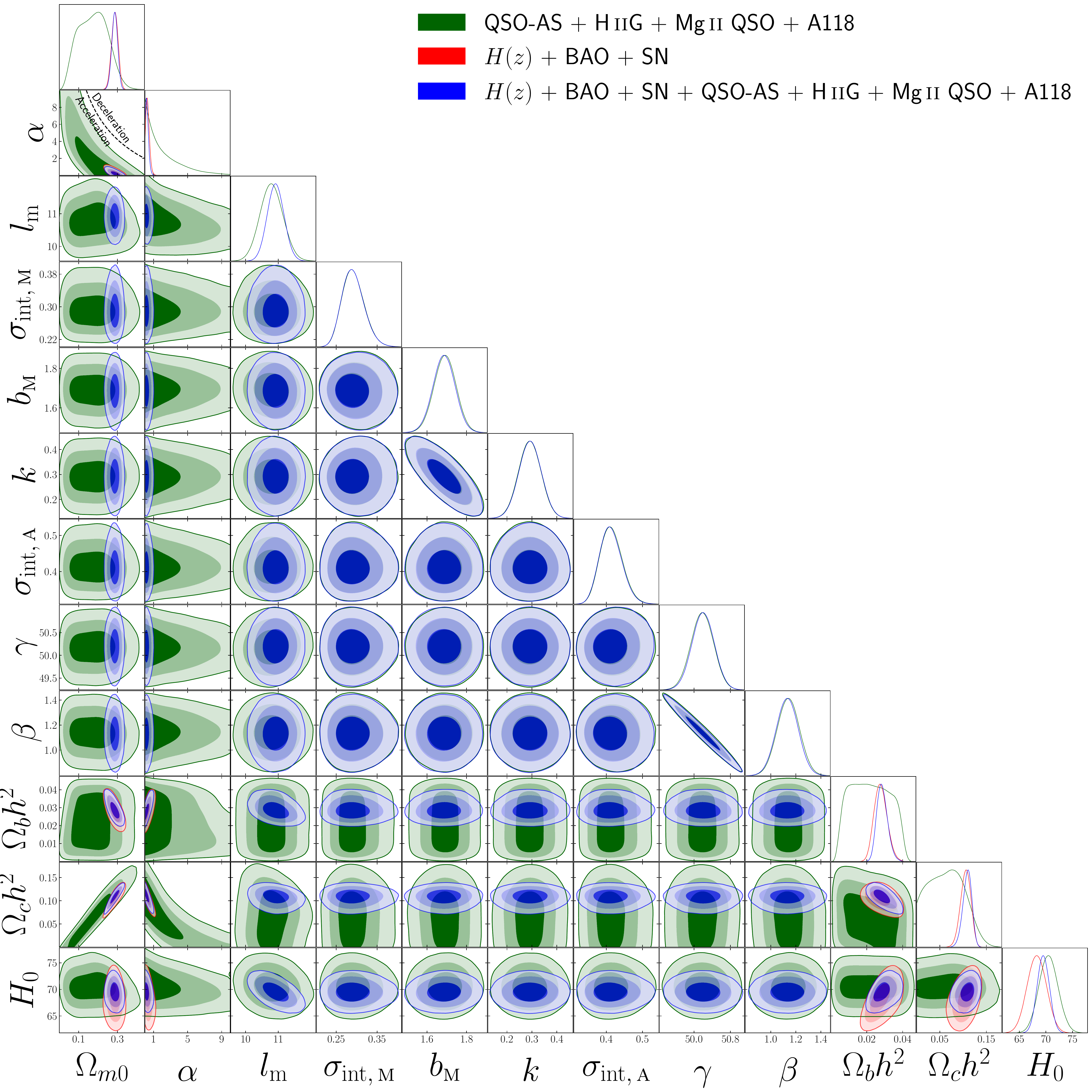}}
 \subfloat[]{%
    \includegraphics[width=0.5\textwidth,height=0.5\textwidth]{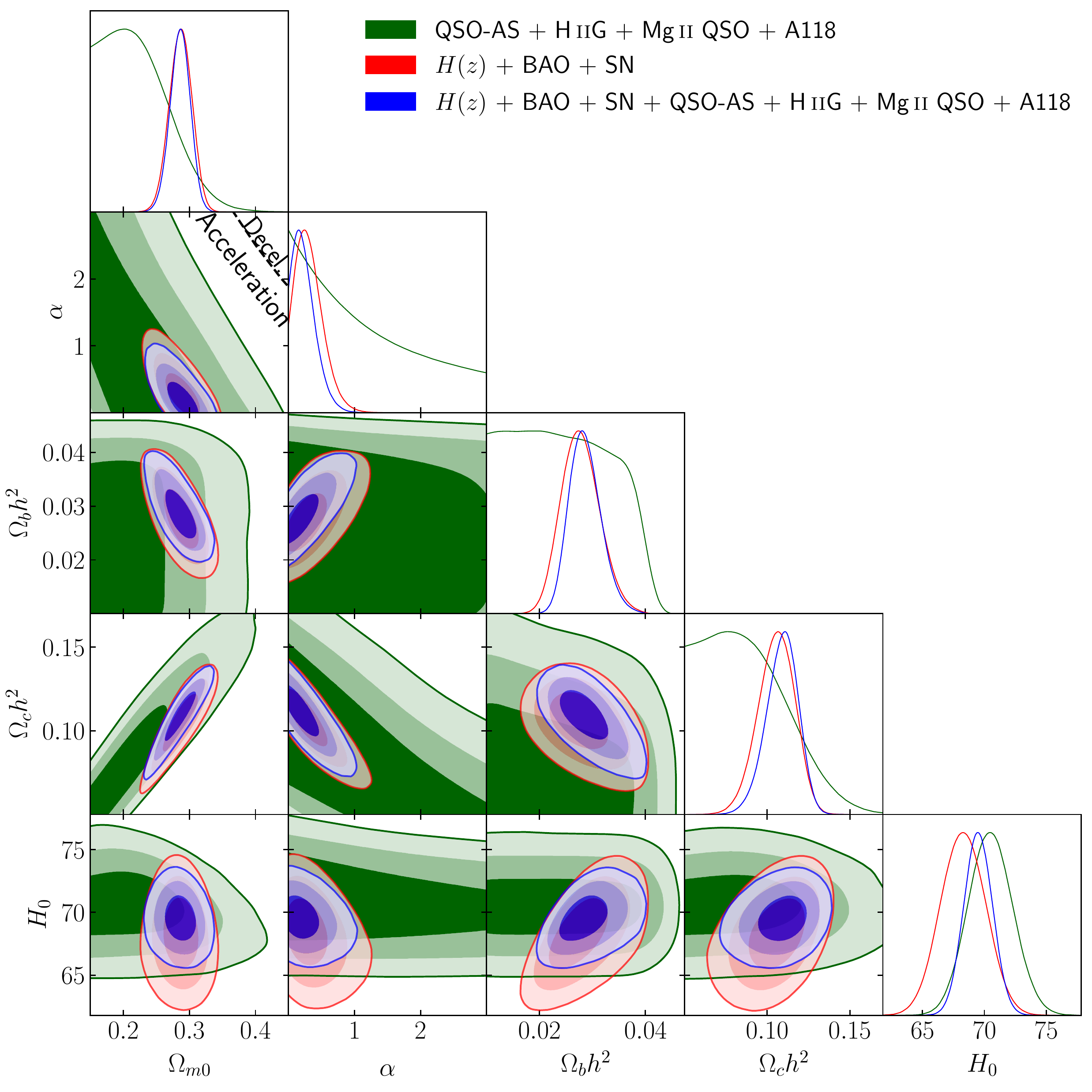}}\\
\caption{One-dimensional likelihood distributions and 1$\sigma$, 2$\sigma$, and 3$\sigma$ two-dimensional likelihood confidence contours for flat \pcdm\ from various combinations of data. The zero-acceleration black dashed lines divide the parameter space into regions associated with currently-accelerating (below left) and currently-decelerating (above right) cosmological expansion. The $\alpha = 0$ axes correspond to flat \lcdm.}
\label{fig5}
\end{figure*}

\begin{figure*}
\centering
 \subfloat[]{%
    \includegraphics[width=0.5\textwidth,height=0.5\textwidth]{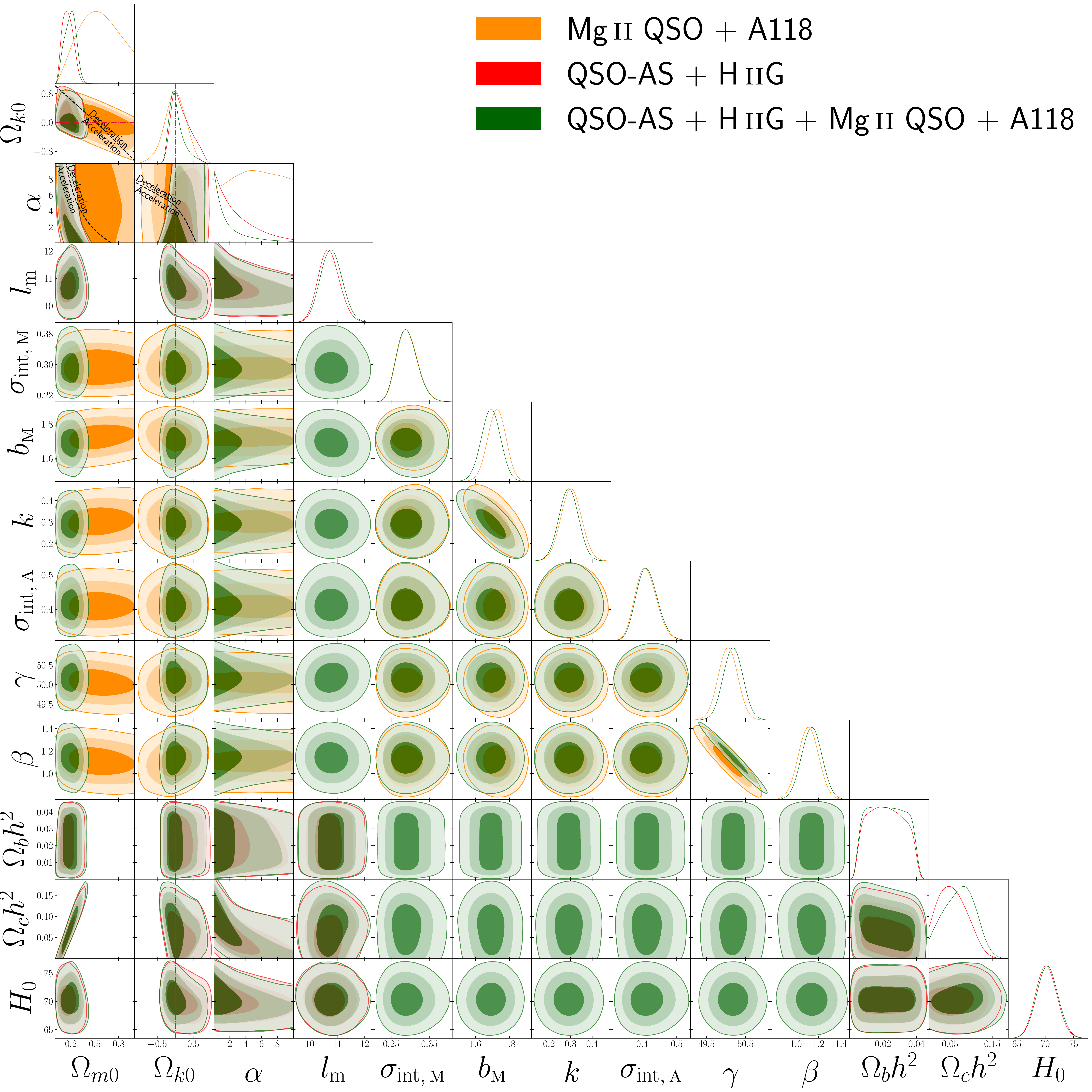}}
 \subfloat[]{%
    \includegraphics[width=0.5\textwidth,height=0.5\textwidth]{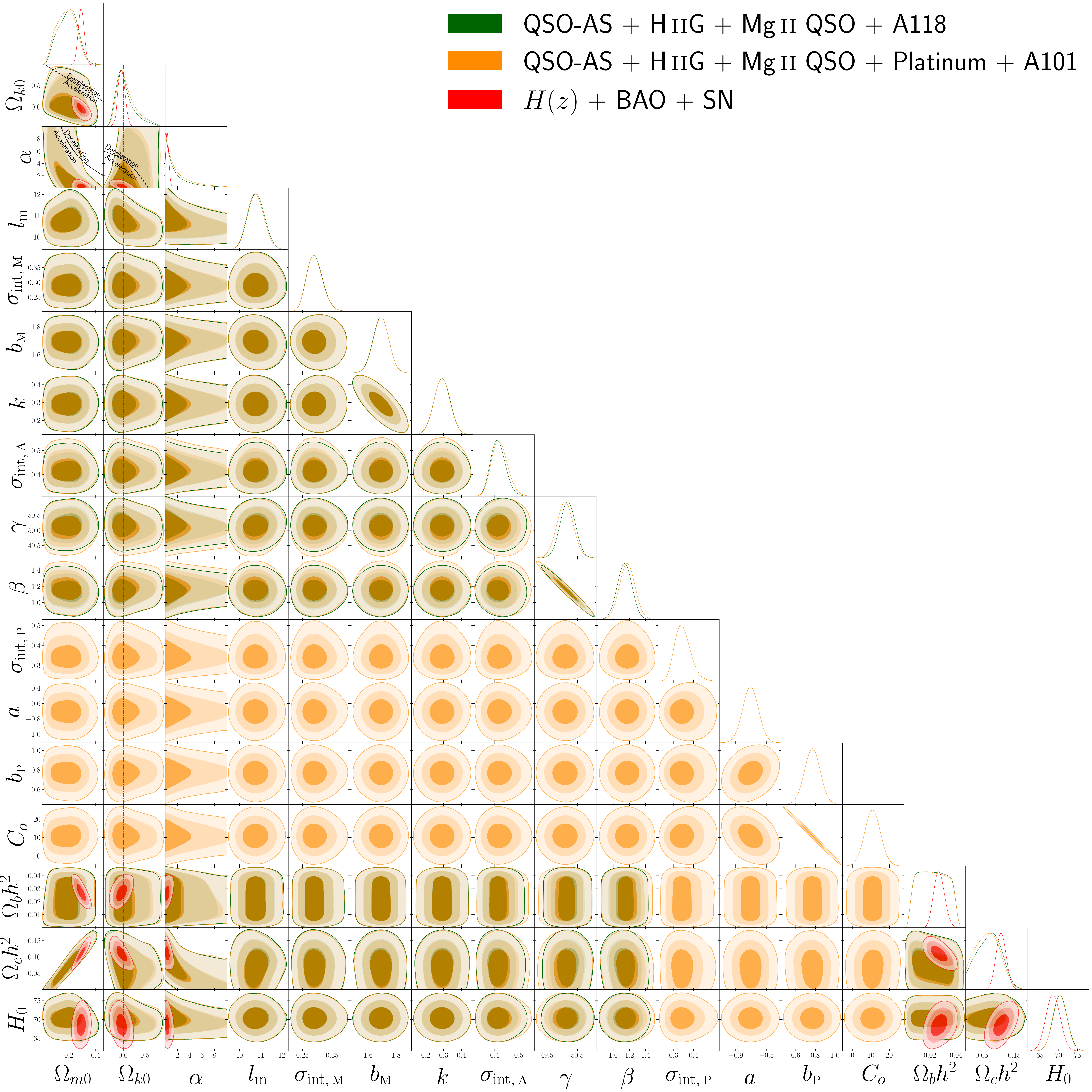}}\\
 \subfloat[]{%
    \includegraphics[width=0.5\textwidth,height=0.5\textwidth]{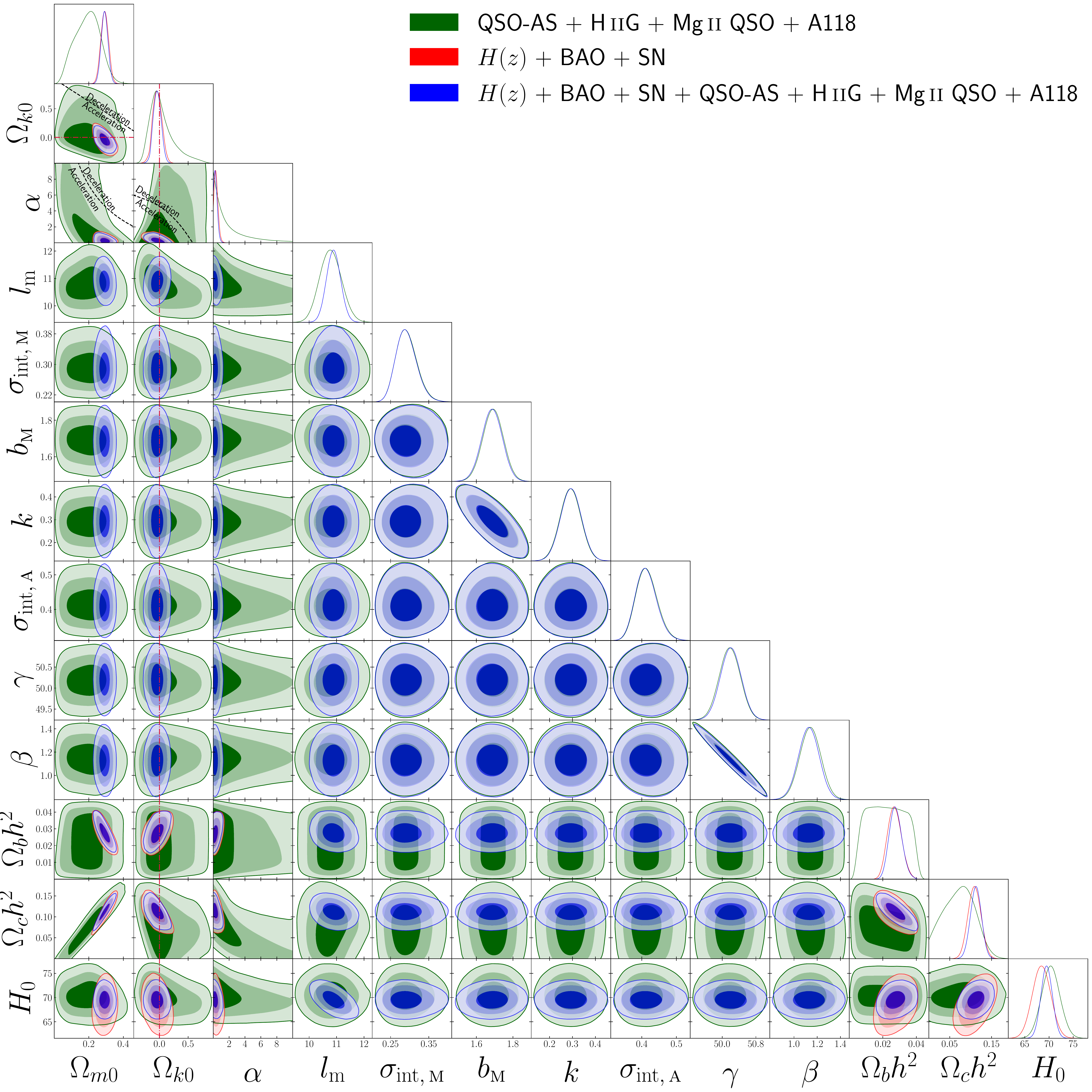}}
 \subfloat[]{%
    \includegraphics[width=0.5\textwidth,height=0.5\textwidth]{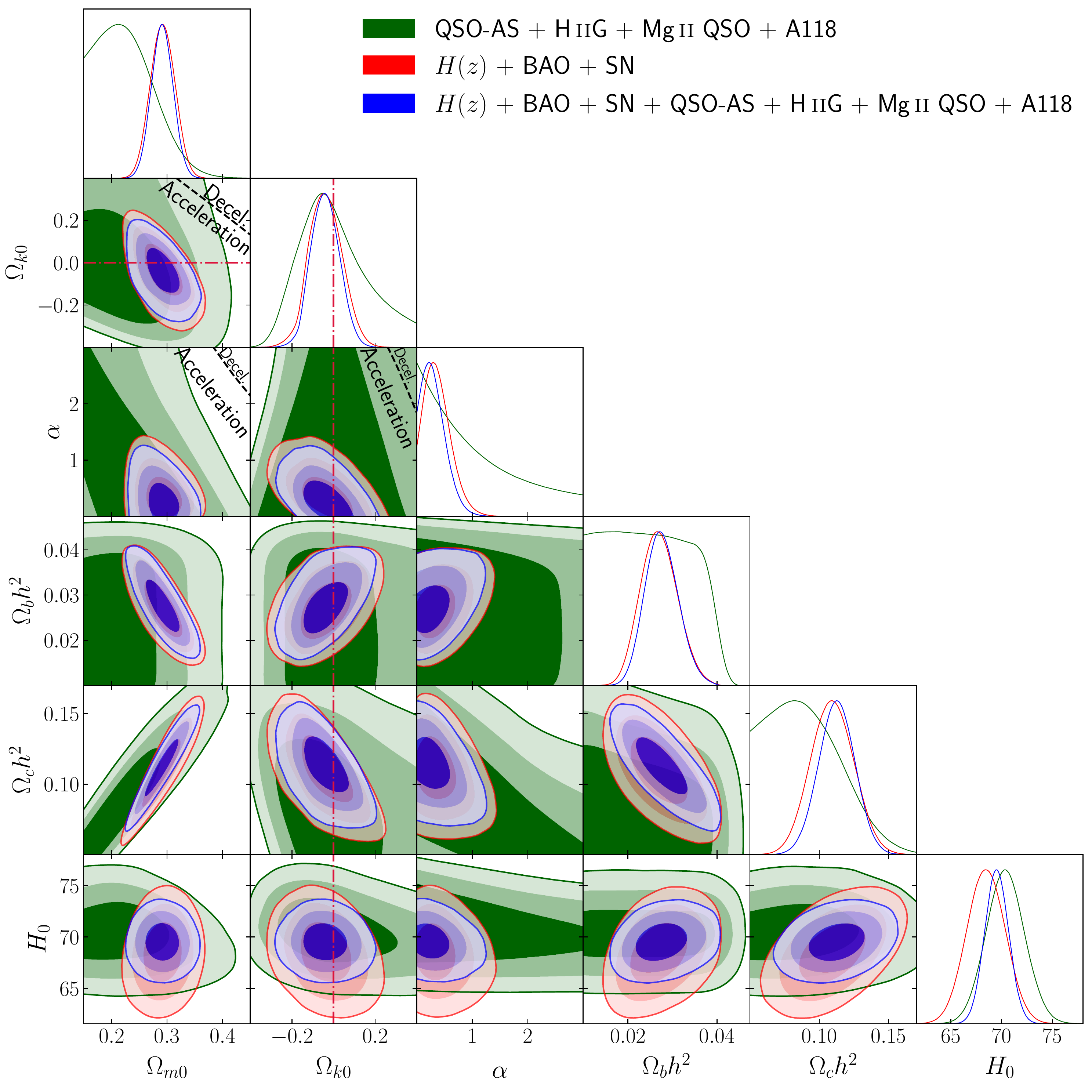}}\\
\caption{Same as Fig.\ \ref{fig5} but for non-flat \pcdm. The zero-acceleration black dashed lines are computed for the third cosmological parameter set to the $H(z)$ + BAO data best-fitting values listed in Table \ref{tab:BFP}, and divide the parameter space into regions associated with currently-accelerating (below left) and currently-decelerating (above right) cosmological expansion. The crimson dash-dot lines represent flat hypersurfaces, with closed spatial hypersurfaces either below or to the left. The $\alpha = 0$ axes correspond to non-flat \lcdm.}
\label{fig6}
\end{figure*}

The posterior one-dimensional probability distributions and two-dimensional confidence regions of the cosmological and non-cosmological parameters are shown in Figs.\ \ref{fig1}--\ref{fig6}, in red (QSO-AS + \hiig\ and $H(z)$ + BAO + SN), green (QSO-AS + \hiig\ + \mq\ + A118), orange (\mq\ + A118 and QSO-AS + \hiig\ + \mq\ + Platinum + A101, QHMPA101), and blue ($H(z)$ + BAO + SN + QSO-AS + \hiig\ + \mq\ + A118, HzBSNQHMA). The unmarginalized best-fitting parameter values, as well as the corresponding $-2\ln\mathcal{L}_{\rm max}$, AIC, BIC, DIC, $\Delta \mathrm{AIC}$, $\Delta \mathrm{BIC}$, and $\Delta \mathrm{DIC}$ values, for all models and data combinations, are listed in Table \ref{tab:BFP}, whereas the marginalized posterior mean parameter values and uncertainties ($\pm 1\sigma$ error bars or $2\sigma$ limits), for all models and data combinations, are listed in Table \ref{tab:1d_BFP}.\footnote{We use \textsc{python} package \textsc{getdist} \citep{Lewis_2019} to analyze the samples and generate the plots.}

In the non-flat \lcdm\ and flat and non-flat \pcdm\ models, \mq\ + A118 data mildly favor currently decelerating cosmological expansion, which is most likely caused by the choice of fixed $\Omega_b$ and $H_0$ values. All other data combinations more favor currently accelerating cosmological expansion.

\subsection{Constraints from $H(z)$, BAO, and SN Ia data}
 \label{subsec:HzBSN}

The updated $H(z)$ + BAO results derived here are quite similar to the $H(z)$ + BAO results given in \cite{CaoDainottiRatra2022}, so we do not discuss them in detail. $H(z)$ + BAO + SN is a more important data combination, so here we discuss these constraints in more detail. While the computation of the $H(z)$ + BAO + SN results reported in \cite{CaoRyanRatra2022} neglected the late-time contribution of non-relativistic neutrinos, in this paper, where we account for the contributions of one massive and two massless neutrino species, we find very similar constraints. 

The constraints from $H(z)$ + BAO + SN data on \om\ range from a low of $0.287\pm0.017$ (flat \pcdm) to a high of $0.304^{+0.014}_{-0.015}$ (flat \lcdm), with a difference of $0.75\sigma$. 

The $H_0$ constraints range from a low of $68.29\pm1.78$ \hunit\ (flat \pcdm) to a high of $69.04\pm1.77$ \hunit\ (flat \lcdm), with a difference of $0.30\sigma$, which are $0.09\sigma$ (flat \pcdm) and $0.31\sigma$ (flat \lcdm) higher than the median statistics estimate of $H_0=68\pm2.8$ \hunit\ \citep{chenratmed}, and $2.23\sigma$ (flat \pcdm) and $1.89\sigma$ (flat \lcdm) lower than the local Hubble constant measurement of $H_0 = 73.2 \pm 1.3$ \hunit\ \citep{Riess_2021}.

The constraints on \ok\ are $0.040\pm0.070$, $-0.001\pm0.098$, and $-0.038^{+0.071}_{-0.085}$ for non-flat \lcdm, XCDM, and \pcdm, respectively. Although non-flat hypersurfaces are mildly favored, flat hypersurfaces are well within 1$\sigma$.

There is a slight preference for dark energy dynamics. For flat (non-flat) XCDM, $w_{\rm X}=-0.941\pm0.064$ ($w_{\rm X}=-0.948^{+0.098}_{-0.068}$), with central values being $0.92\sigma$ ($0.76\sigma$) higher than $w_{\rm X}=-1$; and for flat (non-flat) \pcdm, $\alpha=0.324^{+0.122}_{-0.264}$ ($\alpha=0.382^{+0.151}_{-0.299}$), with central values being $1.23\sigma$ ($1.28\sigma$) away from $\alpha=0$.

\subsection{Constraints from QSO-AS, \hiig, \mq, A118, and Platinum + A101 data}
\label{subsec:QHMA}

Given our improved treatment of neutrinos in this paper, compared to our earlier analyses, we have reanalyzed data we had previously studied. 

As shown in \cite{CaoRyanRatra2022}, QSO-AS data alone do not deal well with $H_0$, so an unbounded prior range for $H_0$ makes it hard for the computation to converge and results in an unreasonably high $H_0$ value and so an unreasonably low $l_{\rm m}$ value. However, we expect constraints on the other cosmological parameters consistent with those given in \cite{CaoRyanRatra2022}. Constraints from \hiig\ data are consistent with what are given in \cite{CaoRyanRatra2022}. Constraints from \mq\ data are consistent with those described in \cite{Khadkaetal_2021a} while those from A118 and Platinum + A101 data are consistent with those in \cite{CaoDainottiRatra2022}. We find that cosmological parameter constraints from those four data sets are mutually consistent so they can be used to do joint analyses. As expected, cosmological parameter constraints from the joint QSO-AS + \hiig\ data and \mq\ + A118 data are indeed mutually consistent, as seen in Tables \ref{tab:BFP} and \ref{tab:1d_BFP}. We do not discuss these results in detail since there are no significant changes compared to those derived in our earlier analyses. We consider the joint analyses results of QSO-AS + \hiig\ + \mq\ + A118 data to be more useful and discuss these in more detail next.\footnote{We note that cosmological parameter constraints from Platinum, A101, and Platinum + A101 data are also consistent with those from QSO-AS, \hiig, and \mq\ data, so we also investigate the joint QSO-AS + \hiig\ + \mq\ + Platinum + A101 (QHPMAP101) data combination. As seen in Tables \ref{tab:BFP} and \ref{tab:1d_BFP}, we find no significant differences between the QHPMAP101 cosmological constraints and those from the QSO-AS + \hiig\ + \mq\ + A118 data combination that contains fewer non-cosmological parameters.}

The constraints from QSO-AS + \hiig\ + \mq\ + A118 data on \om\ range from a low of $0.175^{+0.075}_{-0.081}$ (flat \pcdm) to a high of $0.314^{+0.051}_{-0.044}$ (flat XCDM), with a difference of $1.60\sigma$. Following the pattern of \hiig\ data, the \om\ difference is relatively large.

The $H_0$ constraints range from a low of $70.38\pm1.84$ \hunit\ (non-flat \pcdm) to a high of $73.14^{+2.14}_{-2.48}$ \hunit\ (flat XCDM), with a difference of $0.89\sigma$, which are $0.71\sigma$ (non-flat \pcdm) and $1.37\sigma$ (flat XCDM) higher than the median statistics estimate of $H_0=68\pm2.8$ \hunit\ \citep{chenratmed}, and $1.25\sigma$ (non-flat \pcdm) and $0.02\sigma$ (flat XCDM) lower than the local Hubble constant measurement of $H_0 = 73.2 \pm 1.3$ \hunit\ \citep{Riess_2021}.

The constraints on \ok\ are $-0.139^{+0.116}_{-0.228}$, $0.054^{+0.227}_{-0.238}$, and $0.044^{+0.104}_{-0.256}$ for non-flat \lcdm, XCDM, and \pcdm, respectively. As opposed to $H(z)$ + BAO + SN results, non-flat \lcdm\ mildly favors closed hypersurfaces, whereas non-flat XCDM and non-flat \pcdm\ mildly favor open hypersurfaces. However, flat hypersurfaces are well within 1$\sigma$.

There are mild preferences for dark energy dynamics. For flat (non-flat) XCDM, $w_{\rm X}=-1.836^{+0.804}_{-0.419}$ ($w_{\rm X}=-2.042^{+1.295}_{-0.451}$), with central values being $1.04\sigma$ ($0.80\sigma$) lower than $w_{\rm X}=-1$; and for flat (non-flat) \pcdm, $\alpha<6.756$ ($\alpha<7.239$), with $\alpha=0$ being within $1\sigma$.

\subsection{Constraints from $H(z)$ + BAO + SN + QSO-AS + \hiig\ + \mq\ + A118 (HzBSNQHMA) data}
\label{subsec:HzBSNQHMA}

Cosmological parameter constraints from $H(z)$ + BAO + SN data are consistent with those from QSO-AS + \hiig\ + \mq\ + A118 data. From model to model, there are differences, ranging from $-0.41\sigma$ (flat XCDM) to $1.45\sigma$ (flat \pcdm), between \om\ constraints from $H(z)$ + BAO + SN data and those from QSO-AS + \hiig\ + \mq\ + A118 data; %0.95, 0.52, -0.41, 0.02, 1.45, 1.24
and there are differences, ranging from $0.73\sigma$ (non-flat \pcdm) to $1.48\sigma$ (flat XCDM), between $H_0$ constraints from QSO-AS + \hiig\ + \mq\ + A118 data and those from $H(z)$ + BAO + SN data. %0.91, 1.18, 1.48, 1.38, 0.88, 0.73
For the XCDM parametrizations, $H(z)$ + BAO + SN data slightly prefer non-phantom dark energy dynamics, whereas QSO-AS + \hiig\ + \mq\ + A118 data prefer phantom dark energy dynamics, however, their differences are within $1\sigma$. As can be seen in the (d) panels of Figs.\ \ref{fig1}--\ref{fig6}, the two-dimensional posterior cosmological constraints from $H(z)$ + BAO + SN data and QSO-AS + \hiig\ + \mq\ + A118 data are significantly more mutually consistent than the less than $1.5\sigma$ differences between the maximum and minimum one-dimensional posterior mean values discussed above. Consequently we can combine these data in a joint HzBSNQHMA data analysis; we discuss the results from this analysis next.

The constraints from HzBSNQHMA data on \om\ range from a low of $0.286\pm0.015$ (flat \pcdm) to a high of $0.300\pm0.012$ (flat \lcdm), with a difference of $0.73\sigma$.

The $H_0$ constraints range from a low of $69.50\pm1.14$ \hunit\ (flat \pcdm) to a high of $69.87\pm1.13$ \hunit\ (flat \lcdm), with a difference of $0.23\sigma$, which are $0.50\sigma$ (flat \pcdm) and $0.62\sigma$ (flat \lcdm) higher than the median statistics estimate of $H_0=68\pm2.8$ \hunit\ \citep{chenratmed}, and $2.14\sigma$ (flat \pcdm) and $1.93\sigma$ (flat \lcdm) lower than the local Hubble constant measurement of $H_0 = 73.2 \pm 1.3$ \hunit\ \citep{Riess_2021}.\footnote{Other local determinations of $H_0$ result in somewhat lower central values with somewhat larger error bars \citep{rigault_etal_2015,zhangetal2017,Dhawan,FernandezArenas,Breuvaletal_2020, Efstathiou_2020, Khetan_et_al_2021,rameez_sarkar_2021, Freedman2021}. Our $H_0$ determinations here are consistent with earlier median statistics estimates \citep{gott_etal_2001, Calabreseetal2012} and with other recent $H_0$ determinations \citep{chen_etal_2017,DES_2018,Gomez-ValentAmendola2018, planck2018b,dominguez_etal_2019,Cuceu_2019,zeng_yan_2019,schoneberg_etal_2019, Blum_et_al_2020, Lyu_et_al_2020, Philcox_et_al_2020, Birrer_et_al_2020, Denzel_et_al_2020,Pogosianetal_2020,Kimetal_2020,Harvey_2020,Boruahetal_2021, Zhang_Huang_2021,lin_ishak_2021,Wuetal2022}.}

The constraints on \ok\ are $0.018\pm0.059$, $-0.009^{+0.077}_{-0.083}$, and $-0.040^{+0.064}_{-0.072}$ for non-flat \lcdm, XCDM, and \pcdm, respectively. Following the same pattern as $H(z)$ + BAO + SN data results, flat hypersurfaces are also well within 1$\sigma$.

There is a slight preference for dark energy dynamics. For flat (non-flat) XCDM, $w_{\rm X}=-0.959\pm0.059$ ($w_{\rm X}=-0.959^{+0.090}_{-0.063}$), with central values being $0.69\sigma$ ($0.65\sigma$) higher than $w_{\rm X}=-1$; and for flat (non-flat) \pcdm, $\alpha=0.249^{+0.069}_{-0.239}$ ($\alpha=0.316^{+0.101}_{-0.292}$), with central values being $1.04\sigma$ ($1.08\sigma$) away from $\alpha=0$.

\subsection{Model Comparison}
 \label{subsec:comp}

From the AIC, BIC, and DIC values listed in Table \ref{tab:BFP}, we find the following results:
\begin{itemize}
    \item[1)]{\bf AIC} $H(z)$ + BAO data favor flat \pcdm\ the most, \mq\ + A118 data favor flat XCDM the most, QSO-AS + \hiig\ data favor non-flat XCDM the most, and the other data combinations favor flat \lcdm\ the most. However the evidence against the rest of the models/parametrizations is either only weak or positive.
    
    \item[2)] {\bf BIC} All data combinations favor flat \lcdm\ the most. $H(z)$ + BAO data only provide weak or positive evidence against other models/parametrizations.
    
    Both \mq\ + A118 and QSO-AS + \hiig\ data provide strong (very strong) evidence against non-flat XCDM (non-flat \pcdm) and positive evidence against the others. 
    
    QSO-AS + \hiig\ + \mq\ + A118 data provide strong (very strong) evidence against flat \pcdm\ (non-flat XCDM and non-flat \pcdm) and positive evidence against non-flat \lcdm\ and flat XCDM. 
    
    $H(z)$ + BAO + SN data provide strong (very strong) evidence against non-flat \lcdm\ (non-flat XCDM and non-flat \pcdm) and positive evidence against flat XCDM and flat \pcdm. 
    
    $H(z)$ + BAO + SN + QSO-AS + \hiig\ + \mq\ + A118 data provide very strong evidence against non-flat XCDM and non-flat \pcdm, and strong evidence against the others. 
    
    QSO-AS + \hiig\ + \mq\ + Platinum + A101 data provide strong (very strong) evidence against flat \pcdm\ (non-flat XCDM and non-flat \pcdm) and positive evidence against non-flat \lcdm\ and flat XCDM.

    \item[3)] {\bf DIC} $H(z)$ + BAO, \mq\ + A118, and $H(z)$ + BAO + SN data favor flat \pcdm\ the most, and the other data combinations favor flat \lcdm\ the most. There is strong evidence against non-flat XCDM from QSO-AS + \hiig\ data, strong evidence against non-flat \pcdm\ from QSO-AS + \hiig, QSO-AS + \hiig\ + \mq\ + A118, and QSO-AS + \hiig\ + \mq\ + Platinum + A101 data, and weak or positive evidence against the others from the remaining data sets.
\end{itemize}

Perhaps the most reliable summary conclusion is that, based on DIC, the $H(z)$ + BAO + SN + QSO-AS + \hiig\ + \mq\ + A118 data combination does not provide strong evidence against any of the cosmological models/parametrizations.

\section{Conclusion}
\label{sec:conc}

In this paper we use many of the most up-to-date available non-CMB data sets to determine cosmological constraints. We analyze 32 $H(z)$, 12 BAO, 1048 Pantheon SN Ia, 20 binned DES-3yr SN Ia, 120 QSO-AS, 181 \hiig, 78 \mq, 118 (101) A118 (A101) GRB, and 50 Platinum GRB measurements and find that the cosmological constraints from each data set are mutually consistent. We find very small differences between cosmological constraints determined from QSO-AS + \hiig\ + \mq\ + A118 data and those from QSO-AS + \hiig\ + \mq\ + Platinum + A101 data, so report only the cosmological constraints from joint $H(z)$ + BAO + SN + QSO-AS + \hiig\ + \mq\ + A118 (HzBSNQHMA) data.

The HzBSNQHMA data provide a fairly restrictive summary value\footnote{As in \cite{CaoRyanRatra2021, CaoRyanRatra2022}, the summary central value is computed from the mean of the two central-most of the six mean values and the summary uncertainty is computed from the quadrature sum of the systematic uncertainty, defined to be half of the difference between the two central-most mean values, and the statistical uncertainty, defined to be the average of the error bars of the two central-most results.} of $\Om=0.295\pm0.017$ that agrees well with many other recent measurements and a fairly restrictive summary value of $H_0=69.7\pm1.2$ \hunit\ that is in better agreement with the result of \cite{chenratmed} than with the result of \cite{Riess_2021}.\footnote{Our model-independent $H_0$ error bar is slightly smaller than that of \cite{Riess_2021} and is a factor of 2.2 larger than that of the flat \lcdm\ model \citep{planck2018b} TT,TE,EE+lowE+lensing CMB anisotropy $H_0$ error bar while our $\Om$ error bar is a factor of 2.3 larger than the corresponding \textit{Planck} flat \lcdm\ one.} Our $H_0$ measurement here lies in the middle of the flat \lcdm\ model result of \cite{planck2018b} and the local expansion rate result of \cite{Riess_2021}, slightly closer to the former. Based on DIC, the HzBSNQHMA data compilation prefers flat \lcdm\ the most, but does not rule out mild dark energy dynamics or a little spatial curvature energy density (evidence against them is either weak or positive). 

We hope that in the near future the quality and amount of the types of lower-redshift, non-CMB, data we have used here will improve enough to result in cosmological parameter error bars comparable to those from \textit{Planck} CMB anisotropy data.

\begin{sidewaystable*}
\centering
\resizebox*{\columnwidth}{0.74\columnwidth}{%
\begin{threeparttable}
\caption{Unmarginalized best-fitting parameter values for all models from various combinations of data.}\label{tab:BFP}
\begin{tabular}{lccccccccccccccccccccccccc}
\toprule
Model & Data set & $\Omega_{b}h^2$ & $\Omega_{c}h^2$ & $\Omega_{m0}$ & $\Omega_{k0}$ & $w_{\mathrm{X}}$/$\alpha$\tnote{a} & $H_0$\tnote{b} & $l_{\mathrm{m}}$\tnote{c} & $\sigma_{\mathrm{int,\,\textsc{m}}}$ & $b_{\mathrm{\textsc{m}}}$ & $k$ & $\sigma_{\mathrm{int,\,\textsc{a}}}$ & $\gamma$ & $\beta$ & $\sigma_{\mathrm{int,\,\textsc{p}}}$ & $a$ & $b_{\mathrm{\textsc{p}}}$ & $C_{o}$ & $-2\ln\mathcal{L}_{\mathrm{max}}$ & AIC & BIC & DIC & $\Delta \mathrm{AIC}$ & $\Delta \mathrm{BIC}$ & $\Delta \mathrm{DIC}$ \\
\midrule
 & $H(z)$ + BAO & 0.0244 & 0.1181 & 0.301 & -- & -- & 68.98 & -- & -- & -- & -- & -- & -- & -- & -- & -- & -- & -- & 25.64 & 31.64 & 36.99 & 32.32 & 0.00 & 0.00 & 0.00\\
 & \mq\ + A118 & -- & 0.2526 & 0.567 & -- & -- & -- & -- & 0.286 & 1.711 & 0.300 & 0.398 & 50.12 & 1.105 & -- & -- & -- & -- & 159.64 & 173.64 & 196.59 & 173.37 & 0.00 & 0.00 & 0.00\\
Flat & QSO-AS + \hiig & 0.0332 & 0.0947 & 0.251 & -- & -- & 71.53 & 11.06 & -- & -- & -- & -- & -- & -- & -- & -- & -- & -- & 786.45 & 794.45 & 809.28 & 792.69 & 0.00 & 0.00 & 0.00\\
\lcdm & QSO-AS + \hiig\ + \mq\ + A118 & 0.0183 & 0.1127 & 0.258 & -- & -- & 71.50 & 11.04 & 0.282 & 1.682 & 0.294 & 0.404 & 50.23 & 1.124 & -- & -- & -- & -- & 947.34 & 967.34 & 1009.43 & 966.23 & 0.00 & 0.00 & 0.00\\
 & $H(z)$ + BAO + SN & 0.0242 & 0.1191 & 0.304 & -- & -- & 68.86 & -- & -- & -- & -- & -- & -- & -- & -- & -- & -- & -- & 1082.39 & 1088.39 & 1103.44 & 1089.92 & 0.00 & 0.00 & 0.00\\
 & HzNBSNQHMA\tnote{e} & 0.0258 & 0.1207 & 0.300 & -- & -- & 70.06 & 10.93 & 0.286 & 1.671 & 0.296 & 0.406 & 50.26 & 1.110 & -- & -- & -- & -- & 2031.30 & 2051.30 & 2105.13 & 2051.86 & 0.00 & 0.00 & 0.00\\
 & QHMPA101\tnote{e} & 0.0197 & 0.1144 & 0.264 & -- & -- & 71.41 & 11.02 & 0.283 & 1.673 & 0.315 & 0.410 & 50.11 & 1.169 & 0.321 & $-0.746$ & 0.773 & 10.81 & 964.62 & 992.62 & 1052.44 & 990.65 & 0.00 & 0.00 & 0.00\\
% \\
\midrule
 & $H(z)$ + BAO & 0.0260 & 0.1098 & 0.292 & 0.048 & -- & 68.35 & -- & -- & -- & -- & -- & -- & -- & -- & -- & -- & -- & 25.30 & 33.30 & 40.43 & 33.87 & 1.66 & 3.44 & 1.54\\
 & \mq\ + A118 & -- & 0.2692 & 0.601 & 0.252 & -- & -- & -- & 0.281 & 1.713 & 0.308 & 0.401 & 50.07 & 1.117 & -- & -- & -- & -- & 159.65 & 175.65 & 201.87 & 174.46 & 2.01 & 5.28 & 1.09\\
Non-flat & QSO-AS + \hiig & 0.0228 & 0.1145 & 0.260 & $-0.360$ & -- & 72.91 & 11.60 & -- & -- & -- & -- & -- & -- & -- & -- & -- & -- & 784.18 & 794.18 & 812.71 & 793.24 & $-0.27$ & 3.43 & 0.55\\
\lcdm & QSO-AS + \hiig\ + \mq\ + A118 & 0.0261 & 0.1186 & 0.278 & $-0.254$ & -- & 72.25 & 11.29 & 0.282 & 1.678 & 0.299 & 0.410 & 50.27 & 1.106 & -- & -- & -- & -- & 945.97 & 967.97 & 1014.27 & 966.91 & 0.63 & 4.84 & 0.69\\
 & $H(z)$ + BAO + SN & 0.0255 & 0.1121 & 0.295 & 0.035 & -- & 68.53 & -- & -- & -- & -- & -- & -- & -- & -- & -- & -- & -- & 1082.11 & 1090.11 & 1110.16 & 1091.17 & 1.72 & 6.72 & 1.24\\
 & HzBSNQHMA\tnote{d} & 0.0261 & 0.1182 & 0.297 & 0.008 & -- & 69.90 & 10.94 & 0.281 & 1.674 & 0.301 & 0.400 & 50.20 & 1.126 & -- & -- & -- & -- & 2031.26 & 2053.26 & 2112.48 & 2053.69 & 1.96 & 7.35 & 1.84\\
 & QHMPA101\tnote{e} & 0.0246 & 0.1161 & 0.276 & $-0.221$ & -- & 71.56 & 11.37 & 0.287 & 1.682 & 0.304 & 0.411 & 50.20 & 1.132 & 0.329 & $-0.674$ & 0.790 & 9.64 & 963.83 & 993.83 & 1057.92 & 992.04 & 1.21 & 5.48 & 1.39\\
% \\
\midrule
 & $H(z)$ + BAO & 0.0296 & 0.0951 & 0.290 & -- & $-0.754$ & 65.79 & -- & -- & -- & -- & -- & -- & -- & -- & -- & -- & -- & 22.39 & 30.39 & 37.52 & 30.63 & $-1.25$ & 0.53 & $-1.69$\\
 & \mq\ + A118 & -- & 0.0448 & 0.143 & -- & $-4.972$ & -- & -- & 0.279 & 1.554 & 0.292 & 0.407 & 50.65 & 1.126 & -- & -- & -- & -- & 156.84 & 172.84 & 199.07 & 174.36 & $-0.80$ & 2.48 & 0.99\\
Flat & QSO-AS + \hiig & 0.0174 & 0.1308 & 0.285 & -- & $-1.280$ & 72.32 & 11.23 & -- & -- & -- & -- & -- & -- & -- & -- & -- & -- & 786.05 & 796.05 & 814.58 & 795.03 & 1.60 & 5.30 & 2.34\\
XCDM & QSO-AS + \hiig\ + \mq\ + A118 & 0.0069 & 0.1558 & 0.307 & -- & $-1.505$ & 72.94 & 11.32 & 0.282 & 1.679 & 0.290 & 0.401 & 50.24 & 1.122 & -- & -- & -- & -- & 946.23 & 968.23 & 1014.52 & 967.53 & 0.89 & 5.09 & 1.31\\
 & $H(z)$ + BAO + SN & 0.0258 & 0.1115 & 0.295 & -- & $-0.940$ & 68.37 & -- & -- & -- & -- & -- & -- & -- & -- & -- & -- & -- & 1081.34 & 1089.34 & 1109.40 & 1090.43 & 0.95 & 5.96 & 0.50\\
 & HzBSNQHMA\tnote{d} & 0.0266 & 0.1162 & 0.296 & -- & $-0.975$ & 69.62 & 10.94 & 0.280 & 1.696 & 0.278 & 0.406 & 50.24 & 1.118 & -- & -- & -- & -- & 2030.88 & 2052.88 & 2112.10 & 2053.30 & 1.58 & 6.97 & 1.44\\
 & QHMPA101\tnote{e} & 0.0176 & 0.1450 & 0.311 & -- & $-1.573$ & 72.46 & 11.41 & 0.276 & 1.686 & 0.294 & 0.417 & 50.12 & 1.168 & 0.326 & $-0.682$ & 0.756 & 11.45 & 963.77 & 993.77 & 1057.86 & 992.26 & 1.15 & 5.42 & 1.61\\
% \\
\midrule
 & $H(z)$ + BAO & 0.0289 & 0.0985 & 0.296 & $-0.053$ & $-0.730$ & 65.76 & -- & -- & -- & -- & -- & -- & -- & -- & -- & -- & -- & 22.13 & 32.13 & 41.05 & 32.51 & 0.49 & 4.06 & 0.19\\
 & \mq\ + A118 & -- & 0.0169 & 0.086 & $-0.027$ & $-5.000$ & -- & -- & 0.279 & 1.473 & 0.298 & 0.396 & 50.91 & 1.118 & -- & -- & -- & -- & 156.52 & 174.52 & 204.03 & 177.50 & 0.88 & 7.44 & 4.13\\
Non-flat & QSO-AS + \hiig & 0.0300 & 0.0031 & 0.065 & $-0.560$ & $-0.651$ & 71.87 & 11.45 & -- & -- & -- & -- & -- & -- & -- & -- & -- & -- & 781.18 & 793.18 & 815.43 & 799.59 & $-1.27$ & 6.15 & 6.90\\
XCDM & QSO-AS + \hiig\ + \mq\ + A118 & 0.0365 & 0.1224 & 0.307 & $-0.087$ & $-1.278$ & 72.14 & 11.33 & 0.292 & 1.677 & 0.287 & 0.398 & 50.23 & 1.123 & -- & -- & -- & -- & 946.08 & 970.08 & 1020.59 & 969.33 & 2.74 & 11.16 & 3.11\\
 & $H(z)$ + BAO + SN & 0.0255 & 0.1155 & 0.300 & $-0.030$ & $-0.922$ & 68.68 & -- & -- & -- & -- & -- & -- & -- & -- & -- & -- & -- & 1081.28 & 1091.28 & 1116.35 & 1092.47 & 2.89 & 12.91 & 2.55\\
 & HzBSNQHMA\tnote{d} & 0.0260 & 0.1158 & 0.295 & $-0.016$ & $-0.947$ & 69.53 & 10.94 & 0.288 & 1.697 & 0.277 & 0.409 & 50.15 & 1.151 & -- & -- & -- & -- & 2030.78 & 2054.78 & 2119.38 & 2055.42 & 3.48 & 14.25 & 3.56\\
 & QHMPA101\tnote{e} & 0.0354 & 0.1154 & 0.292 & $-0.048$ & $-1.221$ & 72.03 & 11.31 & 0.272 & 1.680 & 0.294 & 0.313 & 50.06 & 1.178 & 0.313 & $-0.736$ & 0.769 & 10.98 & 963.91 & 995.91 & 1064.28 & 993.76 & 3.29 & 11.84 & 3.11\\
% \\
\midrule
 & $H(z)$ + BAO & 0.0310 & 0.0900 & 0.280 & -- & 1.010 & 65.89 & -- & -- & -- & -- & -- & -- & -- & -- & -- & -- & -- & 22.31 & 30.31 & 37.45 & 29.90 & $-1.33$ & 0.46 & $-2.42$\\
 & \mq\ + A118 & -- & 0.2760 & 0.615 & -- & 0.353 & -- & -- & 0.279 & 1.714 & 0.308 & 0.403 & 50.09 & 1.105 & -- & -- & -- & -- & 159.69 & 175.69 & 201.91 & 172.93 & 2.05 & 5.32 & $-0.44$\\
Flat & QSO-AS + \hiig & 0.0198 & 0.1066 & 0.249 & -- & 0.000 & 71.42 & 11.10 & -- & -- & -- & -- & -- & -- & -- & -- & -- & -- & 786.46 & 796.46 & 815.00 & 796.31 & 2.01 & 5.72 & 3.62\\
\pcdm & QSO-AS + \hiig\ + \mq\ + A118 & 0.0098 & 0.1196 & 0.253 & -- & 0.020 & 71.63 & 10.98 & 0.283 & 1.683 & 0.289 & 0.416 & 50.18 & 1.145 & -- & -- & -- & -- & 947.60 & 969.60 & 1015.89 & 970.30 & 2.26 & 6.46 & 4.07\\
 & $H(z)$ + BAO + SN & 0.0263 & 0.1097 & 0.292 & -- & 0.203 & 68.39 & -- & -- & -- & -- & -- & -- & -- & -- & -- & -- & -- & 1081.22 & 1089.22 & 1109.28 & 1089.91 & 0.83 & 5.84 & $-0.01$\\
 & HzBSNQHMA\tnote{d} & 0.0274 & 0.1116 & 0.289 & -- & 0.150 & 69.51 & 10.97 & 0.280 & 1.685 & 0.292 & 0.409 & 50.23 & 1.121 & -- & -- & -- & -- & 2030.52 & 2052.52 & 2111.74 & 2053.18 & 1.22 & 6.61 & 1.33\\
 & QHMPA101\tnote{e} & 0.0388 & 0.0884 & 0.251 & -- & 0.105 & 71.31 & 11.13 & 0.287 & 1.679 & 0.293 & 0.402 & 50.11 & 1.168 & 0.317 & $-0.664$ & 0.775 & 10.43 & 964.97 & 994.97 & 1059.07 & 994.25 & 2.35 & 6.63 & 3.60\\
% \\
\midrule
 & $H(z)$ + BAO & 0.0306 & 0.0920 & 0.284 & $-0.058$ & 1.200 & 65.91 & -- & -- & -- & -- & -- & -- & -- & -- & -- & -- & -- & 22.05 & 32.05 & 40.97 & 31.30 & 0.41 & 3.98 & $-1.02$\\
 & \mq\ + A118 & -- & 0.2233 & 0.507 & 0.101 & 6.181 & -- & -- & 0.280 & 1.723 & 0.303 & 0.401 & 50.04 & 1.115 & -- & -- & -- & -- & 159.63 & 177.63 & 207.13 & 173.84 & 3.99 & 10.54 & 0.47\\
Non-flat & QSO-AS + \hiig & 0.0338 & 0.0979 & 0.251 & $-0.250$ & 0.000 & 72.53 & 11.47 & -- & -- & -- & -- & -- & -- & -- & -- & -- & -- & 784.61 & 796.61 & 818.85 & 801.32 & 2.16 & 9.57 & 8.63\\
\pcdm & QSO-AS + \hiig\ + \mq\ + A118 & 0.0073 & 0.1290 & 0.265 & $-0.228$ & 0.115 & 71.91 & 11.29 & 0.288 & 1.673 & 0.300 & 0.412 & 50.23 & 1.120 & -- & -- & -- & -- & 946.09 & 970.09 & 1020.59 & 972.93 & 2.75 & 11.16 & 6.70\\
 & $H(z)$ + BAO + SN & 0.0261 & 0.1119 & 0.295 & $-0.023$ & 0.253 & 68.56 & -- & -- & -- & -- & -- & -- & -- & -- & -- & -- & -- & 1081.12 & 1091.12 & 1116.19 & 1091.27 & 2.73 & 12.75 & 1.35\\
 & HzBSNQHMA\tnote{d} & 0.0251 & 0.1207 & 0.300 & $-0.056$ & 0.195 & 69.84 & 10.91 & 0.284 & 1.699 & 0.279 & 0.411 & 50.19 & 1.132 & -- & -- & -- & -- & 2030.76 & 2054.76 & 2119.36 & 2055.13 & 3.46 & 14.23 & 3.28\\
 & QHMPA101\tnote{e} & 0.0370 & 0.0977 & 0.260 & $-0.217$ & 0.066 & 72.15 & 11.23 & 0.274 & 1.689 & 0.277 & 0.408 & 50.08 & 1.180 & 0.335 & $-0.703$ & 0.777 & 10.42 & 963.77 & 995.77 & 1064.14 & 997.14 & 3.15 & 11.70 & 6.49\\
\bottomrule
\end{tabular}
%}
\begin{tablenotes}[flushleft]
\item [a] \wx\ corresponds to flat/non-flat XCDM and $\alpha$ corresponds to flat/non-flat \pcdm.
\item [b] \hunit. In these GRB cases, $\Omega_b$ and $H_0$ are set to be 0.05 and 70 \hunit, respectively.
\item [c] pc.
\item [d] $H(z)$ + BAO + SN + QSO-AS + \hiig\ + \mq\ + A118.
\item [e] QSO-AS + \hiig\ + \mq\ + Platinum + A101.
\end{tablenotes}
\end{threeparttable}%
}
\end{sidewaystable*}

\begin{sidewaystable*}
\centering
\resizebox*{\columnwidth}{0.74\columnwidth}{%
\begin{threeparttable}
\caption{One-dimensional marginalized posterior mean values and uncertainties ($\pm 1\sigma$ error bars or $2\sigma$ limits) of the parameters for all models from various combinations of data.}\label{tab:1d_BFP}
\begin{tabular}{lcccccccccccccccccc}
\toprule
Model & Data set & $\Omega_{b}h^2$ & $\Omega_{c}h^2$ & $\Omega_{m0}$ & $\Omega_{k0}$ & $w_{\mathrm{X}}$/$\alpha$\tnote{a} & $H_0$\tnote{b} & $l_{\mathrm{m}}$\tnote{c} & $\sigma_{\mathrm{int,\,\textsc{m}}}$ & $b_{\mathrm{\textsc{m}}}$ & $k$ & $\sigma_{\mathrm{int,\,\textsc{a}}}$ & $\gamma$ & $\beta$ & $\sigma_{\mathrm{int,\,\textsc{p}}}$ & $a$ & $b_{\mathrm{\textsc{p}}}$ & $C_{o}$ \\
\midrule
 & $H(z)$ + BAO & $0.0247\pm0.0030$ & $0.1186^{+0.0076}_{-0.0083}$ & $0.301^{+0.016}_{-0.018}$ & -- & -- & $69.14\pm1.85$ & -- & -- & -- & -- & -- & -- & -- & -- & -- & -- & -- \\
 & \mq\ + A118 & -- & -- & $0.609^{+0.294}_{-0.210}$ & -- & -- & -- & -- & $0.293^{+0.023}_{-0.030}$ & $1.712\pm0.056$ & $0.301\pm0.047$ & $0.411^{+0.027}_{-0.033}$ & $50.10\pm0.25$ & $1.112\pm0.088$ & -- & -- & -- & -- \\%$>0.230$
Flat & QSO-AS + \hiig & $0.0225\pm0.0113$ & $0.1076^{+0.0197}_{-0.0224}$ & $0.257^{+0.037}_{-0.047}$ & -- & -- & $71.52\pm1.79$ & $11.04\pm0.34$ & -- & -- & -- & -- & -- & -- & -- & -- & -- & -- \\
\lcdm & QSO-AS + \hiig\ + \mq\ + A118 & $0.0224\pm0.0116$ & $0.1118^{+0.0197}_{-0.0223}$ & $0.266^{+0.037}_{-0.047}$ & -- & -- & $71.28\pm1.73$ & $11.00\pm0.33$ & $0.292^{+0.023}_{-0.029}$ & $1.686\pm0.055$ & $0.291\pm0.044$ & $0.414^{+0.027}_{-0.033}$ & $50.20\pm0.24$ & $1.135\pm0.087$ & -- & -- & -- & -- \\
 & $H(z)$ + BAO + SN & $0.0244\pm0.0027$ & $0.1199\pm0.0076$ & $0.304^{+0.014}_{-0.015}$ & -- & -- & $69.04\pm1.77$ & -- & -- & -- & -- & -- & -- & -- & -- & -- & -- & -- \\
 & HzBSNQHMA\tnote{d} & $0.0256\pm0.0020$ & $0.1201\pm0.0061$ & $0.300\pm0.012$ & -- & -- & $69.87\pm1.13$ & $10.96\pm0.25$ & $0.292^{+0.023}_{-0.029}$ & $1.684\pm0.055$ & $0.293\pm0.044$ & $0.413^{+0.026}_{-0.033}$ & $50.20\pm0.24$ & $1.131\pm0.087$ & -- & -- & -- & -- \\
 & QHMPA101\tnote{e} & $0.0225^{+0.0117}_{-0.0118}$ & $0.1105^{+0.0197}_{-0.0223}$ & $0.264^{+0.036}_{-0.047}$ & -- & -- & $71.34^{+1.73}_{-1.74}$ & $11.02\pm0.34$ & $0.291^{+0.023}_{-0.029}$ & $1.686\pm0.055$ & $0.291\pm0.044$ & $0.421^{+0.029}_{-0.036}$ & $50.12\pm0.26$ & $1.164\pm0.094$ & $0.346^{+0.032}_{-0.045}$ & $-0.710\pm0.101$ & $0.765\pm0.078$ & $11.06\pm4.19$ \\
% \\
\midrule
 & $H(z)$ + BAO & $0.0266^{+0.0039}_{-0.0045}$ & $0.1088\pm0.0166$ & $0.291\pm0.023$ & $0.059^{+0.081}_{-0.091}$ & -- & $68.37\pm2.10$ & -- & -- & -- & -- & -- & -- & -- & -- & -- & -- & -- \\
 & \mq\ + A118 & -- & -- & $>0.274$ & $0.521^{+0.512}_{-0.872}$ & -- & -- & -- & $0.294^{+0.023}_{-0.030}$ & $1.730^{+0.058}_{-0.053}$ & $0.302\pm0.047$ & $0.411^{+0.027}_{-0.033}$ & $50.03\pm0.26$ & $1.118\pm0.089$ & -- & -- & -- & -- \\
Non-flat & QSO-AS + \hiig & $0.0224\pm0.0111$ & $0.1122^{+0.0223}_{-0.0218}$ & $0.260^{+0.039}_{-0.045}$ & $-0.196^{+0.112}_{-0.295}$ & -- & $72.25\pm1.99$ & $11.35\pm0.49$ & -- & -- & -- & -- & -- & -- & -- & -- & -- & -- \\
\lcdm & QSO-AS + \hiig\ + \mq\ + A118 & $0.0225\pm0.0116$ & $0.1162\pm0.0218$ & $0.271^{+0.038}_{-0.045}$ & $-0.139^{+0.116}_{-0.228}$ & -- & $71.77\pm1.87$ & $11.19\pm0.42$ & $0.291^{+0.022}_{-0.029}$ & $1.680\pm0.055$ & $0.292\pm0.044$ & $0.415^{+0.027}_{-0.033}$ & $50.23\pm0.24$ & $1.122\pm0.088$ & -- & -- & -- & -- \\
 & $H(z)$ + BAO + SN & $0.0260^{+0.0037}_{-0.0043}$ & $0.1119\pm0.0157$ & $0.294\pm0.022$ & $0.040\pm0.070$ & -- & $68.62\pm1.90$ & -- & -- & -- & -- & -- & -- & -- & -- & -- & -- & -- \\
 & HzBSNQHMA\tnote{d} & $0.0265^{+0.0032}_{-0.0038}$ & $0.1168\pm0.0127$ & $0.295\pm0.019$ & $0.018\pm0.059$ & -- & $69.79\pm1.14$ & $10.96\pm0.25$ & $0.292^{+0.023}_{-0.029}$ & $1.685\pm0.055$ & $0.293\pm0.044$ & $0.413^{+0.027}_{-0.033}$ & $50.20\pm0.24$ & $1.131\pm0.086$ & -- & -- & -- & -- \\
 & QHMPA101\tnote{e} & $0.0224\pm0.0117$ & $0.1138^{+0.0226}_{-0.0224}$ & $0.267^{+0.039}_{-0.046}$ & $-0.096^{+0.127}_{-0.251}$ & -- & $71.65^{+1.86}_{-1.85}$ & $11.15\pm0.42$ & $0.291^{+0.023}_{-0.029}$ & $1.682\pm0.055$ & $0.292\pm0.044$ & $0.424^{+0.029}_{-0.036}$ & $50.14\pm0.26$ & $1.154\pm0.095$ & $0.346^{+0.032}_{-0.045}$ & $-0.712\pm0.102$ & $0.760\pm0.080$ & $11.37\pm4.34$ \\
% \\
\midrule
 & $H(z)$ + BAO & $0.0295^{+0.0042}_{-0.0050}$ & $0.0969^{+0.0178}_{-0.0152}$ & $0.289\pm0.020$ & -- & $-0.784^{+0.140}_{-0.107}$ & $66.22^{+2.31}_{-2.54}$ & -- & -- & -- & -- & -- & -- & -- & -- & -- & -- & -- \\
 & \mq\ + A118 & -- & -- & $0.493^{+0.209}_{-0.353}$ & -- & $<-0.218$ & -- & -- & $0.291^{+0.023}_{-0.030}$ & $1.670^{+0.095}_{-0.062}$ & $0.301\pm0.046$ & $0.412^{+0.027}_{-0.033}$ & $50.25^{+0.28}_{-0.36}$ & $1.110\pm0.088$ \\
Flat & QSO-AS + \hiig & $0.0224\pm0.0112$ & $0.1391^{+0.0333}_{-0.0256}$ & $0.305^{+0.056}_{-0.047}$ & -- & $-1.683^{+0.712}_{-0.387}$ & $72.92^{+2.15}_{-2.40}$ & $11.31\pm0.43$ & -- & -- & -- & -- & -- & -- & -- & -- & -- & -- \\
XCDM & QSO-AS + \hiig\ + \mq\ + A118 & $0.0224\pm0.0117$ & $0.1449^{+0.0314}_{-0.0246}$ & $0.314^{+0.051}_{-0.044}$ & -- & $-1.836^{+0.804}_{-0.419}$ & $73.14^{+2.14}_{-2.48}$ & $11.34\pm0.43$ & $0.291^{+0.023}_{-0.029}$ & $1.675\pm0.056$ & $0.294\pm0.044$ & $0.413^{+0.027}_{-0.033}$ & $50.24\pm0.24$ & $1.124\pm0.087$ & -- & -- & -- & -- \\
 & $H(z)$ + BAO + SN & $0.0262^{+0.0033}_{-0.0037}$ & $0.1120\pm0.0110$ & $0.295\pm0.016$ & -- & $-0.941\pm0.064$ & $68.55\pm1.85$ & -- & -- & -- & -- & -- & -- & -- & -- & -- & -- & -- \\
 & HzBSNQHMA\tnote{d} & $0.0271^{+0.0027}_{-0.0031}$ & $0.1449^{+0.0098}_{-0.0097}$ & $0.294\pm0.015$ & -- & $-0.959\pm0.059$ & $69.66\pm1.16$ & $10.94\pm0.25$ & $0.292^{+0.023}_{-0.029}$ & $1.685\pm0.055$ & $0.293\pm0.044$ & $0.413^{+0.027}_{-0.033}$ & $50.20\pm0.24$ & $1.131\pm0.087$ & -- & -- & -- & -- \\
 & QHMPA101\tnote{e} & $0.0224\pm0.0118$ & $0.1429^{+0.0333}_{-0.0248}$ & $0.310^{+0.054}_{-0.044}$ & -- & $-1.801^{+0.814}_{-0.421}$ & $73.09^{+2.13}_{-2.48}$ & $11.33\pm0.43$ & $0.290^{+0.022}_{-0.029}$ & $1.675\pm0.056$ & $0.294\pm0.044$ & $0.421^{+0.029}_{-0.036}$ & $50.16\pm0.26$ & $1.152\pm0.094$ & $0.346^{+0.032}_{-0.045}$ & $-0.713\pm0.102$ & $0.758\pm0.081$ & $11.45\pm4.35$ \\
% \\
\midrule
 & $H(z)$ + BAO & $0.0294^{+0.0047}_{-0.0050}$ & $0.0980^{+0.0186}_{-0.0187}$ & $0.292\pm0.025$ & $-0.027\pm0.109$ & $-0.770^{+0.149}_{-0.098}$ & $66.13^{+2.35}_{-2.36}$ & -- & -- & -- & -- & -- & -- & -- & -- & -- & -- & -- \\
 & \mq\ + A118 & -- & -- & $>0.170$ & $0.309^{+0.347}_{-0.650}$ & $<-0.132$ & -- & -- & $0.293^{+0.023}_{-0.030}$ & $1.711^{+0.081}_{-0.058}$ & $0.302\pm0.047$ & $0.413^{+0.027}_{-0.033}$ & $50.10^{+0.28}_{-0.32}$ & $1.114\pm0.091$ & -- & -- & -- & -- \\
Non-flat & QSO-AS + \hiig & $0.0224\pm0.0114$ & $0.1122^{+0.0473}_{-0.0326}$ & $0.258^{+0.086}_{-0.057}$ & $0.018^{+0.345}_{-0.383}$ & $-1.670^{+1.063}_{-0.245}$ & $72.34\pm2.16$ & $11.20\pm0.49$ & -- & -- & -- & -- & -- & -- & -- & -- & -- & -- \\
XCDM & QSO-AS + \hiig\ + \mq\ + A118 & $0.0225\pm0.0118$ & $0.1327^{+0.0348}_{-0.0286}$ & $0.294^{+0.059}_{-0.051}$ & $0.054^{+0.227}_{-0.238}$ & $-2.042^{+1.295}_{-0.451}$ & $72.79^{+2.18}_{-2.42}$ & $11.23^{+0.44}_{-0.49}$ & $0.291^{+0.023}_{-0.030}$ & $1.678\pm0.057$ & $0.294\pm0.045$ & $0.414^{+0.027}_{-0.033}$ & $50.23\pm0.25$ & $1.127\pm0.089$ & -- & -- & -- & -- \\
 & $H(z)$ + BAO + SN & $0.0262^{+0.0037}_{-0.0043}$ & $0.1119\pm0.0157$ & $0.295\pm0.022$ & $-0.001\pm0.098$ & $-0.948^{+0.098}_{-0.068}$ & $68.53\pm1.90$ & -- & -- & -- & -- & -- & -- & -- & -- & -- & -- & -- \\
 & HzBSNQHMA\tnote{d} & $0.0269^{+0.0033}_{-0.0039}$ & $0.1155^{+0.0128}_{-0.0127}$ & $0.295\pm0.019$ & $-0.009^{+0.077}_{-0.083}$ & $-0.959^{+0.090}_{-0.063}$ & $69.65\pm1.16$ & $10.93\pm0.26$ & $0.292^{+0.023}_{-0.029}$ & $1.685\pm0.055$ & $0.293\pm0.044$ & $0.413^{+0.027}_{-0.033}$ & $50.20\pm0.24$ & $1.130\pm0.087$ & -- & -- & -- & -- \\
 & QHMPA101\tnote{e} & $0.0225\pm0.0118$ & $0.1283^{+0.0354}_{-0.0290}$ & $0.286^{+0.060}_{-0.052}$ & $0.104^{+0.232}_{-0.244}$ & $-2.131^{+1.385}_{-0.524}$ & $72.70^{+2.19}_{-2.43}$ & $11.19\pm0.46$ & $0.291^{+0.023}_{-0.029}$ & $1.679\pm0.056$ & $0.293\pm0.044$ & $0.421^{+0.029}_{-0.036}$ & $50.14\pm0.26$ & $1.159\pm0.095$ & $0.346^{+0.032}_{-0.045}$ & $-0.711\pm0.102$ & $0.761\pm0.080$ & $11.28\pm4.32$ \\
% \\
\midrule
 & $H(z)$ + BAO & $0.0320^{+0.0054}_{-0.0041}$ & $0.0855^{+0.0175}_{-0.0174}$ & $0.275\pm0.023$ & -- & $1.267^{+0.536}_{-0.807}$ & $65.47^{+2.22}_{-2.21}$ & -- & -- & -- & -- & -- & -- & -- & -- & -- & -- & -- \\
 & \mq\ + A118 & -- & -- & $0.561^{+0.332}_{-0.241}$ & -- & -- & -- & -- & $0.293^{+0.023}_{-0.030}$ & $1.725\pm0.052$ & $0.303\pm0.046$ & $0.411^{+0.026}_{-0.032}$ & $50.05\pm0.25$ & $1.110\pm0.087$ & -- & -- & -- & -- \\%$>0.142$
Flat & QSO-AS + \hiig & $0.0217^{+0.0081}_{-0.0138}$ & $0.0543^{+0.0225}_{-0.0471}$ & $0.154^{+0.053}_{-0.086}$ & -- & $<6.506$ & $70.64\pm1.80$ & $10.81\pm0.34$ & -- & -- & -- & -- & -- & -- & -- & -- & -- & -- \\
\pcdm & QSO-AS + \hiig\ + \mq\ + A118 & $0.0219^{+0.0076}_{-0.0165}$ & $0.0644^{+0.0384}_{-0.0410}$ & $0.175^{+0.075}_{-0.081}$ & -- & $<6.756$ & $70.50\pm1.76$ & $10.80\pm0.34$ & $0.292^{+0.023}_{-0.029}$ & $1.691\pm0.055$ & $0.291\pm0.044$ & $0.414^{+0.027}_{-0.033}$ & $50.18\pm0.24$ & $1.140\pm0.088$ & -- & -- & -- & -- \\
 & $H(z)$ + BAO + SN & $0.0278^{+0.0032}_{-0.0039}$ & $0.1054^{+0.0117}_{-0.0100}$ & $0.287\pm0.017$ & -- & $0.324^{+0.122}_{-0.264}$ & $68.29\pm1.78$ & -- & -- & -- & -- & -- & -- & -- & -- & -- & -- & -- \\
 & HzBSNQHMA\tnote{d} & $0.0286^{+0.0025}_{-0.0033}$ & $0.1089^{+0.0103}_{-0.0083}$ & $0.286\pm0.015$ & -- & $0.249^{+0.069}_{-0.239}$ & $69.50\pm1.14$ & $10.92\pm0.25$ & $0.292^{+0.023}_{-0.029}$ & $1.686\pm0.054$ & $0.293\pm0.044$ & $0.413^{+0.027}_{-0.033}$ & $50.20\pm0.24$ & $1.132\pm0.086$ & -- & -- & -- & -- \\
 & QHMPA101\tnote{e} & $0.0217^{+0.0098}_{-0.0164}$ & $0.0607^{+0.0296}_{-0.0493}$ & $0.167^{+0.067}_{-0.088}$ & -- & $<7.149$ & $70.53\pm1.76$ & $10.80\pm0.34$ & $0.292^{+0.023}_{-0.029}$ & $1.691\pm0.054$ & $0.291\pm0.044$ & $0.421^{+0.029}_{-0.036}$ & $50.10\pm0.26$ & $1.172\pm0.093$ & $0.347^{+0.032}_{-0.045}$ & $-0.707\pm0.102$ & $0.768\pm0.078$ & $10.89\pm4.21$ \\
% \\
\midrule
 & $H(z)$ + BAO & $0.0320^{+0.0057}_{-0.0038}$ & $0.0865^{+0.0172}_{-0.0198}$ & $0.277^{+0.023}_{-0.026}$ & $-0.034^{+0.087}_{-0.098}$ & $1.360^{+0.584}_{-0.819}$ & $65.53\pm2.19$ & -- & -- & -- & -- & -- & -- & -- & -- & -- & -- & -- \\
 & \mq\ + A118 & -- & -- & $0.549^{+0.241}_{-0.268}$ & $-0.001^{+0.296}_{-0.286}$ & -- & -- & -- & $0.293^{+0.023}_{-0.030}$ & $1.725\pm0.052$ & $0.303\pm0.047$ & $0.412^{+0.027}_{-0.033}$ & $50.05\pm0.25$ & $1.111\pm0.089$ & -- & -- & -- & -- \\%$>0.171$
Non-flat & QSO-AS + \hiig & $0.0219^{+0.0093}_{-0.0130}$ & $0.0576^{+0.0268}_{-0.0424}$ & $0.163^{+0.058}_{-0.081}$ & $0.181^{+0.180}_{-0.339}$ & $<7.875$ & $70.21\pm1.83$ & $10.70^{+0.36}_{-0.41}$ & -- & -- & -- & -- & -- & -- & -- & -- & -- & -- \\
\pcdm & QSO-AS + \hiig\ + \mq\ + A118 & $0.0222^{+0.0098}_{-0.0135}$ & $0.0725^{+0.0398}_{-0.0368}$ & $0.193^{+0.077}_{-0.071}$ & $0.044^{+0.104}_{-0.256}$ & $<7.239$ & $70.38\pm1.84$ & $10.79^{+0.38}_{-0.41}$ & $0.292^{+0.024}_{-0.030}$ & $1.691\pm0.055$ & $0.292\pm0.045$ & $0.415^{+0.028}_{-0.033}$ & $50.18\pm0.25$ & $1.140\pm0.089$ & -- & -- & -- & -- \\
 & $H(z)$ + BAO + SN & $0.0271^{+0.0038}_{-0.0043}$ & $0.1095\pm0.0152$ & $0.292\pm0.022$ & $-0.038^{+0.071}_{-0.085}$ & $0.382^{+0.151}_{-0.299}$ & $68.48\pm1.85$ & -- & -- & -- & -- & -- & -- & -- & -- & -- & -- & -- \\
 & HzBSNQHMA\tnote{d} & $0.0277^{+0.0034}_{-0.0040}$ & $0.1126\pm0.0128$ & $0.291\pm0.019$ & $-0.040^{+0.064}_{-0.072}$ & $0.316^{+0.101}_{-0.292}$ & $69.52\pm1.15$ & $10.89\pm0.25$ & $0.292^{+0.023}_{-0.030}$ & $1.685\pm0.055$ & $0.294\pm0.044$ & $0.413^{+0.027}_{-0.033}$ & $50.20\pm0.24$ & $1.128\pm0.087$ & -- & -- & -- & -- \\
 & QHMPA101\tnote{e} & $0.0221^{+0.0094}_{-0.0138}$ & $0.0677^{+0.0368}_{-0.0389}$ & $0.183^{+0.072}_{-0.076}$ & $0.084^{+0.119}_{-0.275}$ & $<7.625$ & $70.33\pm1.83$ & $10.76^{+0.37}_{-0.41}$ & $0.292^{+0.023}_{-0.030}$ & $1.692\pm0.054$ & $0.291\pm0.044$ & $0.421^{+0.029}_{-0.036}$ & $50.09\pm0.26$ & $1.172\pm0.094$ & $0.347^{+0.033}_{-0.046}$ & $-0.707\pm0.102$ & $0.769\pm0.078$ & $10.87\pm4.22$ \\
\bottomrule
\end{tabular}
%}
\begin{tablenotes}[flushleft]
\item [a] \wx\ corresponds to flat/non-flat XCDM and $\alpha$ corresponds to flat/non-flat \pcdm.
\item [b] \hunit. In \mq, A118, and \mq\ + A118 cases, $\Omega_b$ and $H_0$ are set to be 0.05 and 70 \hunit, respectively.
\item [c] pc.
\item [d] $H(z)$ + BAO + SN + QSO-AS + \hiig\ + \mq\ + A118.
\item [e] QSO-AS + \hiig\ + \mq\ + Platinum + A101.
\end{tablenotes}
\end{threeparttable}%
}
\end{sidewaystable*}

\section*{Acknowledgements}
We thank Javier de Cruz P\'{e}rez and Chan-Gyung Park for useful discussions about BAO data. This research was supported in part by DOE grant DE-SC0011840. The computations for this project were performed on the Beocat Research Cluster at Kansas State University, which is funded in part by NSF grants CNS-1006860, EPS-1006860, EPS-0919443, ACI-1440548, CHE-1726332, and NIH P20GM113109.

\section*{Data availability}

The \hiig\ data were provided to us by the authors of \cite{GM2021} and will be shared on request to the corresponding author with the permission of the authors of \cite{GM2021}. All other data we use are publicly available in the papers cited in Sec.\  \ref{sec:data}.

%%%%%%%%%%%%%%%%%%%%%%%%%%%%%%%%%%%%%%%%%%%%%%%%%%



\bibliographystyle{mnras}
\bibliography{mybibfile} % if your bibtex file is called example.bib

\begin{thebibliography}{}
\makeatletter
\relax
\def\mn@urlcharsother{\let\do\@makeother \do\$\do\&\do\#\do\^\do\_\do\%\do\~}
\def\mn@doi{\begingroup\mn@urlcharsother \@ifnextchar [ {\mn@doi@}
  {\mn@doi@[]}}
\def\mn@doi@[#1]#2{\def\@tempa{#1}\ifx\@tempa\@empty \href
  {http://dx.doi.org/#2} {doi:#2}\else \href {http://dx.doi.org/#2} {#1}\fi
  \endgroup}
\def\mn@eprint#1#2{\mn@eprint@#1:#2::\@nil}
\def\mn@eprint@arXiv#1{\href {http://arxiv.org/abs/#1} {{\tt arXiv:#1}}}
\def\mn@eprint@dblp#1{\href {http://dblp.uni-trier.de/rec/bibtex/#1.xml}
  {dblp:#1}}
\def\mn@eprint@#1:#2:#3:#4\@nil{\def\@tempa {#1}\def\@tempb {#2}\def\@tempc
  {#3}\ifx \@tempc \@empty \let \@tempc \@tempb \let \@tempb \@tempa \fi \ifx
  \@tempb \@empty \def\@tempb {arXiv}\fi \@ifundefined
  {mn@eprint@\@tempb}{\@tempb:\@tempc}{\expandafter \expandafter \csname
  mn@eprint@\@tempb\endcsname \expandafter{\@tempc}}}

\bibitem[\protect\citeauthoryear{{Abdalla} et~al.,}{{Abdalla}
  et~al.}{2022}]{Abdallaetal2022}
{Abdalla} E.,  et~al., 2022, preprint, \href
  {https://ui.adsabs.harvard.edu/abs/2022arXiv220306142A} {} (\mn@eprint {}
  {2203.06142})

\bibitem[\protect\citeauthoryear{{Amati}, {D'Agostino}, {Luongo}, {Muccino}  \&
  {Tantalo}}{{Amati} et~al.}{2019}]{Amati2019}
{Amati} L.,  {D'Agostino} R.,  {Luongo} O.,  {Muccino} M.,   {Tantalo} M.,
  2019, \mn@doi [\mnras] {10.1093/mnrasl/slz056}, \href
  {https://ui.adsabs.harvard.edu/abs/2019MNRAS.486L..46A} {486, L46}

\bibitem[\protect\citeauthoryear{{Arjona} \& {Nesseris}}{{Arjona} \&
  {Nesseris}}{2021}]{ArjonaNesseris2021}
{Arjona} R.,  {Nesseris} S.,  2021, \mn@doi [\prd]
  {10.1103/PhysRevD.103.103539}, \href
  {https://ui.adsabs.harvard.edu/abs/2021PhRvD.103j3539A} {103, 103539}

\bibitem[\protect\citeauthoryear{{Audren}, {Lesgourgues}, {Benabed}  \&
  {Prunet}}{{Audren} et~al.}{2013}]{Audren:2012wb}
{Audren} B.,  {Lesgourgues} J.,  {Benabed} K.,   {Prunet} S.,  2013, \mn@doi
  [\jcap] {10.1088/1475-7516/2013/02/001}, \href
  {https://ui.adsabs.harvard.edu/abs/2013JCAP...02..001A} {2013, 001}

\bibitem[\protect\citeauthoryear{{Bautista} et~al.,}{{Bautista}
  et~al.}{2021}]{eBOSSL_2021}
{Bautista} J.~E.,  et~al., 2021, \mn@doi [\mnras] {10.1093/mnras/staa2800},
  \href {https://ui.adsabs.harvard.edu/abs/2021MNRAS.500..736B} {500, 736}

\bibitem[\protect\citeauthoryear{{Birrer} et~al.,}{{Birrer}
  et~al.}{2020}]{Birrer_et_al_2020}
{Birrer} S.,  et~al., 2020, \mn@doi [\aap] {10.1051/0004-6361/202038861}, \href
  {https://ui.adsabs.harvard.edu/abs/2020A&A...643A.165B} {643, A165}

\bibitem[\protect\citeauthoryear{{Blas}, {Lesgourgues}  \& {Tram}}{{Blas}
  et~al.}{2011}]{class}
{Blas} D.,  {Lesgourgues} J.,   {Tram} T.,  2011, \mn@doi [\jcap]
  {10.1088/1475-7516/2011/07/034}, \href
  {https://ui.adsabs.harvard.edu/abs/2011JCAP...07..034B} {2011, 034}

\bibitem[\protect\citeauthoryear{{Blum}, {Castorina}  \&
  {Simonovi{\'c}}}{{Blum} et~al.}{2020}]{Blum_et_al_2020}
{Blum} K.,  {Castorina} E.,   {Simonovi{\'c}} M.,  2020, \mn@doi [\apjl]
  {10.3847/2041-8213/ab8012}, \href
  {https://ui.adsabs.harvard.edu/abs/2020ApJ...892L..27B} {892, L27}

\bibitem[\protect\citeauthoryear{{Borghi}, {Moresco}  \& {Cimatti}}{{Borghi}
  et~al.}{2022}]{Borghi_etal_2022}
{Borghi} N.,  {Moresco} M.,   {Cimatti} A.,  2022, \mn@doi [\apjl]
  {10.3847/2041-8213/ac3fb2}, \href
  {https://ui.adsabs.harvard.edu/abs/2022ApJ...928L...4B} {928, L4}

\bibitem[\protect\citeauthoryear{{Boruah}, {Hudson}  \& {Lavaux}}{{Boruah}
  et~al.}{2021}]{Boruahetal_2021}
{Boruah} S.~S.,  {Hudson} M.~J.,   {Lavaux} G.,  2021, \mn@doi [\mnras]
  {10.1093/mnras/stab2320}, \href
  {https://ui.adsabs.harvard.edu/abs/2021MNRAS.507.2697B} {507, 2697}

\bibitem[\protect\citeauthoryear{{Breuval} et~al.,}{{Breuval}
  et~al.}{2020}]{Breuvaletal_2020}
{Breuval} L.,  et~al., 2020, \mn@doi [\aap] {10.1051/0004-6361/202038633},
  \href {https://ui.adsabs.harvard.edu/abs/2020A&A...643A.115B} {643, A115}

\bibitem[\protect\citeauthoryear{{Calabrese}, {Archidiacono}, {Melchiorri}  \&
  {Ratra}}{{Calabrese} et~al.}{2012}]{Calabreseetal2012}
{Calabrese} E.,  {Archidiacono} M.,  {Melchiorri} A.,   {Ratra} B.,  2012,
  \mn@doi [\prd] {10.1103/PhysRevD.86.043520}, \href
  {https://ui.adsabs.harvard.edu/abs/2012PhRvD..86d3520C} {86, 043520}

\bibitem[\protect\citeauthoryear{{Cao}, {Zheng}, {Biesiada}, {Qi}, {Chen}  \&
  {Zhu}}{{Cao} et~al.}{2017}]{Cao_et_al2017b}
{Cao} S.,  {Zheng} X.,  {Biesiada} M.,  {Qi} J.,  {Chen} Y.,   {Zhu} Z.-H.,
  2017, \mn@doi [\aap] {10.1051/0004-6361/201730551}, \href
  {https://ui.adsabs.harvard.edu/abs/2017A&A...606A..15C} {606, A15}

\bibitem[\protect\citeauthoryear{{Cao}, {Ryan}  \& {Ratra}}{{Cao}
  et~al.}{2020}]{CaoRyanRatra2020}
{Cao} S.,  {Ryan} J.,   {Ratra} B.,  2020, \mn@doi [\mnras]
  {10.1093/mnras/staa2190}, \href
  {https://ui.adsabs.harvard.edu/abs/2020MNRAS.tmp.2278C} {497, 3191}

\bibitem[\protect\citeauthoryear{{Cao}, {Ryan}, {Khadka}  \& {Ratra}}{{Cao}
  et~al.}{2021a}]{Caoetal_2021}
{Cao} S.,  {Ryan} J.,  {Khadka} N.,   {Ratra} B.,  2021a, \mn@doi [\mnras]
  {10.1093/mnras/staa3748}, \href
  {https://ui.adsabs.harvard.edu/abs/2020MNRAS.tmp.3537C} {501, 1520}

\bibitem[\protect\citeauthoryear{{Cao}, {Ryan}  \& {Ratra}}{{Cao}
  et~al.}{2021b}]{CaoRyanRatra2021}
{Cao} S.,  {Ryan} J.,   {Ratra} B.,  2021b, \mn@doi [\mnras]
  {10.1093/mnras/stab942}, \href
  {https://ui.adsabs.harvard.edu/abs/2021MNRAS.504..300C} {504, 300}

\bibitem[\protect\citeauthoryear{{Cao}, {Dainotti}  \& {Ratra}}{{Cao}
  et~al.}{2022a}]{CaoDainottiRatra2022b}
{Cao} S.,  {Dainotti} M.,   {Ratra} B.,  2022a, preprint, \href
  {https://ui.adsabs.harvard.edu/abs/2022arXiv220408710C} {} (\mn@eprint {}
  {2204.08710})

\bibitem[\protect\citeauthoryear{{Cao}, {Ryan}  \& {Ratra}}{{Cao}
  et~al.}{2022b}]{CaoRyanRatra2022}
{Cao} S.,  {Ryan} J.,   {Ratra} B.,  2022b, \mn@doi [\mnras]
  {10.1093/mnras/stab3304}, \href
  {https://ui.adsabs.harvard.edu/abs/2022MNRAS.509.4745C} {509, 4745}

\bibitem[\protect\citeauthoryear{{Cao}, {Khadka}  \& {Ratra}}{{Cao}
  et~al.}{2022c}]{CaoKhadkaRatra2021}
{Cao} S.,  {Khadka} N.,   {Ratra} B.,  2022c, \mn@doi [\mnras]
  {10.1093/mnras/stab3559}, \href
  {https://ui.adsabs.harvard.edu/abs/2021MNRAS.tmp.3230C} {510, 2928}

\bibitem[\protect\citeauthoryear{{Cao}, {Dainotti}  \& {Ratra}}{{Cao}
  et~al.}{2022d}]{CaoDainottiRatra2022}
{Cao} S.,  {Dainotti} M.,   {Ratra} B.,  2022d, \mn@doi [\mnras]
  {10.1093/mnras/stac517}, \href
  {https://ui.adsabs.harvard.edu/abs/2022MNRAS.tmp..503C} {512, 439}

\bibitem[\protect\citeauthoryear{{Carter}, {Beutler}, {Percival}, {Blake},
  {Koda}  \& {Ross}}{{Carter} et~al.}{2018}]{Carter_2018}
{Carter} P.,  {Beutler} F.,  {Percival} W.~J.,  {Blake} C.,  {Koda} J.,
  {Ross} A.~J.,  2018, \mn@doi [\mnras] {10.1093/mnras/sty2405}, \href
  {https://ui.adsabs.harvard.edu/abs/2018MNRAS.481.2371C} {481, 2371}

\bibitem[\protect\citeauthoryear{{Ch{\'a}vez}, {Terlevich}, {Terlevich},
  {Bresolin}, {Melnick}, {Plionis}  \& {Basilakos}}{{Ch{\'a}vez}
  et~al.}{2014}]{Chavez_2014}
{Ch{\'a}vez} R.,  {Terlevich} R.,  {Terlevich} E.,  {Bresolin} F.,  {Melnick}
  J.,  {Plionis} M.,   {Basilakos} S.,  2014, \mn@doi [\mnras]
  {10.1093/mnras/stu987}, \href
  {https://ui.adsabs.harvard.edu/abs/2014MNRAS.442.3565C} {442, 3565}

\bibitem[\protect\citeauthoryear{{Chen} \& {Ratra}}{{Chen} \&
  {Ratra}}{2011}]{chenratmed}
{Chen} G.,  {Ratra} B.,  2011, \mn@doi [\pasp] {10.1086/662131}, \href
  {http://adsabs.harvard.edu/abs/2011PASP..123.1127C} {123, 1127}

\bibitem[\protect\citeauthoryear{{Chen}, {Ratra}, {Biesiada}, {Li}  \&
  {Zhu}}{{Chen} et~al.}{2016}]{Chen_et_al_2016}
{Chen} Y.,  {Ratra} B.,  {Biesiada} M.,  {Li} S.,   {Zhu} Z.-H.,  2016, \mn@doi
  [\apj] {10.3847/0004-637X/829/2/61}, \href
  {http://adsabs.harvard.edu/abs/2016ApJ...829...61C} {829, 61}

\bibitem[\protect\citeauthoryear{{Chen}, {Kumar}  \& {Ratra}}{{Chen}
  et~al.}{2017}]{chen_etal_2017}
{Chen} Y.,  {Kumar} S.,   {Ratra} B.,  2017, \mn@doi [\apj]
  {10.3847/1538-4357/835/1/86}, \href
  {http://adsabs.harvard.edu/abs/2017ApJ...835...86C} {835, 86}

\bibitem[\protect\citeauthoryear{{Cuceu}, {Farr}, {Lemos}  \&
  {Font-Ribera}}{{Cuceu} et~al.}{2019}]{Cuceu_2019}
{Cuceu} A.,  {Farr} J.,  {Lemos} P.,   {Font-Ribera} A.,  2019, \mn@doi [\jcap]
  {10.1088/1475-7516/2019/10/044}, \href
  {https://ui.adsabs.harvard.edu/abs/2019JCAP...10..044C} {2019, 044}

\bibitem[\protect\citeauthoryear{{Czerny} et~al.,}{{Czerny}
  et~al.}{2021}]{Czernyetal2021}
{Czerny} B.,  et~al., 2021, \mn@doi [Acta Physica Polonica A]
  {10.12693/APhysPolA.139.389}, \href
  {https://ui.adsabs.harvard.edu/abs/2021AcPPA.139..389C} {139, 389}

\bibitem[\protect\citeauthoryear{{DES Collaboration}}{{DES
  Collaboration}}{2018}]{DES_2018}
{DES Collaboration} 2018, \mn@doi [\mnras] {10.1093/mnras/sty1939}, \href
  {http://adsabs.harvard.edu/abs/2018MNRAS.480.3879A} {480, 3879}

\bibitem[\protect\citeauthoryear{{DES Collaboration}}{{DES
  Collaboration}}{2019a}]{DESCollaboration2019}
{DES Collaboration} 2019a, \mn@doi [\prd] {10.1103/PhysRevD.99.123505}, \href
  {https://ui.adsabs.harvard.edu/abs/2019PhRvD..99l3505A} {99, 123505}

\bibitem[\protect\citeauthoryear{{DES Collaboration}}{{DES
  Collaboration}}{2019b}]{DES_2019b}
{DES Collaboration} 2019b, \mn@doi [\mnras] {10.1093/mnras/sty3351}, \href
  {https://ui.adsabs.harvard.edu/abs/2019MNRAS.483.4866A} {483, 4866}

\bibitem[\protect\citeauthoryear{{DES Collaboration}}{{DES
  Collaboration}}{2019c}]{DES_2019d}
{DES Collaboration} 2019c, \mn@doi [\apj] {10.3847/1538-4357/ab08a0}, \href
  {https://ui.adsabs.harvard.edu/abs/2019ApJ...874..150B} {874, 150}

\bibitem[\protect\citeauthoryear{{Dai}, {Zheng}, {Li}, {Gao}  \& {Zhu}}{{Dai}
  et~al.}{2021}]{Daietal_2021}
{Dai} Y.,  {Zheng} X.-G.,  {Li} Z.-X.,  {Gao} H.,   {Zhu} Z.-H.,  2021, \mn@doi
  [\aap] {10.1051/0004-6361/202140895}, \href
  {https://ui.adsabs.harvard.edu/abs/2021A&A...651L...8D} {651, L8}

\bibitem[\protect\citeauthoryear{{Dainotti}, {Postnikov}, {Hernandez}  \&
  {Ostrowski}}{{Dainotti} et~al.}{2016}]{Dainottietal2016}
{Dainotti} M.~G.,  {Postnikov} S.,  {Hernandez} X.,   {Ostrowski} M.,  2016,
  \mn@doi [\apjl] {10.3847/2041-8205/825/2/L20}, \href
  {https://ui.adsabs.harvard.edu/abs/2016ApJ...825L..20D} {825, L20}

\bibitem[\protect\citeauthoryear{{Dainotti}, {Nagataki}, {Maeda}, {Postnikov}
  \& {Pian}}{{Dainotti} et~al.}{2017}]{Dainottietal2017}
{Dainotti} M.~G.,  {Nagataki} S.,  {Maeda} K.,  {Postnikov} S.,   {Pian} E.,
  2017, \mn@doi [\aap] {10.1051/0004-6361/201628384}, \href
  {https://ui.adsabs.harvard.edu/abs/2017A&A...600A..98D} {600, A98}

\bibitem[\protect\citeauthoryear{{Dainotti}, {Lenart}, {Sarracino}, {Nagataki},
  {Capozziello}  \& {Fraija}}{{Dainotti} et~al.}{2020}]{Dainottietal2020}
{Dainotti} M.~G.,  {Lenart} A.~{\L}.,  {Sarracino} G.,  {Nagataki} S.,
  {Capozziello} S.,   {Fraija} N.,  2020, \mn@doi [\apj]
  {10.3847/1538-4357/abbe8a}, \href
  {https://ui.adsabs.harvard.edu/abs/2020ApJ...904...97D} {904, 97}

\bibitem[\protect\citeauthoryear{{Dainotti}, {Nielson}, {Sarracino}, {Rinaldi},
  {Nagataki}, {Capozziello}, {Gnedin}  \& {Bargiacchi}}{{Dainotti}
  et~al.}{2022b}]{DainottiNielson2022}
{Dainotti} M.~G.,  {Nielson} V.,  {Sarracino} G.,  {Rinaldi} E.,  {Nagataki}
  S.,  {Capozziello} S.,  {Gnedin} O.~Y.,   {Bargiacchi} G.,  2022b, preprint,
  \href {https://ui.adsabs.harvard.edu/abs/2022arXiv220315538D} {} (\mn@eprint
  {} {2203.15538})

\bibitem[\protect\citeauthoryear{{Dainotti}, {Bardiacchi}, {Lukasz Lenart},
  {Capozziello}, {Colgain}, {Solomon}, {Stojkovic}  \&
  {Sheikh-Jabbari}}{{Dainotti} et~al.}{2022a}]{DainottiBardiacchi2022}
{Dainotti} M.~G.,  {Bardiacchi} G.,  {Lukasz Lenart} A.,  {Capozziello} S.,
  {Colgain} E.~O.,  {Solomon} R.,  {Stojkovic} D.,   {Sheikh-Jabbari} M.~M.,
  2022a, preprint, \href
  {https://ui.adsabs.harvard.edu/abs/2022arXiv220312914D} {} (\mn@eprint {}
  {2203.12914})

\bibitem[\protect\citeauthoryear{{\MakeLowercase{D}e Cruz Perez}, {Sola
  Peracaula}, {Gomez-Valent}  \& {Moreno-Pulido}}{{\MakeLowercase{D}e Cruz
  Perez} et~al.}{2021}]{deCruzetal2021}
{\MakeLowercase{D}e Cruz Perez} J.,  {Sola Peracaula} J.,  {Gomez-Valent} A.,
  {Moreno-Pulido} C.,  2021, preprint, \href
  {https://ui.adsabs.harvard.edu/abs/2021arXiv211007569D} {} (\mn@eprint {}
  {2110.07569})

\bibitem[\protect\citeauthoryear{{Demianski}, {Piedipalumbo}, {Sawant}  \&
  {Amati}}{{Demianski} et~al.}{2021}]{Demianskietal_2021}
{Demianski} M.,  {Piedipalumbo} E.,  {Sawant} D.,   {Amati} L.,  2021, \mn@doi
  [\mnras] {10.1093/mnras/stab1669}, \href
  {https://ui.adsabs.harvard.edu/abs/2021MNRAS.506..903D} {506, 903}

\bibitem[\protect\citeauthoryear{{Denzel}, {Coles}, {Saha}  \&
  {Williams}}{{Denzel} et~al.}{2021}]{Denzel_et_al_2020}
{Denzel} P.,  {Coles} J.~P.,  {Saha} P.,   {Williams} L. L.~R.,  2021, \mn@doi
  [\mnras] {10.1093/mnras/staa3603}, \href
  {https://ui.adsabs.harvard.edu/abs/2021MNRAS.501..784D} {501, 784}

\bibitem[\protect\citeauthoryear{{Dhawan}, {Jha}  \& {Leibundgut}}{{Dhawan}
  et~al.}{2018}]{Dhawan}
{Dhawan} S.,  {Jha} S.~W.,   {Leibundgut} B.,  2018, \mn@doi [\aap]
  {10.1051/0004-6361/201731501}, \href
  {http://adsabs.harvard.edu/abs/2018A%26A...609A..72D} {609, A72}

\bibitem[\protect\citeauthoryear{{Dhawan}, {Alsing}  \& {Vagnozzi}}{{Dhawan}
  et~al.}{2021}]{Dhawanetal2021}
{Dhawan} S.,  {Alsing} J.,   {Vagnozzi} S.,  2021, \mn@doi [\mnras]
  {10.1093/mnrasl/slab058}, \href
  {https://ui.adsabs.harvard.edu/abs/2021MNRAS.506L...1D} {506, L1}

\bibitem[\protect\citeauthoryear{{Di Valentino} et~al.,}{{Di Valentino}
  et~al.}{2021a}]{DiValentinoetal2021b}
{Di Valentino} E.,  et~al., 2021a, \mn@doi [Classical and Quantum Gravity]
  {10.1088/1361-6382/ac086d}, \href
  {https://ui.adsabs.harvard.edu/abs/2021CQGra..38o3001D} {38, 153001}

\bibitem[\protect\citeauthoryear{{Di Valentino}, {Melchiorri}  \& {Silk}}{{Di
  Valentino} et~al.}{2021b}]{DiValentinoetal2021a}
{Di Valentino} E.,  {Melchiorri} A.,   {Silk} J.,  2021b, \mn@doi [\apjl]
  {10.3847/2041-8213/abe1c4}, \href
  {https://ui.adsabs.harvard.edu/abs/2021ApJ...908L...9D} {908, L9}

\bibitem[\protect\citeauthoryear{{Dom{\'\i}nguez} et~al.,}{{Dom{\'\i}nguez}
  et~al.}{2019}]{dominguez_etal_2019}
{Dom{\'\i}nguez} A.,  et~al., 2019, \mn@doi [\apj] {10.3847/1538-4357/ab4a0e},
  \href {https://ui.adsabs.harvard.edu/abs/2019ApJ...885..137D} {885, 137}

\bibitem[\protect\citeauthoryear{{\MakeLowercase{D}u Mas des Bourboux}
  et~al.,}{{\MakeLowercase{D}u Mas des Bourboux} et~al.}{2020}]{duMas2020}
{\MakeLowercase{D}u Mas des Bourboux} H.,  et~al., 2020, \mn@doi [\apj]
  {10.3847/1538-4357/abb085}, \href
  {https://ui.adsabs.harvard.edu/abs/2020ApJ...901..153D} {901, 153}

\bibitem[\protect\citeauthoryear{{\MakeLowercase{E}BOSS
  Collaboration}}{{\MakeLowercase{E}BOSS Collaboration}}{2021}]{eBOSS_2020}
{\MakeLowercase{E}BOSS Collaboration} 2021, \mn@doi [\prd]
  {10.1103/PhysRevD.103.083533}, \href
  {https://ui.adsabs.harvard.edu/abs/2021PhRvD.103h3533A} {103, 083533}

\bibitem[\protect\citeauthoryear{{Efstathiou}}{{Efstathiou}}{2020}]{Efstathiou_2020}
{Efstathiou} G.,  2020, preprint, \href
  {https://ui.adsabs.harvard.edu/abs/2020arXiv200710716E} {} (\mn@eprint
  {arXiv} {2007.10716})

\bibitem[\protect\citeauthoryear{{Efstathiou} \& {Gratton}}{{Efstathiou} \&
  {Gratton}}{2020}]{EfstathiouGratton2020}
{Efstathiou} G.,  {Gratton} S.,  2020, \mn@doi [\mnras]
  {10.1093/mnrasl/slaa093}, \href
  {https://ui.adsabs.harvard.edu/abs/2020MNRAS.496L..91E} {496, L91}

\bibitem[\protect\citeauthoryear{{Fana Dirirsa} et~al.,}{{Fana Dirirsa}
  et~al.}{2019}]{Dirirsa2019}
{Fana Dirirsa} F.,  et~al., 2019, \mn@doi [\apj] {10.3847/1538-4357/ab4e11},
  \href {https://ui.adsabs.harvard.edu/abs/2019ApJ...887...13F} {887, 13}

\bibitem[\protect\citeauthoryear{{Farooq}, {Ranjeet Madiyar}, {Crandall}  \&
  {Ratra}}{{Farooq} et~al.}{2017}]{Farooq_Ranjeet_Crandall_Ratra_2017}
{Farooq} O.,  {Ranjeet Madiyar} F.,  {Crandall} S.,   {Ratra} B.,  2017,
  \mn@doi [\apj] {10.3847/1538-4357/835/1/26}, \href
  {http://adsabs.harvard.edu/abs/2017ApJ...835...26F} {835, 26}

\bibitem[\protect\citeauthoryear{{Fern{\'a}ndez Arenas} et~al.,}{{Fern{\'a}ndez
  Arenas} et~al.}{2018}]{FernandezArenas}
{Fern{\'a}ndez Arenas} D.,  et~al., 2018, \mn@doi [\mnras]
  {10.1093/mnras/stx2710}, \href
  {http://adsabs.harvard.edu/abs/2018MNRAS.474.1250F} {474, 1250}

\bibitem[\protect\citeauthoryear{{Freedman}}{{Freedman}}{2021}]{Freedman2021}
{Freedman} W.~L.,  2021, \mn@doi [\apj] {10.3847/1538-4357/ac0e95}, \href
  {https://ui.adsabs.harvard.edu/abs/2021ApJ...919...16F} {919, 16}

\bibitem[\protect\citeauthoryear{{Geng}, {Hsu}  \& {Lu}}{{Geng}
  et~al.}{2022}]{Gengetal2022}
{Geng} C.-Q.,  {Hsu} Y.-T.,   {Lu} J.-R.,  2022, \mn@doi [\apj]
  {10.3847/1538-4357/ac4495}, \href
  {https://ui.adsabs.harvard.edu/abs/2022ApJ...926...74G} {926, 74}

\bibitem[\protect\citeauthoryear{{Gil-Mar{\'\i}n} et~al.,}{{Gil-Mar{\'\i}n}
  et~al.}{2020}]{eBOSSG_2020}
{Gil-Mar{\'\i}n} H.,  et~al., 2020, \mn@doi [\mnras] {10.1093/mnras/staa2455},
  \href {https://ui.adsabs.harvard.edu/abs/2020MNRAS.498.2492G} {498, 2492}

\bibitem[\protect\citeauthoryear{{G{\'o}mez-Valent} \&
  {Amendola}}{{G{\'o}mez-Valent} \&
  {Amendola}}{2018}]{Gomez-ValentAmendola2018}
{G{\'o}mez-Valent} A.,  {Amendola} L.,  2018, \mn@doi [\jcap]
  {10.1088/1475-7516/2018/04/051}, \href
  {http://adsabs.harvard.edu/abs/2018JCAP...04..051G} {4, 051}

\bibitem[\protect\citeauthoryear{{Gonz{\'a}lez-Mor{\'a}n}
  et~al.,}{{Gonz{\'a}lez-Mor{\'a}n} et~al.}{2019}]{G-M_2019}
{Gonz{\'a}lez-Mor{\'a}n} A.~L.,  et~al., 2019, \mn@doi [\mnras]
  {10.1093/mnras/stz1577}, \href
  {https://ui.adsabs.harvard.edu/abs/2019MNRAS.487.4669G} {487, 4669}

\bibitem[\protect\citeauthoryear{{Gonz{\'a}lez-Mor{\'a}n}
  et~al.,}{{Gonz{\'a}lez-Mor{\'a}n} et~al.}{2021}]{GM2021}
{Gonz{\'a}lez-Mor{\'a}n} A.~L.,  et~al., 2021, \mn@doi [\mnras]
  {10.1093/mnras/stab1385}, \href
  {https://ui.adsabs.harvard.edu/abs/2021MNRAS.505.1441G} {505, 1441}

\bibitem[\protect\citeauthoryear{{Gott}, {Vogeley}, {Podariu}  \&
  {Ratra}}{{Gott} et~al.}{2001}]{gott_etal_2001}
{Gott} III J.~R.,  {Vogeley} M.~S.,  {Podariu} S.,   {Ratra} B.,  2001, \mn@doi
  [\apj] {10.1086/319055}, \href
  {http://adsabs.harvard.edu/abs/2001ApJ...549....1G} {549, 1}

\bibitem[\protect\citeauthoryear{{Handley}}{{Handley}}{2019}]{Handley2019}
{Handley} W.,  2019, \mn@doi [\prd] {10.1103/PhysRevD.100.123517}, \href
  {https://ui.adsabs.harvard.edu/abs/2019PhRvD.100l3517H} {100, 123517}

\bibitem[\protect\citeauthoryear{{Harvey}}{{Harvey}}{2020}]{Harvey_2020}
{Harvey} D.,  2020, \mn@doi [\mnras] {10.1093/mnras/staa2522}, 498, 2871

\bibitem[\protect\citeauthoryear{{Hou} et~al.,}{{Hou}
  et~al.}{2021}]{eBOSSQ_2021}
{Hou} J.,  et~al., 2021, \mn@doi [\mnras] {10.1093/mnras/staa3234}, \href
  {https://ui.adsabs.harvard.edu/abs/2021MNRAS.500.1201H} {500, 1201}

\bibitem[\protect\citeauthoryear{{Hu}, {Wang}  \& {Dai}}{{Hu}
  et~al.}{2021}]{Huetal_2021}
{Hu} J.~P.,  {Wang} F.~Y.,   {Dai} Z.~G.,  2021, \mn@doi [\mnras]
  {10.1093/mnras/stab2180}, \href
  {https://ui.adsabs.harvard.edu/abs/2021MNRAS.507..730H} {507, 730}

\bibitem[\protect\citeauthoryear{{Jesus}, {Valentim}, {Escobal}, {Pereira}  \&
  {Benndorf}}{{Jesus} et~al.}{2021}]{Jesusetal2021}
{Jesus} J.~F.,  {Valentim} R.,  {Escobal} A.~A.,  {Pereira} S.~H.,   {Benndorf}
  D.,  2021, preprint, \href
  {https://ui.adsabs.harvard.edu/abs/2021arXiv211209722J} {} (\mn@eprint {}
  {2112.09722})

\bibitem[\protect\citeauthoryear{{Johnson}, {Sangwan}  \&
  {Shankaranarayanan}}{{Johnson} et~al.}{2022}]{Johnsonetal2022}
{Johnson} J.~P.,  {Sangwan} A.,   {Shankaranarayanan} S.,  2022, \mn@doi
  [\jcap] {10.1088/1475-7516/2022/01/024}, \href
  {https://ui.adsabs.harvard.edu/abs/2022JCAP...01..024J} {2022, 024}

\bibitem[\protect\citeauthoryear{{Khadka} \& {Ratra}}{{Khadka} \&
  {Ratra}}{2020a}]{KhadkaRatra2020a}
{Khadka} N.,  {Ratra} B.,  2020a, \mn@doi [\mnras] {10.1093/mnras/staa101},
  \href {https://ui.adsabs.harvard.edu/abs/2020MNRAS.492.4456K} {492, 4456}

\bibitem[\protect\citeauthoryear{{Khadka} \& {Ratra}}{{Khadka} \&
  {Ratra}}{2020b}]{KhadkaRatra2020b}
{Khadka} N.,  {Ratra} B.,  2020b, \mn@doi [\mnras] {10.1093/mnras/staa1855},
  \href {https://ui.adsabs.harvard.edu/abs/2020MNRAS.497..263K} {497, 263}

\bibitem[\protect\citeauthoryear{{Khadka} \& {Ratra}}{{Khadka} \&
  {Ratra}}{2020c}]{KhadkaRatra2020c}
{Khadka} N.,  {Ratra} B.,  2020c, \mn@doi [\mnras] {10.1093/mnras/staa2779},
  \href {https://ui.adsabs.harvard.edu/abs/2020MNRAS.499..391K} {499, 391}

\bibitem[\protect\citeauthoryear{{Khadka} \& {Ratra}}{{Khadka} \&
  {Ratra}}{2021}]{KhadkaRatra2021}
{Khadka} N.,  {Ratra} B.,  2021, \mn@doi [\mnras] {10.1093/mnras/stab486},
  \href {https://ui.adsabs.harvard.edu/abs/2021MNRAS.502.6140K} {502, 6140}

\bibitem[\protect\citeauthoryear{{Khadka} \& {Ratra}}{{Khadka} \&
  {Ratra}}{2022}]{KhadkaRatra2022}
{Khadka} N.,  {Ratra} B.,  2022, \mn@doi [\mnras] {10.1093/mnras/stab3678},
  \href {https://ui.adsabs.harvard.edu/abs/2021MNRAS.tmp.3383K} {510, 2753}

\bibitem[\protect\citeauthoryear{{Khadka}, {Mart{\'\i}nez-Aldama},
  {Zaja{\v{c}}ek}, {Czerny}  \& {Ratra}}{{Khadka}
  et~al.}{2021a}]{Khadkaetal2021c}
{Khadka} N.,  {Mart{\'\i}nez-Aldama} M.~L.,  {Zaja{\v{c}}ek} M.,  {Czerny} B.,
   {Ratra} B.,  2021a, preprint, \href
  {https://ui.adsabs.harvard.edu/abs/2021arXiv211200052K} {} (\mn@eprint {}
  {2112.00052})

\bibitem[\protect\citeauthoryear{{Khadka}, {Yu}, {Zaja{\v{c}}ek},
  {Martinez-Aldama}, {Czerny}  \& {Ratra}}{{Khadka}
  et~al.}{2021b}]{Khadkaetal_2021a}
{Khadka} N.,  {Yu} Z.,  {Zaja{\v{c}}ek} M.,  {Martinez-Aldama} M.~L.,  {Czerny}
  B.,   {Ratra} B.,  2021b, \mn@doi [\mnras] {10.1093/mnras/stab2807}, \href
  {https://ui.adsabs.harvard.edu/abs/2021MNRAS.508.4722K} {508, 4722}

\bibitem[\protect\citeauthoryear{{Khadka}, {Luongo}, {Muccino}  \&
  {Ratra}}{{Khadka} et~al.}{2021c}]{Khadkaetal_2021b}
{Khadka} N.,  {Luongo} O.,  {Muccino} M.,   {Ratra} B.,  2021c, \mn@doi [\jcap]
  {10.1088/1475-7516/2021/09/042}, \href
  {https://ui.adsabs.harvard.edu/abs/2021JCAP...09..042K} {2021, 042}

\bibitem[\protect\citeauthoryear{{Khetan} et~al.,}{{Khetan}
  et~al.}{2021}]{Khetan_et_al_2021}
{Khetan} N.,  et~al., 2021, \mn@doi [\aap] {10.1051/0004-6361/202039196}, \href
  {https://ui.adsabs.harvard.edu/abs/2021A&A...647A..72K} {647, A72}

\bibitem[\protect\citeauthoryear{{KiDS Collaboration}}{{KiDS
  Collaboration}}{2021}]{KiDSCollaboration2021}
{KiDS Collaboration} 2021, \mn@doi [\aap] {10.1051/0004-6361/202039805}, \href
  {https://ui.adsabs.harvard.edu/abs/2021A&A...649A..88T} {649, A88}

\bibitem[\protect\citeauthoryear{{Kim}, {Kang}, {Lee}  \& {Jang}}{{Kim}
  et~al.}{2020}]{Kimetal_2020}
{Kim} Y.~J.,  {Kang} J.,  {Lee} M.~G.,   {Jang} I.~S.,  2020, \mn@doi [\apj]
  {10.3847/1538-4357/abbd97}, \href
  {https://ui.adsabs.harvard.edu/abs/2020ApJ...905..104K} {905, 104}

\bibitem[\protect\citeauthoryear{{Lewis}}{{Lewis}}{2019}]{Lewis_2019}
{Lewis} A.,  2019, preprint, \href
  {https://ui.adsabs.harvard.edu/abs/2019arXiv191013970L} {} (\mn@eprint
  {arXiv} {1910.13970})

\bibitem[\protect\citeauthoryear{{Li}, {Du}  \& {Xu}}{{Li}
  et~al.}{2020}]{Lietal2020}
{Li} E.-K.,  {Du} M.,   {Xu} L.,  2020, \mn@doi [\mnras]
  {10.1093/mnras/stz3308}, \href
  {https://ui.adsabs.harvard.edu/abs/2020MNRAS.491.4960L} {491, 4960}

\bibitem[\protect\citeauthoryear{{Li}, {Keeley}, {Shafieloo}, {Zheng}, {Cao},
  {Biesiada}  \& {Zhu}}{{Li} et~al.}{2021}]{Lietal2021}
{Li} X.,  {Keeley} R.~E.,  {Shafieloo} A.,  {Zheng} X.,  {Cao} S.,  {Biesiada}
  M.,   {Zhu} Z.-H.,  2021, \mn@doi [\mnras] {10.1093/mnras/stab2154}, \href
  {https://ui.adsabs.harvard.edu/abs/2021MNRAS.507..919L} {507, 919}

\bibitem[\protect\citeauthoryear{{Lian}, {Cao}, {Biesiada}, {Chen}, {Zhang}  \&
  {Guo}}{{Lian} et~al.}{2021}]{Lian_etal_2021}
{Lian} Y.,  {Cao} S.,  {Biesiada} M.,  {Chen} Y.,  {Zhang} Y.,   {Guo} W.,
  2021, \mn@doi [\mnras] {10.1093/mnras/stab1373}, \href
  {https://ui.adsabs.harvard.edu/abs/2021MNRAS.505.2111L} {505, 2111}

\bibitem[\protect\citeauthoryear{{Lin} \& {Ishak}}{{Lin} \&
  {Ishak}}{2021}]{lin_ishak_2021}
{Lin} W.,  {Ishak} M.,  2021, \mn@doi [\jcap] {10.1088/1475-7516/2021/05/009},
  \href {https://ui.adsabs.harvard.edu/abs/2021JCAP...05..009L} {2021, 009}

\bibitem[\protect\citeauthoryear{{Liu}, {Chen}, {Liang}, {Yuan}, {Yu}  \&
  {Wu1}}{{Liu} et~al.}{2022}]{Liuetal2022}
{Liu} Y.,  {Chen} F.,  {Liang} N.,  {Yuan} Z.,  {Yu} H.,   {Wu1} P.,  2022,
  preprint, \href {https://ui.adsabs.harvard.edu/abs/2022arXiv220303178L} {}
  (\mn@eprint {} {2203.03178})

\bibitem[\protect\citeauthoryear{{Luongo} \& {Muccino}}{{Luongo} \&
  {Muccino}}{2021}]{LuongoMuccino2021}
{Luongo} O.,  {Muccino} M.,  2021, \mn@doi [Galaxies]
  {10.3390/galaxies9040077}, \href
  {https://ui.adsabs.harvard.edu/abs/2021Galax...9...77L} {9, 77}

\bibitem[\protect\citeauthoryear{{Luongo}, {Muccino}, {Colg{\'a}in},
  {Sheikh-Jabbari}  \& {Yin}}{{Luongo} et~al.}{2021}]{Luongoetal2021}
{Luongo} O.,  {Muccino} M.,  {Colg{\'a}in} E.~{\'O}.,  {Sheikh-Jabbari} M.~M.,
   {Yin} L.,  2021, preprint, \href
  {https://ui.adsabs.harvard.edu/abs/2021arXiv210813228L} {} (\mn@eprint {}
  {2108.13228})

\bibitem[\protect\citeauthoryear{{Lusso} et~al.,}{{Lusso}
  et~al.}{2020}]{Lussoetal2020}
{Lusso} E.,  et~al., 2020, \mn@doi [\aap] {10.1051/0004-6361/202038899}, \href
  {https://ui.adsabs.harvard.edu/abs/2020A&A...642A.150L} {642, A150}

\bibitem[\protect\citeauthoryear{{Lyu}, {Haridasu}, {Viel}  \& {Xia}}{{Lyu}
  et~al.}{2020}]{Lyu_et_al_2020}
{Lyu} M.-Z.,  {Haridasu} B.~S.,  {Viel} M.,   {Xia} J.-Q.,  2020, \mn@doi
  [\apj] {10.3847/1538-4357/aba756}, \href
  {https://ui.adsabs.harvard.edu/abs/2020ApJ...900..160L} {900, 160}

\bibitem[\protect\citeauthoryear{{Mania} \& {Ratra}}{{Mania} \&
  {Ratra}}{2012}]{Mania_2012}
{Mania} D.,  {Ratra} B.,  2012, \mn@doi [Physics Letters B]
  {10.1016/j.physletb.2012.07.011}, \href
  {https://ui.adsabs.harvard.edu/abs/2012PhLB..715....9M} {715, 9}

\bibitem[\protect\citeauthoryear{{Mehrabi} et~al.,}{{Mehrabi}
  et~al.}{2022}]{Mehrabietal2022}
{Mehrabi} A.,  et~al., 2022, \mn@doi [\mnras] {10.1093/mnras/stab2915}, \href
  {https://ui.adsabs.harvard.edu/abs/2022MNRAS.509..224M} {509, 224}

\bibitem[\protect\citeauthoryear{{Moresco}}{{Moresco}}{2015}]{72}
{Moresco} M.,  2015, \mn@doi [\mnras] {10.1093/mnrasl/slv037}, \href
  {http://adsabs.harvard.edu/abs/2015MNRAS.450L..16M} {450, L16}

\bibitem[\protect\citeauthoryear{{Moresco} et~al.,}{{Moresco}
  et~al.}{2012}]{70}
{Moresco} M.,  et~al., 2012, \mn@doi [\jcap] {10.1088/1475-7516/2012/08/006},
  \href {http://adsabs.harvard.edu/abs/2012JCAP...08..006M} {8, 006}

\bibitem[\protect\citeauthoryear{{Moresco} et~al.,}{{Moresco}
  et~al.}{2016}]{moresco_et_al_2016}
{Moresco} M.,  et~al., 2016, \mn@doi [\jcap] {10.1088/1475-7516/2016/05/014},
  \href {http://adsabs.harvard.edu/abs/2016JCAP...05..014M} {5, 014}

\bibitem[\protect\citeauthoryear{{Mukherjee} \& {Banerjee}}{{Mukherjee} \&
  {Banerjee}}{2022}]{MukherjeeBanerjee2022}
{Mukherjee} P.,  {Banerjee} N.,  2022, \mn@doi [\prd]
  {10.1103/PhysRevD.105.063516}, \href
  {https://ui.adsabs.harvard.edu/abs/2022PhRvD.105f3516M} {105, 063516}

\bibitem[\protect\citeauthoryear{{Neveux} et~al.,}{{Neveux}
  et~al.}{2020}]{eBOSSQ_2020}
{Neveux} R.,  et~al., 2020, \mn@doi [\mnras] {10.1093/mnras/staa2780}, \href
  {https://ui.adsabs.harvard.edu/abs/2020MNRAS.499..210N} {499, 210}

\bibitem[\protect\citeauthoryear{{Ooba}, {Ratra}  \& {Sugiyama}}{{Ooba}
  et~al.}{2018a}]{Oobaetal2018a}
{Ooba} J.,  {Ratra} B.,   {Sugiyama} N.,  2018a, \mn@doi [\apj]
  {10.3847/1538-4357/aad633}, \href
  {https://ui.adsabs.harvard.edu/abs/2018ApJ...864...80O} {864, 80}

\bibitem[\protect\citeauthoryear{{Ooba}, {Ratra}  \& {Sugiyama}}{{Ooba}
  et~al.}{2018b}]{ooba_etal_2018b}
{Ooba} J.,  {Ratra} B.,   {Sugiyama} N.,  2018b, \mn@doi [\apj]
  {10.3847/1538-4357/aadcf3}, \href
  {http://adsabs.harvard.edu/abs/2018ApJ...866...68O} {866, 68}

\bibitem[\protect\citeauthoryear{{Ooba}, {Ratra}  \& {Sugiyama}}{{Ooba}
  et~al.}{2018c}]{Oobaetal2018b}
{Ooba} J.,  {Ratra} B.,   {Sugiyama} N.,  2018c, \mn@doi [\apj]
  {10.3847/1538-4357/aaec6f}, \href
  {https://ui.adsabs.harvard.edu/abs/2018ApJ...869...34O} {869, 34}

\bibitem[\protect\citeauthoryear{{Ooba}, {Ratra}  \& {Sugiyama}}{{Ooba}
  et~al.}{2019}]{ooba_etal_2019}
{Ooba} J.,  {Ratra} B.,   {Sugiyama} N.,  2019, \mn@doi [\apss]
  {10.1007/s10509-019-3663-4}, \href
  {https://ui.adsabs.harvard.edu/abs/2019Ap&SS.364..176O} {364, 176}

\bibitem[\protect\citeauthoryear{{Park} \& {Ratra}}{{Park} \&
  {Ratra}}{2018}]{park_ratra_2018}
{Park} C.-G.,  {Ratra} B.,  2018, \mn@doi [\apj] {10.3847/1538-4357/aae82d},
  \href {http://adsabs.harvard.edu/abs/2018ApJ...868...83P} {868, 83}

\bibitem[\protect\citeauthoryear{{Park} \& {Ratra}}{{Park} \&
  {Ratra}}{2019a}]{ParkRatra2019b}
{Park} C.-G.,  {Ratra} B.,  2019a, \mn@doi [\apss] {10.1007/s10509-019-3567-3},
  \href {https://ui.adsabs.harvard.edu/abs/2019Ap&SS.364...82P} {364, 82}

\bibitem[\protect\citeauthoryear{{Park} \& {Ratra}}{{Park} \&
  {Ratra}}{2019b}]{park_ratra_2019b}
{Park} C.-G.,  {Ratra} B.,  2019b, \mn@doi [\apss] {10.1007/s10509-019-3627-8},
  \href {https://ui.adsabs.harvard.edu/abs/2019Ap&SS.364..134P} {364, 134}

\bibitem[\protect\citeauthoryear{{Park} \& {Ratra}}{{Park} \&
  {Ratra}}{2019c}]{ParkRatra2019a}
{Park} C.-G.,  {Ratra} B.,  2019c, \mn@doi [\apj] {10.3847/1538-4357/ab3641},
  \href {https://ui.adsabs.harvard.edu/abs/2019ApJ...882..158P} {882, 158}

\bibitem[\protect\citeauthoryear{{Park} \& {Ratra}}{{Park} \&
  {Ratra}}{2020}]{park_ratra_2020}
{Park} C.-G.,  {Ratra} B.,  2020, \mn@doi [\prd] {10.1103/PhysRevD.101.083508},
  \href {https://ui.adsabs.harvard.edu/abs/2020PhRvD.101h3508P} {101, 083508}

\bibitem[\protect\citeauthoryear{{Pavlov}, {Westmoreland}, {Saaidi}  \&
  {Ratra}}{{Pavlov} et~al.}{2013}]{pavlov13}
{Pavlov} A.,  {Westmoreland} S.,  {Saaidi} K.,   {Ratra} B.,  2013, \mn@doi
  [\prd] {10.1103/PhysRevD.88.123513}, \href
  {http://adsabs.harvard.edu/abs/2013PhRvD..88l3513P} {88, 123513}

\bibitem[\protect\citeauthoryear{{Peebles}}{{Peebles}}{1984}]{peeb84}
{Peebles} P.~J.~E.,  1984, \mn@doi [\apj] {10.1086/162425}, \href
  {http://adsabs.harvard.edu/abs/1984ApJ...284..439P} {284, 439}

\bibitem[\protect\citeauthoryear{{Peebles} \& {Ratra}}{{Peebles} \&
  {Ratra}}{1988}]{peebrat88}
{Peebles} P.~J.~E.,  {Ratra} B.,  1988, \mn@doi [\apjl] {10.1086/185100}, \href
  {http://adsabs.harvard.edu/abs/1988ApJ...325L..17P} {325, L17}

\bibitem[\protect\citeauthoryear{{Perivolaropoulos} \&
  {Skara}}{{Perivolaropoulos} \& {Skara}}{2021}]{PerivolaropoulosSkara2021}
{Perivolaropoulos} L.,  {Skara} F.,  2021, preprint, \href
  {https://ui.adsabs.harvard.edu/abs/2021arXiv210505208P} {} (\mn@eprint {}
  {2105.05208})

\bibitem[\protect\citeauthoryear{{Philcox}, {Ivanov}, {Simonovi{\'c}}  \&
  {Zaldarriaga}}{{Philcox} et~al.}{2020}]{Philcox_et_al_2020}
{Philcox} O. H.~E.,  {Ivanov} M.~M.,  {Simonovi{\'c}} M.,   {Zaldarriaga} M.,
  2020, \mn@doi [\jcap] {10.1088/1475-7516/2020/05/032}, \href
  {https://ui.adsabs.harvard.edu/abs/2020JCAP...05..032P} {2020, 032}

\bibitem[\protect\citeauthoryear{{Planck Collaboration}}{{Planck
  Collaboration}}{2020}]{planck2018b}
{Planck Collaboration} 2020, \mn@doi [\aap] {10.1051/0004-6361/201833910},
  \href {https://ui.adsabs.harvard.edu/abs/2020A&A...641A...6P} {641, A6}

\bibitem[\protect\citeauthoryear{{Pogosian}, {Zhao}  \& {Jedamzik}}{{Pogosian}
  et~al.}{2020}]{Pogosianetal_2020}
{Pogosian} L.,  {Zhao} G.-B.,   {Jedamzik} K.,  2020, \mn@doi [\apjl]
  {10.3847/2041-8213/abc6a8}, \href
  {https://ui.adsabs.harvard.edu/abs/2020ApJ...904L..17P} {904, L17}

\bibitem[\protect\citeauthoryear{{Rameez} \& {Sarkar}}{{Rameez} \&
  {Sarkar}}{2021}]{rameez_sarkar_2021}
{Rameez} M.,  {Sarkar} S.,  2021, \mn@doi [Classical and Quantum Gravity]
  {10.1088/1361-6382/ac0f39}, \href
  {https://ui.adsabs.harvard.edu/abs/2021CQGra..38o4005R} {38, 154005}

\bibitem[\protect\citeauthoryear{{Rana}, {Jain}, {Mahajan}  \&
  {Mukherjee}}{{Rana} et~al.}{2017}]{Ranaetal2017}
{Rana} A.,  {Jain} D.,  {Mahajan} S.,   {Mukherjee} A.,  2017, \mn@doi [\jcap]
  {10.1088/1475-7516/2017/03/028}, \href
  {https://ui.adsabs.harvard.edu/abs/2017JCAP...03..028R} {2017, 028}

\bibitem[\protect\citeauthoryear{{Ratra} \& {Peebles}}{{Ratra} \&
  {Peebles}}{1988}]{ratpeeb88}
{Ratra} B.,  {Peebles} P.~J.~E.,  1988, \mn@doi [\prd]
  {10.1103/PhysRevD.37.3406}, \href
  {http://adsabs.harvard.edu/abs/1988PhRvD..37.3406R} {37, 3406}

\bibitem[\protect\citeauthoryear{{Ratsimbazafy}, {Loubser}, {Crawford},
  {Cress}, {Bassett}, {Nichol}  \& {V{\"a}is{\"a}nen}}{{Ratsimbazafy}
  et~al.}{2017}]{15}
{Ratsimbazafy} A.~L.,  {Loubser} S.~I.,  {Crawford} S.~M.,  {Cress} C.~M.,
  {Bassett} B.~A.,  {Nichol} R.~C.,   {V{\"a}is{\"a}nen} P.,  2017, \mn@doi
  [\mnras] {10.1093/mnras/stx301}, \href
  {http://adsabs.harvard.edu/abs/2017MNRAS.467.3239R} {467, 3239}

\bibitem[\protect\citeauthoryear{{Renzi}, {Hogg}  \& {Giar{\`e}}}{{Renzi}
  et~al.}{2022}]{Renzietal2021}
{Renzi} F.,  {Hogg} N.~B.,   {Giar{\`e}} W.,  2022, preprint, \href
  {https://ui.adsabs.harvard.edu/abs/2022MNRAS.tmp..996R} {} (\mn@eprint
  {arXiv} {2112.05701})

\bibitem[\protect\citeauthoryear{{Rezaei}, {Sol{\`a} Peracaula}  \&
  {Malekjani}}{{Rezaei} et~al.}{2022}]{Rezaeietal2022}
{Rezaei} M.,  {Sol{\`a} Peracaula} J.,   {Malekjani} M.,  2022, \mn@doi
  [\mnras] {10.1093/mnras/stab3117}, \href
  {https://ui.adsabs.harvard.edu/abs/2022MNRAS.509.2593R} {509, 2593}

\bibitem[\protect\citeauthoryear{{Riess}, {Casertano}, {Yuan}, {Bowers},
  {Macri}, {Zinn}  \& {Scolnic}}{{Riess} et~al.}{2021}]{Riess_2021}
{Riess} A.~G.,  {Casertano} S.,  {Yuan} W.,  {Bowers} J.~B.,  {Macri} L.,
  {Zinn} J.~C.,   {Scolnic} D.,  2021, \mn@doi [\apjl]
  {10.3847/2041-8213/abdbaf}, \href
  {https://ui.adsabs.harvard.edu/abs/2021ApJ...908L...6R} {908, L6}

\bibitem[\protect\citeauthoryear{{Rigault} et~al.,}{{Rigault}
  et~al.}{2015}]{rigault_etal_2015}
{Rigault} M.,  et~al., 2015, \mn@doi [\apj] {10.1088/0004-637X/802/1/20}, \href
  {http://adsabs.harvard.edu/abs/2015ApJ...802...20R} {802, 20}

\bibitem[\protect\citeauthoryear{{Risaliti} \& {Lusso}}{{Risaliti} \&
  {Lusso}}{2015}]{RisalitiLusso2015}
{Risaliti} G.,  {Lusso} E.,  2015, \mn@doi [\apj] {10.1088/0004-637X/815/1/33},
  \href {https://ui.adsabs.harvard.edu/abs/2015ApJ...815...33R} {815, 33}

\bibitem[\protect\citeauthoryear{{Risaliti} \& {Lusso}}{{Risaliti} \&
  {Lusso}}{2019}]{RisalitiLusso2019}
{Risaliti} G.,  {Lusso} E.,  2019, \mn@doi [Nature Astronomy]
  {10.1038/s41550-018-0657-z}, \href
  {https://ui.adsabs.harvard.edu/abs/2019NatAs...3..272R} {3, 272}

\bibitem[\protect\citeauthoryear{{Ryan}, {Doshi}  \& {Ratra}}{{Ryan}
  et~al.}{2018}]{Ryan_1}
{Ryan} J.,  {Doshi} S.,   {Ratra} B.,  2018, \mn@doi [\mnras]
  {10.1093/mnras/sty1922}, \href
  {https://ui.adsabs.harvard.edu/abs/2018MNRAS.480..759R} {480, 759}

\bibitem[\protect\citeauthoryear{{Ryan}, {Chen}  \& {Ratra}}{{Ryan}
  et~al.}{2019}]{Ryanetal2019}
{Ryan} J.,  {Chen} Y.,   {Ratra} B.,  2019, \mn@doi [\mnras]
  {10.1093/mnras/stz1966}, \href
  {https://ui.adsabs.harvard.edu/abs/2019MNRAS.488.3844R} {488, 3844}

\bibitem[\protect\citeauthoryear{{Sangwan}, {Tripathi}  \& {Jassal}}{{Sangwan}
  et~al.}{2018}]{Sangwanetal2018}
{Sangwan} A.,  {Tripathi} A.,   {Jassal} H.~K.,  2018, preprint, \href
  {https://ui.adsabs.harvard.edu/abs/2018arXiv180409350S} {} (\mn@eprint {}
  {1804.09350})

\bibitem[\protect\citeauthoryear{{Sch{\"o}neberg}, {Lesgourgues}  \&
  {Hooper}}{{Sch{\"o}neberg} et~al.}{2019}]{schoneberg_etal_2019}
{Sch{\"o}neberg} N.,  {Lesgourgues} J.,   {Hooper} D.~C.,  2019, \mn@doi
  [\jcap] {10.1088/1475-7516/2019/10/029}, \href
  {https://ui.adsabs.harvard.edu/abs/2019JCAP...10..029S} {2019, 029}

\bibitem[\protect\citeauthoryear{{Scolnic} et~al.,}{{Scolnic}
  et~al.}{2018}]{scolnic_et_al_2018}
{Scolnic} D.~M.,  et~al., 2018, \mn@doi [\apj] {10.3847/1538-4357/aab9bb},
  \href {http://adsabs.harvard.edu/abs/2018ApJ...859..101S} {859, 101}

\bibitem[\protect\citeauthoryear{{Simon}, {Verde}  \& {Jimenez}}{{Simon}
  et~al.}{2005}]{69}
{Simon} J.,  {Verde} L.,   {Jimenez} R.,  2005, \mn@doi [\prd]
  {10.1103/PhysRevD.71.123001}, \href
  {http://adsabs.harvard.edu/abs/2005PhRvD..71l3001S} {71, 123001}

\bibitem[\protect\citeauthoryear{{Singh}, {Sangwan}  \& {Jassal}}{{Singh}
  et~al.}{2019}]{Singhetal2019}
{Singh} A.,  {Sangwan} A.,   {Jassal} H.~K.,  2019, \mn@doi [\jcap]
  {10.1088/1475-7516/2019/04/047}, \href
  {https://ui.adsabs.harvard.edu/abs/2019JCAP...04..047S} {2019, 047}

\bibitem[\protect\citeauthoryear{{Sinha} \& {Banerjee}}{{Sinha} \&
  {Banerjee}}{2021}]{SinhaBanerjee2021}
{Sinha} S.,  {Banerjee} N.,  2021, \mn@doi [\jcap]
  {10.1088/1475-7516/2021/04/060}, \href
  {https://ui.adsabs.harvard.edu/abs/2021JCAP...04..060S} {2021, 060}

\bibitem[\protect\citeauthoryear{{Sol{\`a} Peracaula}, {G{\'o}mez-Valent}  \&
  {de Cruz P{\'e}rez}}{{Sol{\`a} Peracaula}
  et~al.}{2019}]{SolaPercaulaetal2019}
{Sol{\`a} Peracaula} J.,  {G{\'o}mez-Valent} A.,   {de Cruz P{\'e}rez} J.,
  2019, \mn@doi [Physics of the Dark Universe] {10.1016/j.dark.2019.100311},
  \href {https://ui.adsabs.harvard.edu/abs/2019PDU....25..311S} {25, 100311}

\bibitem[\protect\citeauthoryear{{Stern}, {Jimenez}, {Verde}, {Kamionkowski}
  \& {Stanford}}{{Stern} et~al.}{2010}]{71}
{Stern} D.,  {Jimenez} R.,  {Verde} L.,  {Kamionkowski} M.,   {Stanford} S.~A.,
   2010, \mn@doi [\jcap] {10.1088/1475-7516/2010/02/008}, \href
  {http://adsabs.harvard.edu/abs/2010JCAP...02..008S} {2, 008}

\bibitem[\protect\citeauthoryear{{Ure{\~n}a-L{\'o}pez} \&
  {Roy}}{{Ure{\~n}a-L{\'o}pez} \& {Roy}}{2020}]{UrenaLopezRoy2020}
{Ure{\~n}a-L{\'o}pez} L.~A.,  {Roy} N.,  2020, \mn@doi [\prd]
  {10.1103/PhysRevD.102.063510}, \href
  {https://ui.adsabs.harvard.edu/abs/2020PhRvD.102f3510U} {102, 063510}

\bibitem[\protect\citeauthoryear{{Vagnozzi}, {Di Valentino}, {Gariazzo},
  {Melchiorri}, {Mena}  \& {Silk}}{{Vagnozzi} et~al.}{2021a}]{Vagnozzietal2020}
{Vagnozzi} S.,  {Di Valentino} E.,  {Gariazzo} S.,  {Melchiorri} A.,  {Mena}
  O.,   {Silk} J.,  2021a, \mn@doi [Physics of the Dark Universe]
  {10.1016/j.dark.2021.100851}, \href
  {https://ui.adsabs.harvard.edu/abs/2021PDU....3300851V} {33, 100851}

\bibitem[\protect\citeauthoryear{{Vagnozzi}, {Loeb}  \& {Moresco}}{{Vagnozzi}
  et~al.}{2021b}]{Vagnozzietal2021}
{Vagnozzi} S.,  {Loeb} A.,   {Moresco} M.,  2021b, \mn@doi [\apj]
  {10.3847/1538-4357/abd4df}, \href
  {https://ui.adsabs.harvard.edu/abs/2021ApJ...908...84V} {908, 84}

\bibitem[\protect\citeauthoryear{{Wang}, {Wang}, {Cheng}  \& {Dai}}{{Wang}
  et~al.}{2016}]{Wang_2016}
{Wang} J.~S.,  {Wang} F.~Y.,  {Cheng} K.~S.,   {Dai} Z.~G.,  2016, \mn@doi
  [\aap] {10.1051/0004-6361/201526485}, \href
  {https://ui.adsabs.harvard.edu/abs/2016A&A...585A..68W} {585, A68}

\bibitem[\protect\citeauthoryear{{Wang}, {Hu}, {Zhang}  \& {Dai}}{{Wang}
  et~al.}{2022}]{Wangetal_2021}
{Wang} F.~Y.,  {Hu} J.~P.,  {Zhang} G.~Q.,   {Dai} Z.~G.,  2022, \mn@doi [\apj]
  {10.3847/1538-4357/ac3755}, \href
  {https://ui.adsabs.harvard.edu/abs/2022ApJ...924...97W} {924, 97}

\bibitem[\protect\citeauthoryear{{Wei}}{{Wei}}{2018}]{Wei2018}
{Wei} J.-J.,  2018, \mn@doi [\apj] {10.3847/1538-4357/aae696}, \href
  {https://ui.adsabs.harvard.edu/abs/2018ApJ...868...29W} {868, 29}

\bibitem[\protect\citeauthoryear{{Wei} \& {Melia}}{{Wei} \&
  {Melia}}{2022}]{WeiMelia2022}
{Wei} J.-J.,  {Melia} F.,  2022, \mn@doi [\apj] {10.3847/1538-4357/ac562c},
  \href {https://ui.adsabs.harvard.edu/abs/2022ApJ...928..165W} {928, 165}

\bibitem[\protect\citeauthoryear{{Wu}, {Zhang}  \& {Wang}}{{Wu}
  et~al.}{2022}]{Wuetal2022}
{Wu} Q.,  {Zhang} G.-Q.,   {Wang} F.-Y.,  2022, preprint, \href
  {https://ui.adsabs.harvard.edu/abs/2022MNRAS.tmpL..23W} {} (\mn@eprint
  {arXiv} {2108.00581})

\bibitem[\protect\citeauthoryear{{Xu}, {Chen}, {Xu}  \& {Cao}}{{Xu}
  et~al.}{2021}]{Xuetal2021}
{Xu} T.,  {Chen} Y.,  {Xu} L.,   {Cao} S.,  2021, preprint, \href
  {https://ui.adsabs.harvard.edu/abs/2021arXiv210902453X} {} (\mn@eprint {}
  {2109.02453})

\bibitem[\protect\citeauthoryear{{Yang}, {Banerjee}  \& {{\'O}
  Colg{\'a}in}}{{Yang} et~al.}{2020}]{Yangetal2020}
{Yang} T.,  {Banerjee} A.,   {{\'O} Colg{\'a}in} E.,  2020, \mn@doi [\prd]
  {10.1103/PhysRevD.102.123532}, \href
  {https://ui.adsabs.harvard.edu/abs/2020PhRvD.102l3532Y} {102, 123532}

\bibitem[\protect\citeauthoryear{{Yu}, {Ratra}  \& {Wang}}{{Yu}
  et~al.}{2018}]{Yuetal2018}
{Yu} H.,  {Ratra} B.,   {Wang} F.-Y.,  2018, \mn@doi [\apj]
  {10.3847/1538-4357/aab0a2}, \href
  {https://ui.adsabs.harvard.edu/abs/2018ApJ...856....3Y} {856, 3}

\bibitem[\protect\citeauthoryear{{Yu} et~al.,}{{Yu} et~al.}{2021}]{Yuetal2021}
{Yu} Z.,  et~al., 2021, \mn@doi [\mnras] {10.1093/mnras/stab2244}, \href
  {https://ui.adsabs.harvard.edu/abs/2021MNRAS.507.3771Y} {507, 3771}

\bibitem[\protect\citeauthoryear{{Zaja{\v{c}}ek} et~al.,}{{Zaja{\v{c}}ek}
  et~al.}{2021}]{Zajaceketal2021}
{Zaja{\v{c}}ek} M.,  et~al., 2021, \mn@doi [\apj] {10.3847/1538-4357/abe9b2},
  \href {https://ui.adsabs.harvard.edu/abs/2021ApJ...912...10Z} {912, 10}

\bibitem[\protect\citeauthoryear{{Zeng} \& {Yan}}{{Zeng} \&
  {Yan}}{2019}]{zeng_yan_2019}
{Zeng} H.,  {Yan} D.,  2019, \mn@doi [\apj] {10.3847/1538-4357/ab35e3}, \href
  {https://ui.adsabs.harvard.edu/abs/2019ApJ...882...87Z} {882, 87}

\bibitem[\protect\citeauthoryear{{Zhai}, {Blanton}, {Slosar}  \&
  {Tinker}}{{Zhai} et~al.}{2017}]{Zhaietal2017}
{Zhai} Z.,  {Blanton} M.,  {Slosar} A.,   {Tinker} J.,  2017, \mn@doi [\apj]
  {10.3847/1538-4357/aa9888}, \href
  {https://ui.adsabs.harvard.edu/abs/2017ApJ...850..183Z} {850, 183}

\bibitem[\protect\citeauthoryear{{Zhang} \& {Huang}}{{Zhang} \&
  {Huang}}{2021}]{Zhang_Huang_2021}
{Zhang} X.,  {Huang} Q.-G.,  2021, \mn@doi [\prd]
  {10.1103/PhysRevD.103.043513}, \href
  {https://ui.adsabs.harvard.edu/abs/2021PhRvD.103d3513Z} {103, 043513}

\bibitem[\protect\citeauthoryear{{Zhang}, {Zhang}, {Yuan}, {Liu}, {Zhang}  \&
  {Sun}}{{Zhang} et~al.}{2014}]{73}
{Zhang} C.,  {Zhang} H.,  {Yuan} S.,  {Liu} S.,  {Zhang} T.-J.,   {Sun} Y.-C.,
  2014, \mn@doi [Research in Astronomy and Astrophysics]
  {10.1088/1674-4527/14/10/002}, \href
  {http://adsabs.harvard.edu/abs/2014RAA....14.1221Z} {14, 1221}

\bibitem[\protect\citeauthoryear{{Zhang}, {Childress}, {Davis}, {Karpenka},
  {Lidman}, {Schmidt}  \& {Smith}}{{Zhang} et~al.}{2017}]{zhangetal2017}
{Zhang} B.~R.,  {Childress} M.~J.,  {Davis} T.~M.,  {Karpenka} N.~V.,  {Lidman}
  C.,  {Schmidt} B.~P.,   {Smith} M.,  2017, \mn@doi [\mnras]
  {10.1093/mnras/stx1600}, \href
  {http://adsabs.harvard.edu/abs/2017MNRAS.471.2254Z} {471, 2254}

\bibitem[\protect\citeauthoryear{{Zhao} \& {Xia}}{{Zhao} \&
  {Xia}}{2021}]{ZhaoXia2021}
{Zhao} D.,  {Xia} J.-Q.,  2021, \mn@doi [European Physical Journal C]
  {10.1140/epjc/s10052-021-09491-0}, \href
  {https://ui.adsabs.harvard.edu/abs/2021EPJC...81..694Z} {81, 694}

\bibitem[\protect\citeauthoryear{{Zheng}, {Cao}, {Biesiada}, {Li}, {Liu}  \&
  {Liu}}{{Zheng} et~al.}{2021}]{Zhengetal2021}
{Zheng} X.,  {Cao} S.,  {Biesiada} M.,  {Li} X.,  {Liu} T.,   {Liu} Y.,  2021,
  \mn@doi [Science China Physics, Mechanics, and Astronomy]
  {10.1007/s11433-020-1664-9}, \href
  {https://ui.adsabs.harvard.edu/abs/2021SCPMA..6459511Z} {64, 259511}

\makeatother
\end{thebibliography}

%\begin{thebibliography}{99}
%\end{thebibliography}

%%%%%%%%%%%%%%%%%%%%%%%%%%%%%%%%%%%%%%%%%%%%%%%%%%

%%%%%%%%%%%%%%%%% APPENDICES %%%%%%%%%%%%%%%%%%%%%

%\appendix

% If you want to present additional material which would interrupt the flow of the main paper,
% it can be placed in an Appendix which appears after the list of references.

%%%%%%%%%%%%%%%%%%%%%%%%%%%%%%%%%%%%%%%%%%%%%%%%%%

% Don't change these lines
\bsp	% typesetting comment
\label{lastpage}
\end{document}